\documentclass[lettersize,journal]{IEEEtran}
\makeatletter
\long\def\@makecaption#1#2{\ifx\@captype\@IEEEtablestring%
\footnotesize\begin{center}{\normalfont\footnotesize #1}\\
{\normalfont\footnotesize\scshape #2}\end{center}%
\@IEEEtablecaptionsepspace
\else
\@IEEEfigurecaptionsepspace
\setbox\@tempboxa\hbox{\normalfont\footnotesize {#1.}~~ #2}%
\ifdim \wd\@tempboxa >\hsize%
\setbox\@tempboxa\hbox{\normalfont\footnotesize {#1.}~~ }%
\parbox[t]{\hsize}{\normalfont\footnotesize \noindent\unhbox\@tempboxa#2}%
\else
\hbox to\hsize{\normalfont\footnotesize\hfil\box\@tempboxa\hfil}\fi\fi}
\makeatother
\makeatletter
\newcommand\notsotiny{\@setfontsize\notsotiny{6.31415}{7.1828}}
\makeatother
\usepackage{amsmath,amsfonts}
\usepackage{algorithmic}
\usepackage{algorithm}
\usepackage{array}
\usepackage[caption=false]{subfig}
\usepackage{textcomp}
\usepackage{stfloats}
\usepackage{url,soul}
\usepackage[hidelinks]{hyperref}
\usepackage{verbatim}
\usepackage{graphicx}
\usepackage[comma,sort&compress,numbers]{natbib}
\usepackage[dvipsnames]{xcolor}
\usepackage{booktabs}
\usepackage[normalem]{ulem}
\useunder{\uline}{\ul}{}

\begin{document}

\title{BVI-VFI: A Video Quality Database for Video Frame Interpolation}

\author{Duolikun Danier,~\IEEEmembership{Student Member,~IEEE,} Fan Zhang,~\IEEEmembership{Member,~IEEE,} and David R. Bull,~\IEEEmembership{Fellow,~IEEE}

\thanks{This work involved collecting data from human participants. The relevant experiments have been approved by the Faculty of Engineering Research Ethics Committee of the University of Bristol (Ref 10739; PI: David R. Bull; Title: Subjective Quality Study on Video Frame Interpolation).}
\thanks{The authors are with the Bristol Vision Institute, University of Bristol, Bristol BS8 1TH, U.K. (e-mail:duolikun.danier@bristol.ac.uk; fan.zhang@bristol.ac.uk; dave.bull@bristol.ac.uk).}
\thanks{The authors acknowledge the funding from China Scholarship Council, the University of Bristol and the
UKRI MyWorld Strength in Places Programme.}
}

\maketitle

\begin{abstract}

Video frame interpolation (VFI) is a fundamental research topic in video processing, which is currently attracting increased attention across the research community. While the development of more advanced VFI algorithms has been extensively researched, there remains little understanding of how humans perceive the quality of interpolated content and how well existing objective quality assessment methods perform when measuring the perceived quality. In order to narrow this research gap, we have developed a new video quality database named BVI-VFI, which contains 540 distorted sequences generated by applying five commonly used VFI algorithms to 36 diverse source videos with various spatial resolutions and frame rates. We collected more than 10,800 quality ratings for these videos through a large scale subjective study involving 189 human subjects. Based on the collected subjective scores, we further analysed the influence of VFI algorithms and frame rates on the perceptual quality of interpolated videos. Moreover, we benchmarked the performance of 33 classic and state-of-the-art objective image/video quality metrics on the new database, and demonstrated the urgent requirement for more accurate bespoke quality assessment methods for VFI. To facilitate further research in this area, we have made BVI-VFI publicly available at \url{https://github.com/danier97/BVI-VFI-database}.
\end{abstract}


\section{Introduction}\label{sec:intro}

Video frame interpolation (VFI) is an important video processing technique which is used to synthesise intermediate frames between every two consecutive frames in a video sequence. {Conventionally known as motion interpolation, VFI has been employed to increase the frame rates of content captured at low frame rates, and to compensate for motion blur in LCD displays~\cite{bao2018high}. Although the ``soap opera effect'' (the unnaturalness in perception caused by the extra non-real frames being displayed) can arise as a byproduct of such a process, various previous studies~\cite{watson2013high, emoto2014high, allison2016expert, wilcox2015evidence, tag2016eye, mackin2018study, nasiri2017perceptual} have confirmed that high-frame-rate (HFR) formats provide improved viewing experience in terms of perceived motion smoothness, perceived realism, and immersiveness. Since VFI enables HFR formats to be generated from low-frame-rate videos, it offers significant promise for improving perceptual video quality.} VFI {also} offers utility, and has been the subject of increased popularity, across many applications {beyond HFR format generation} in recent years; these include slow motion generation~\cite{jiang2018super}, video compression~\cite{usman2016frame}, medical imaging~\cite{karani2017temporal} and animation production~\cite{siyao2021deep}. 

{Other than frame repetition and averaging, early VFI methods employed in televisions and other display devices were mainly based on motion estimation and compensation, where motion vectors between frames were used to interpolate the intermediate pixels~\cite{bao2018high}. Recently,} driven by the development of various deep learning techniques and more powerful computational hardware, there has been a surge in the reporting of new video frame interpolation methods, which are generally classified into two groups: flow- and kernel-based. While flow-based methods rely on optical flow to warp reference frames, kernel-based approaches estimate local interpolation kernels to synthesise output pixels. Although these VFI approaches have delivered significant improvements in terms of interpolation performance~\cite{xu2019quadratic, danier2022spatio, lu2022video, argaw2022long, hu2022many, liu2022jnmr}, challenging scenarios still exist that cause interpolation failure; these often relate to content containing  large motions, dynamic textures, and occlusions~\cite{danier2022spatio}. 

While there is ongoing activity to develop new VFI methods that tackle these challenges, the perceptual quality assessment of frame interpolated content remains underinvestigated. Currently, the most widely adopted approach for assessing VFI performance is to calculate the distortion between the interpolated frames and their original ground-truth counterparts using image quality assessment (IQA) models including PSNR, SSIM~\cite{wang2004image}, and LPIPS~\cite{zhang2018unreasonable}. More recently, new perceptually oriented image and video quality metrics have been developed for other applications such as video compression, with notable examples including VSI~\cite{zhang2014vsi}, DISTS~\cite{ding2020image}, VMAF~\cite{li2016toward}, FAST~\cite{wu2019quality} and C3DVQA~\cite{xu2020c3dvqa}. However, none of these models has been fully evaluated on frame interpolated videos against subjective ground truth. Due to this concern, in order to accurately assess VFI performance, some researchers~\cite{kalluri2020flavr, danier2022spatio} have resorted to benchmarking based on subjective opinion scores through psychophysical experiments; these however are very time consuming and resource-heavy. In this context, there is an urgent need to develop a video quality database containing diverse VFI content alongside reliable subjective score metadata, which can be employed to investigate the competence of existing quality metrics for the VFI task.

Although there have been little reports of research in this area~\cite{men2019visual, men2020visual}, the associated databases either contain human opinion scores only on single interpolated images, or only focus on slow-motion videos at a fixed frame rate. Also, the video sequences in \cite{men2020visual} suffer from compression artefacts in addition to VFI-related distortions,  making it difficult to decouple these during assessment. We have previously addressed these issues in \cite{danier2022subjective}, where a small video quality database for VFI was developed based on a limited subjective study and a benchmark experiment only involving a few objective quality metrics. To overcome these limits and make further progress in understanding the perceptual quality of frame interpolated videos, in this paper we extend our previous work~\cite{danier2022subjective} and present a new video quality database, BVI-VFI, which contains 540 interpolated videos generated by various VFI algorithms, covering different frame rates, spatial resolutions and diverse content types. The database also includes subjective quality scores for all videos collected through a large-scale subjective experiment. Based on the subjective data, we performed a much more comprehensive evaluation of existing objective quality metrics, involving {33} conventional and learning-based image and video quality models. {This work differs from our previous work~\cite{danier2022subjective} in the following aspects.}
{
\begin{itemize}
    \item Compared to the original 36 reference and 180 distorted sequences, the new BVI-VFI database contains 108 reference and 540 distorted sequences.
    \item While the study in \cite{danier2022subjective} concerns only HD videos, in this work we cover three resolutions: $960\times 540$, $1920\times 1080$, and $3840\times 2160$.
    \item The study on the subjective data in \cite{danier2022subjective} was limited to the effect of frame rate. In this work we additionally analyse the impact of various video features on the perceived quality of frame interpolated videos.
    \item While in \cite{danier2022subjective} only eight full-reference quality metrics were benchmarked, in this work we evaluate 33 image/video quality metrics, covering both full- and no-reference categories. Additionally, we perform cross-validation experiments to better reflect the performance of learning-based metrics.
\end{itemize}
}

The primary contributions are summarised below. 

\begin{itemize}
    \item We developed the first bespoke video quality database for frame interpolation, BVI-VFI, that covers multiple frame rates (30-120fps) and spatial resolutions (540p-2160p). It contains 540 distorted sequences generated by five different video frame interpolation methods from 36 source videos, which uniformly cover a wide range of video features.
    \item We conducted a large-scale laboratory-based psychophysical experiment to collect subjective quality ground truths for all the videos in BVI-VFI.
    \item We performed quantitative comparison of 33 classic and state-of-the-art image/video quality assessment methods on the BVI-VFI database. Cross-validation experiments were also performed for learning-based metrics.
    \item The proposed new database serves as an important platform for developing and validating new quality metrics for video frame interpolation. It can also be used as a test dataset for benchmarking VFI algorithms due to its content diversity.
\end{itemize}

The rest of the paper is organised as follows. We first briefly review the relevant literature in Section~\ref{sec:relatedwork}, and then describe the process of source video collection and test sequence generation in Section~\ref{sec:database}. The subjective experiment, the data processing procedures and analysis of the collected subjective opinions are presented in Section~\ref{sec:subjective}. Section~\ref{sec:objective} summarises the comparative study results for 33 quality assessment methods. Finally, Section~\ref{sec:conclusion} draws conclusions and outlines potential future work.

\section{Related Work}\label{sec:relatedwork}

In this section, we first describe previous works in video frame interpolation, and then summarise the related research work on objective video quality metrics and subjective quality assessment in the context of VFI.

\subsection{Video Frame Interpolation}

Early attempts~\cite{barron1994performance, baker2011database} to perform video frame interpolation typically used estimated optical flow maps to warp input frames. This paradigm, referred to as flow-based VFI, was further developed in the learning-based VFI literature~\cite{park2020bmbc}. These methods adopted various techniques to enhance interpolation quality, including the use of contextual information~\cite{niklaus2018context}, designing bespoke flow estimation module~\cite{liu2017video, jiang2018super, xue2019video, park2020bmbc, huang2020rife, park2021asymmetric, argaw2022long, lu2022video}, employing a coarse-to-fine refinement strategy~\cite{zhang2020flexible, sim2021xvfi, kong2022ifrnet, reda2022film}, developing new warping operations~\cite{niklaus2020softmax, niklaus2022splatting, hu2022many} and adopting higher-order motion modelling with additional input frames~\cite{xu2019quadratic, liu2020enhanced, liu2022jnmr}. Some researchers argue that the imposition of a one-to-one mapping between the target and source pixels can limit the ability of flow-based methods to handle complex motions. This has led to the development  of kernel-based methods~\cite{adaconv, niklaus2017video, lee2020adacof, gui2020featureflow, shi2020video, ding2021cdfi, cheng2021multiple, chen2021pdwn, danier2022enhancing, shi2022video} that predict adaptive local interpolation kernels to synthesis the output pixels. This creates a many-to-one mapping between the source and target pixels, supporting additional degrees of freedom. Moreover, other researchers reported the limitations of predicting fixed-shaped kernels~\cite{adaconv, niklaus2017video}, and introduced deformable kernels~\cite{dai2017deformable} to achieve improved interpolation performance. Finally, observing that fixed kernel sizes can limit the captured motion magnitude, some VFI methods~\cite{bao2019depth, bao2019memc, danier2022spatio} combine flow-based and kernel-based approaches in a single framework to benefit from both model types.

Besides the flow-/kernel-based classes, other VFI paradigms exist, for example based on pixel hallucination~\cite{choi2020channel, kalluri2020flavr}, phase information~\cite{meyer2015phase, meyer2018phasenet}, event cameras~\cite{tulyakov2021time, tulyakov2022time, he2022timereplayer}, unsupervised learning~\cite{reda2019unsupervised, cheng2022unsupervised}, and meta-learning~\cite{choi2021test}. More recently, the joint problem of deblurring and interpolation has also been addressed in \cite{shen2020blurry, zhang2020video}. 

\subsection{Objective Quality Assessment for VFI}

In the current VFI literature, the commonly adopted approach for benchmarking interpolation performance is to measure the distortion between an interpolated video and its ground-truth version. The most popular methods are PSNR, SSIM~\cite{wang2004image} and LPIPS~\cite{zhang2018unreasonable}, all of which are applied at the level of a single image or frame. There are also many image/video quality metrics developed for other applications, including approaches based on classic signal processing methods, e.g., MS-SSIM~\cite{wang2003multiscale}, VIF~\cite{sheikh2005information}, VSI~\cite{zhang2014vsi}, FAST~\cite{wu2019quality}, SpEED~\cite{bampis2017speed}, VIQE~\cite{zheng2022completely} and ST-RRED~\cite{soundararajan2012video}. More recently, machine learning techniques have been employed in the development of perceptual metrics including VMAF~\cite{li2016toward}, C3DVQA~\cite{xu2020c3dvqa} and CONTRIQUE~\cite{madhusudana2022image}. Alongside these generic quality metrics, assessment methods that were designed to specifically model the effect of frame rate/spatial resolution down-sampling or frame interpolation have been reported, including FRQM~\cite{zhang2017frame}, ST-GREED~\cite{madhusudana2021st}, VSTR~\cite{lee2021space}, FAVER~\cite{zheng2022faver} and FloLPIPS. However, none of these methods have been rigorously benchmarked due to the lack of databases with diverse content and ground-truth metadata.

\subsection{Subjective Quality Assessment for VFI}
Various subjective quality databases exist that support studies into how the human vision system (HVS) perceives video quality. These include those developed in the context of video compression, e.g., the early VQEG FR-TV Phase I~\cite{antkowiak2000final}, LIVE VQA~\cite{seshadrinathan2010study}, LIVE Mobile~\cite{moorthy2012video}, CSIQ-VQA~\cite{vu2014vis3}, BVI-HD~\cite{zhang2018bvi}, LIVE-SJTU~\cite{min2020study}, TVG~\cite{9746547}, CSCVQ~\cite{li2020subjective} and LIVE Livestream~\cite{shang2021study}. Video quality databases also exist that support investigations into distortions due to video parameter variations (e.g, frame rate, spatial resolution and bit depth, with or without video compression artefacts), including MCL-V~\cite{lin2015mcl}, MCML-4K-UHD~\cite{cheon2017subjective}, BVI-SR~\cite{mackin2018study}, BVI-HFR~\cite{mackin2018study}, BVI-BD~\cite{mackin2021subjective}, FRD-VQA~\cite{huang2016perceptual}, LIVE-YT-HFR~\cite{madhusudana2021subjective}, AVT-VQDB-UHD-1~\cite{rao2019avt} and ETRI-LIVE STSVQ~\cite{lee2021subjective}. 

There are, however, very few examples of databases that contain content distorted by different VFI methods; the most relevant contributions being \cite{men2019visual} and \cite{men2020visual}. In \cite{men2019visual}, subjective quality scores were collected by showing viewers individual interpolated frames instead of video sequences. This methodology is limited since it does not consider temporal artefacts which could significantly influence the perceived video quality~\cite{bull2021intelligent}. The later database, KosMo-1k~\cite{men2020visual}, addresses this issue by collecting subjective opinions when viewing interpolated videos. However, the video sequences were played in slow-motion (one specific use case of VFI) and, in addition,  all the distorted sequences were contaminated by video compression artefacts, making it difficult for subjects to isolate VFI artefacts. Due to their disadvantages, none of these databases can be recommended for  evaluating the performance of VFI quality metrics. Hence there  is an urgent requirement for a bespoke video quality database. 

\section{The BVI-VFI Database}\label{sec:database}

This section describes the approach used to select the source sequences in the BVI-VFI database, and how the test videos were generated. 

\subsection{Reference Sequences}
 
\begin{figure*}[t]
\renewcommand*\thesubfigure{\arabic{subfigure}} 
    \centering
    \subfloat[]{\includegraphics[width=0.110\linewidth]{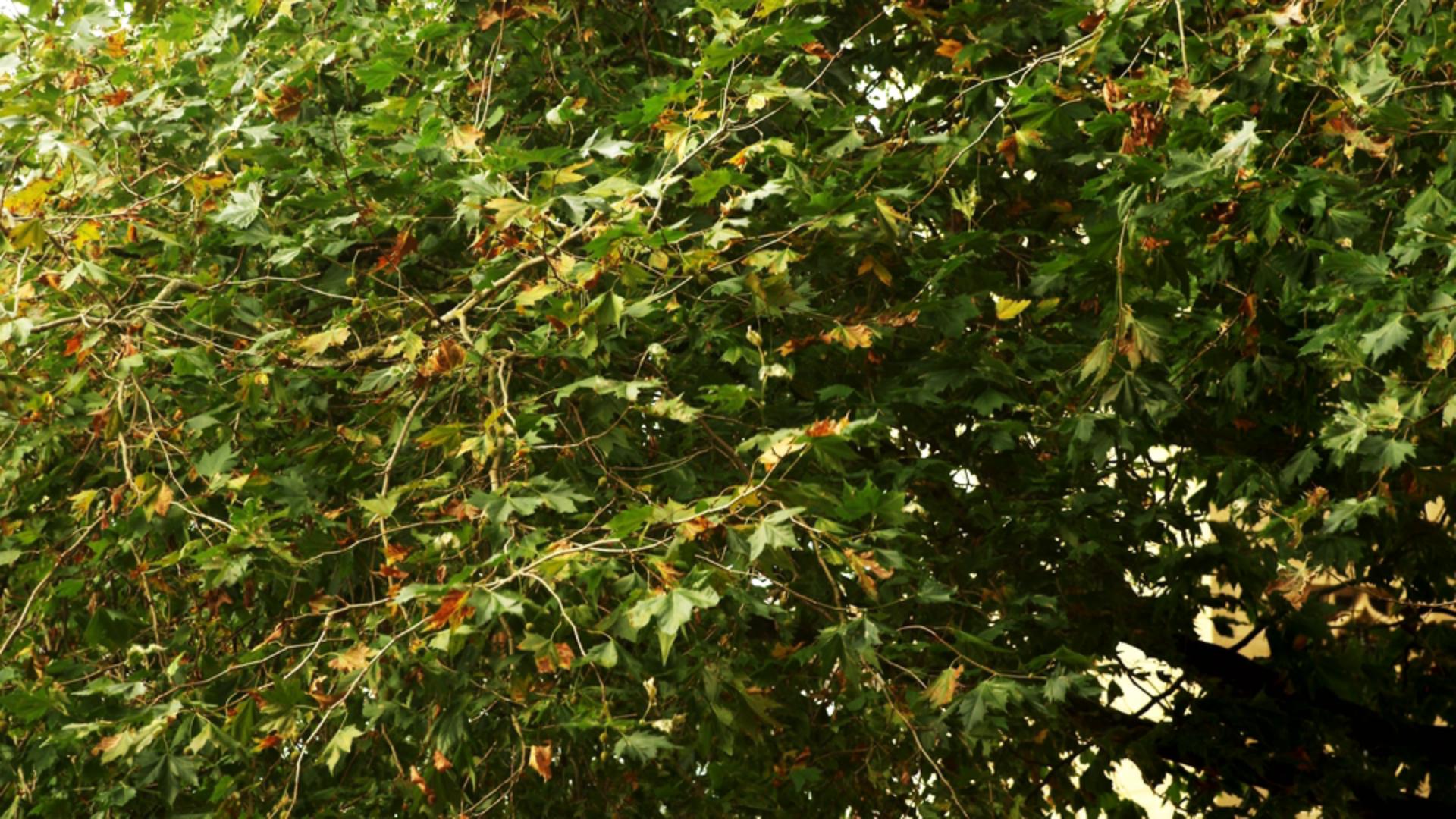}}\;\!\!\!\!
    \subfloat[]{\includegraphics[width=0.110\linewidth]{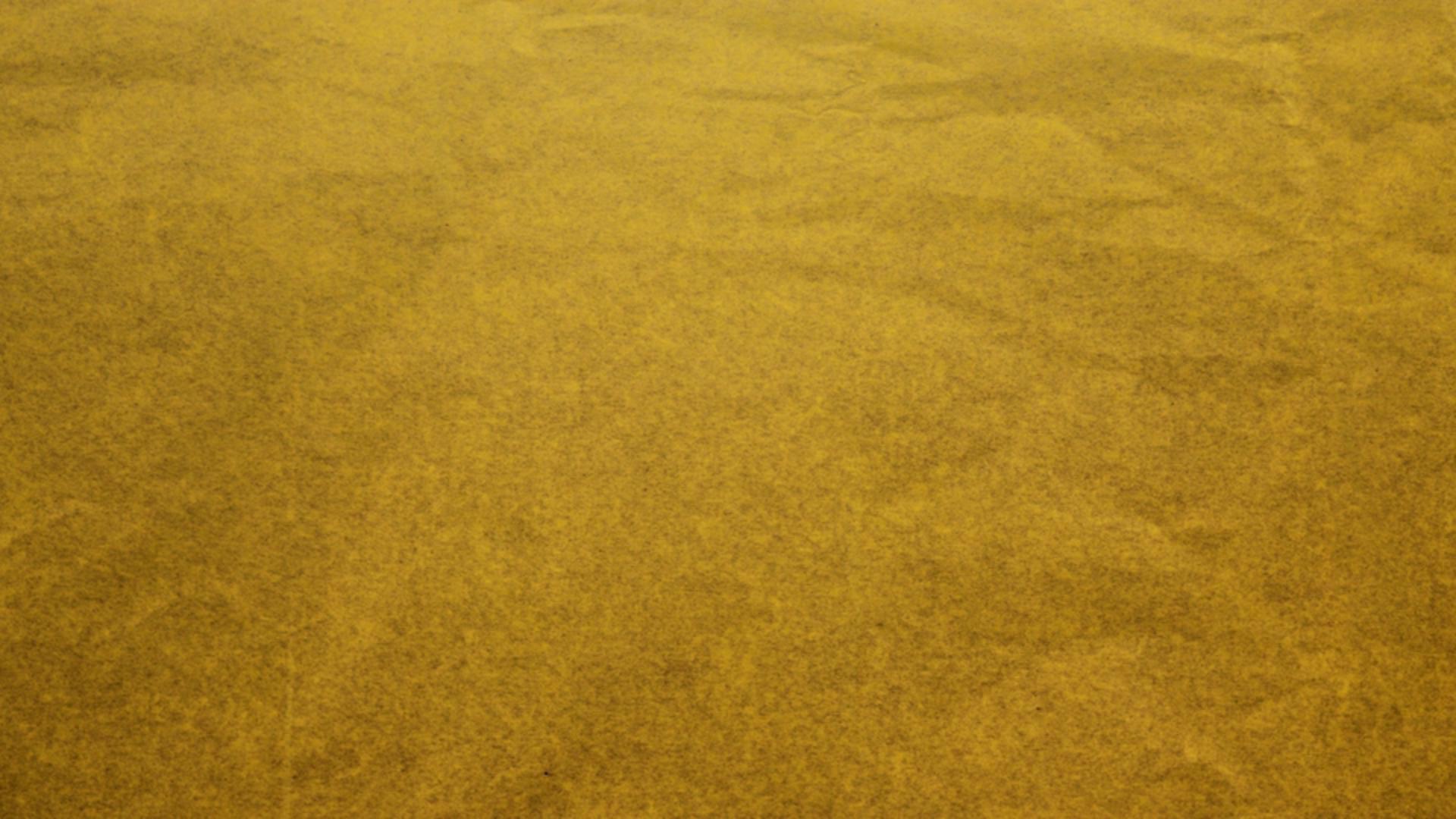}}\;\!\!\!\!
    \subfloat[]{\includegraphics[width=0.110\linewidth]{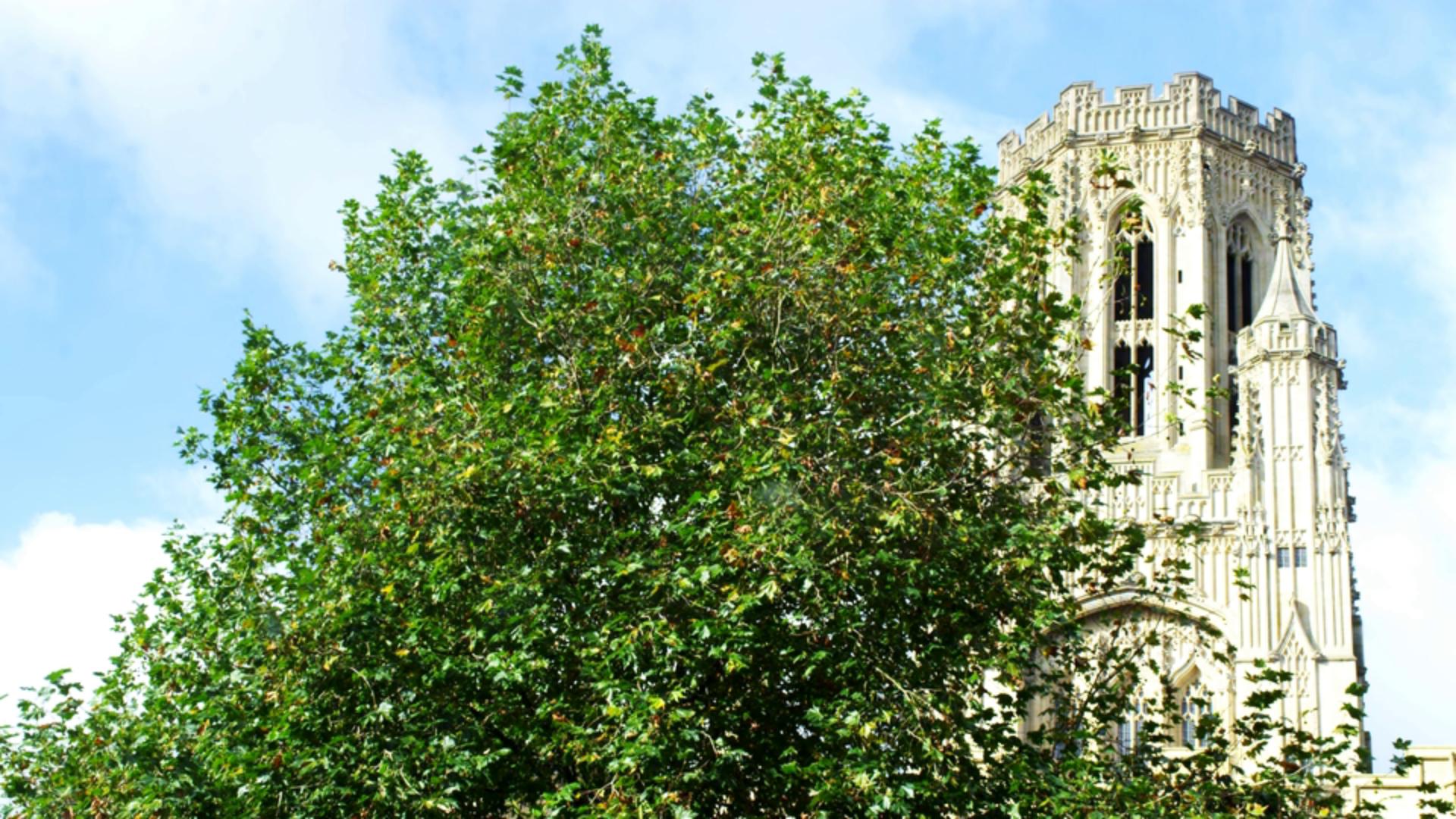}}\;\!\!\!\!
    \subfloat[]{\includegraphics[width=0.110\linewidth]{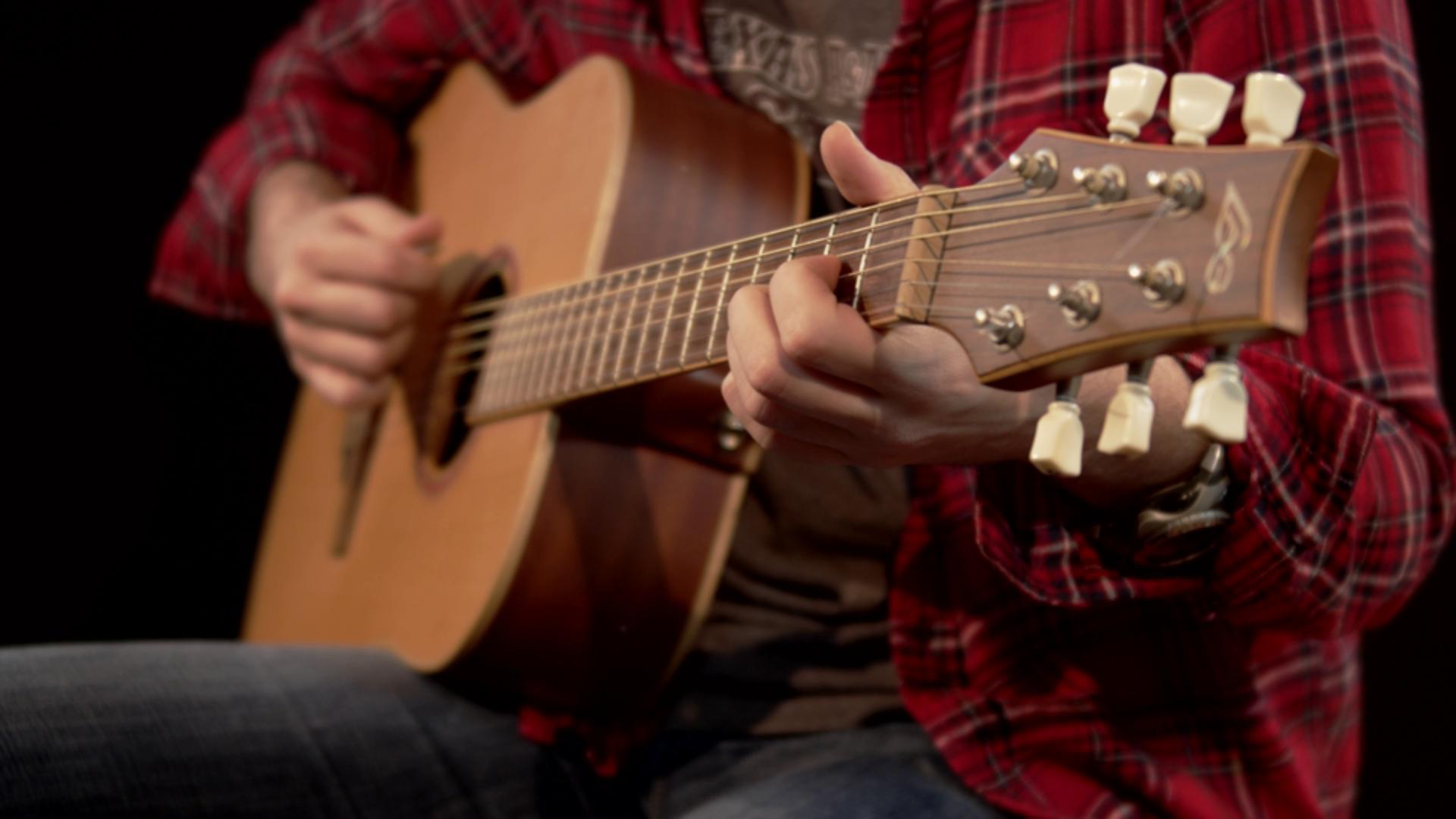}}\;\!\!\!\!
    \subfloat[]{\includegraphics[width=0.110\linewidth]{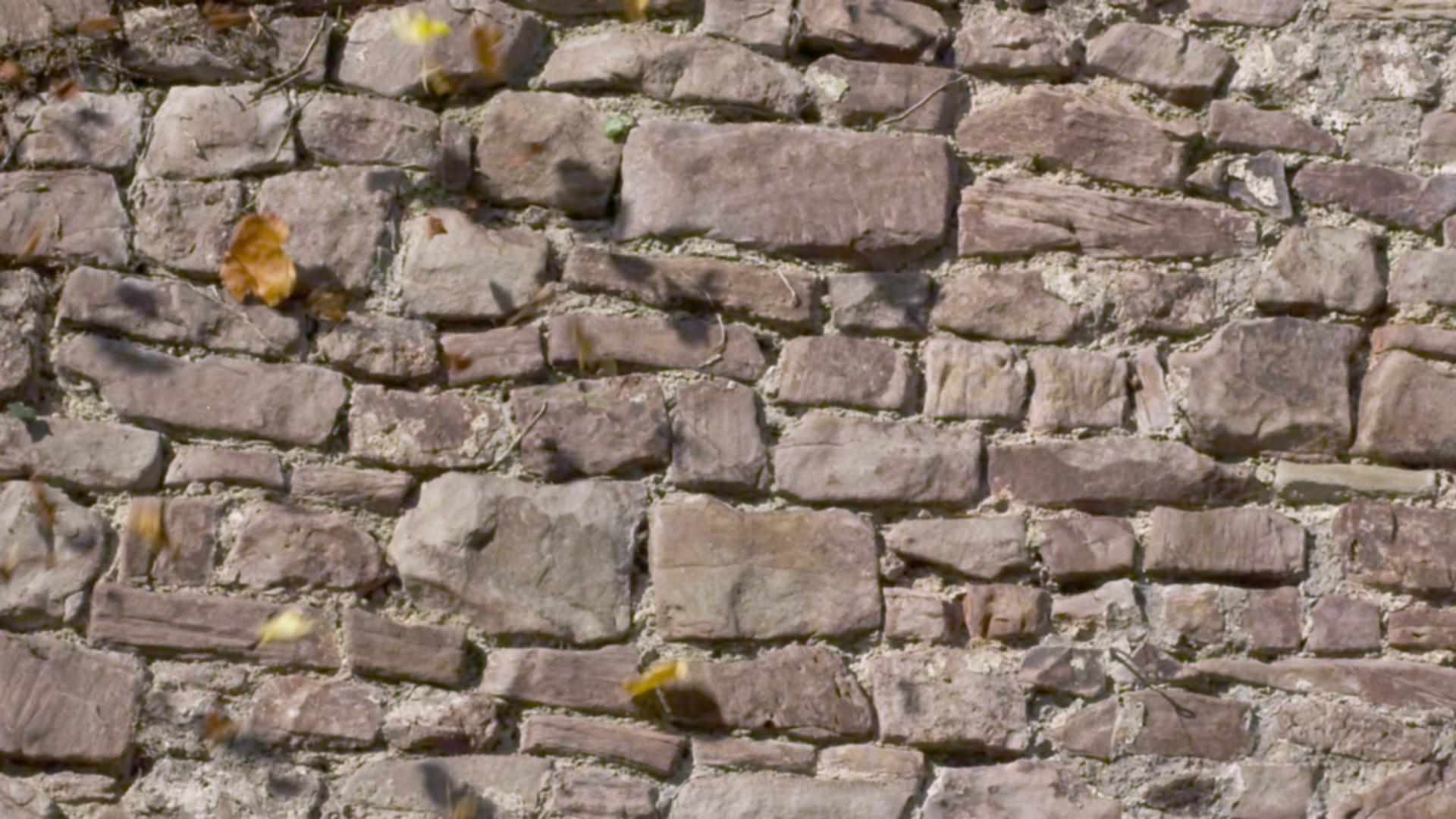}}\;\!\!\!\!
    \subfloat[]{\includegraphics[width=0.110\linewidth]{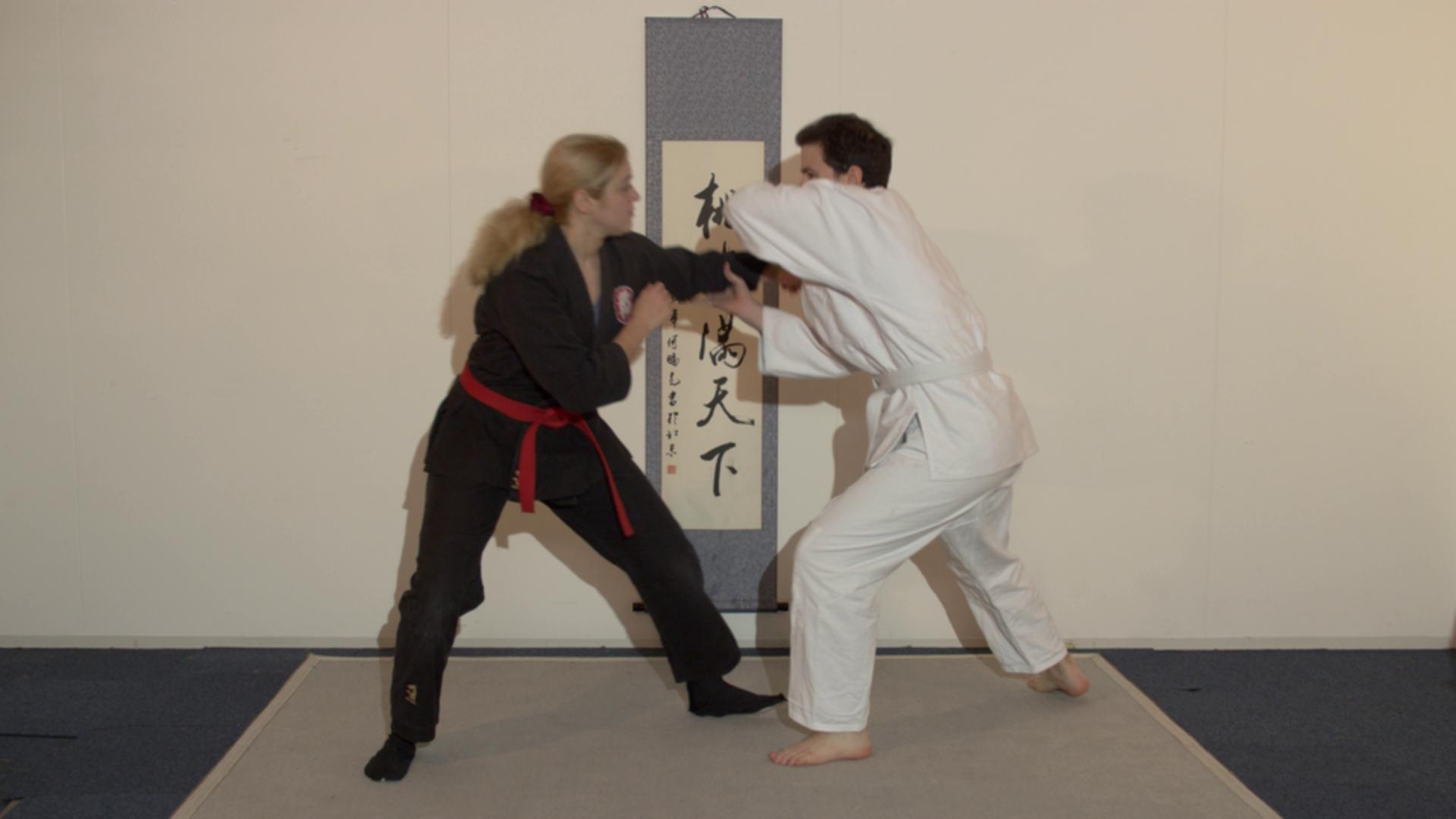}}\;\!\!\!\!
    \subfloat[]{\includegraphics[width=0.110\linewidth]{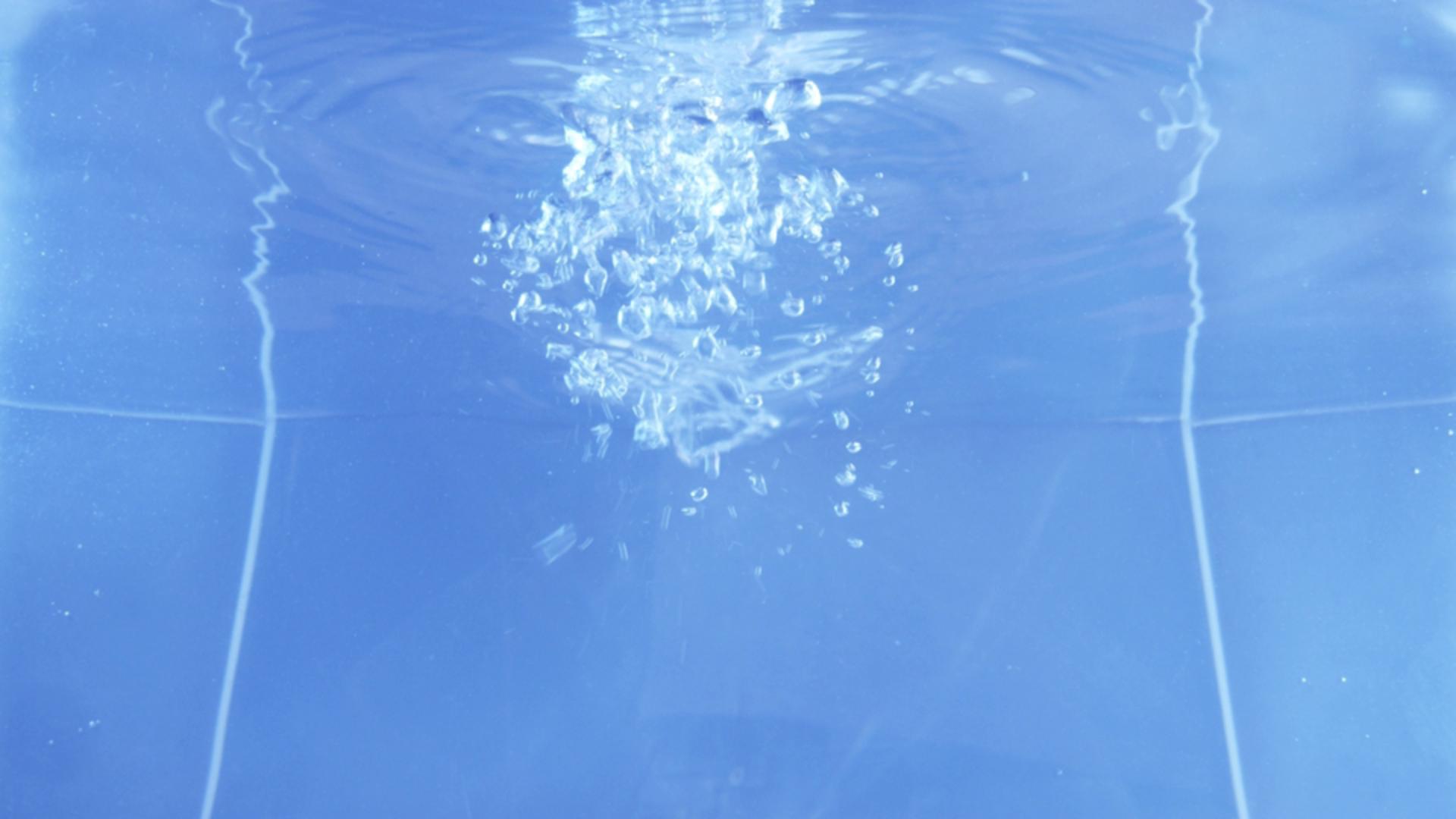}}\;\!\!\!\!
    \subfloat[]{\includegraphics[width=0.110\linewidth]{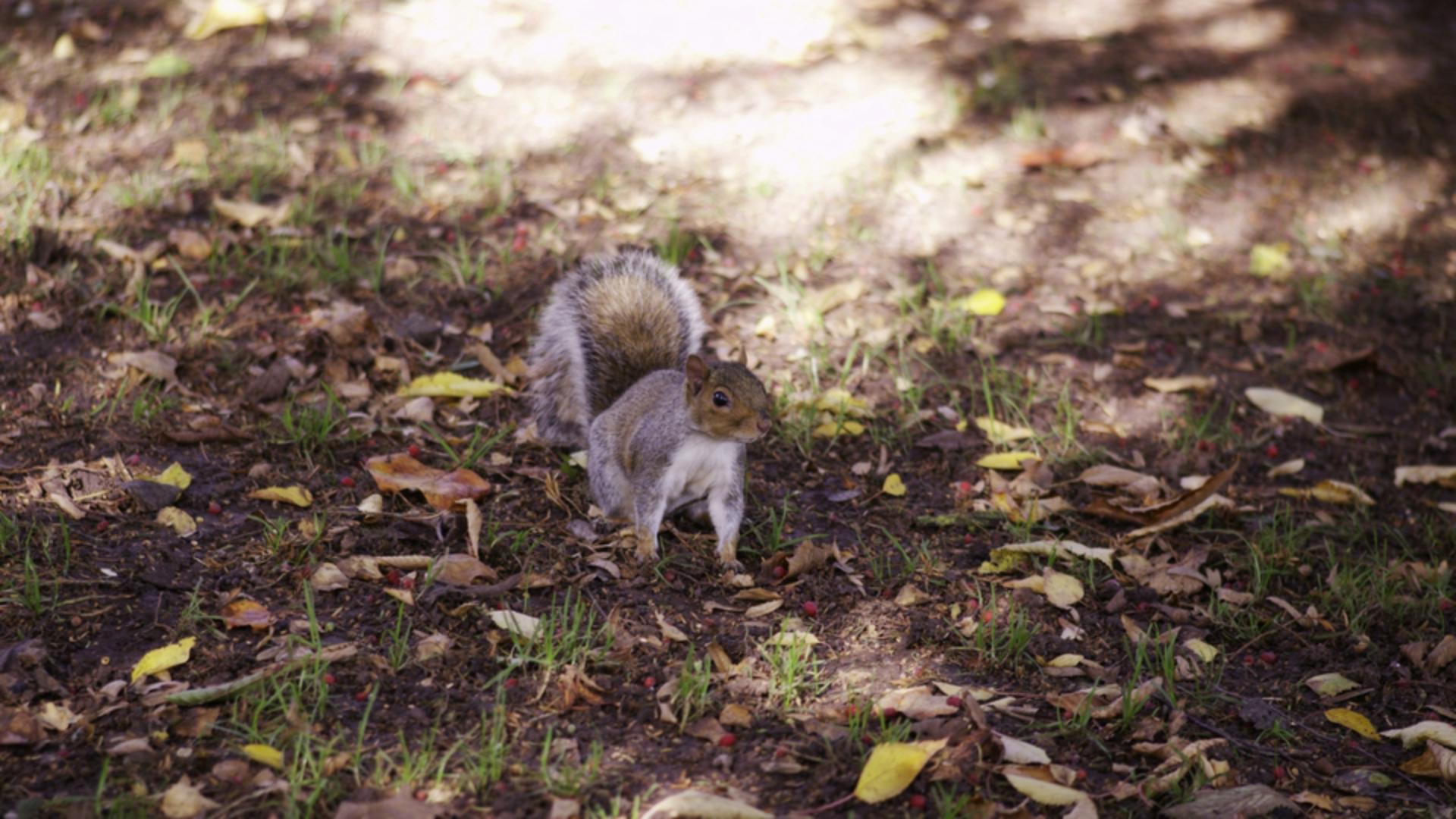}}\;\!\!\!\!
    \subfloat[]{\includegraphics[width=0.110\linewidth]{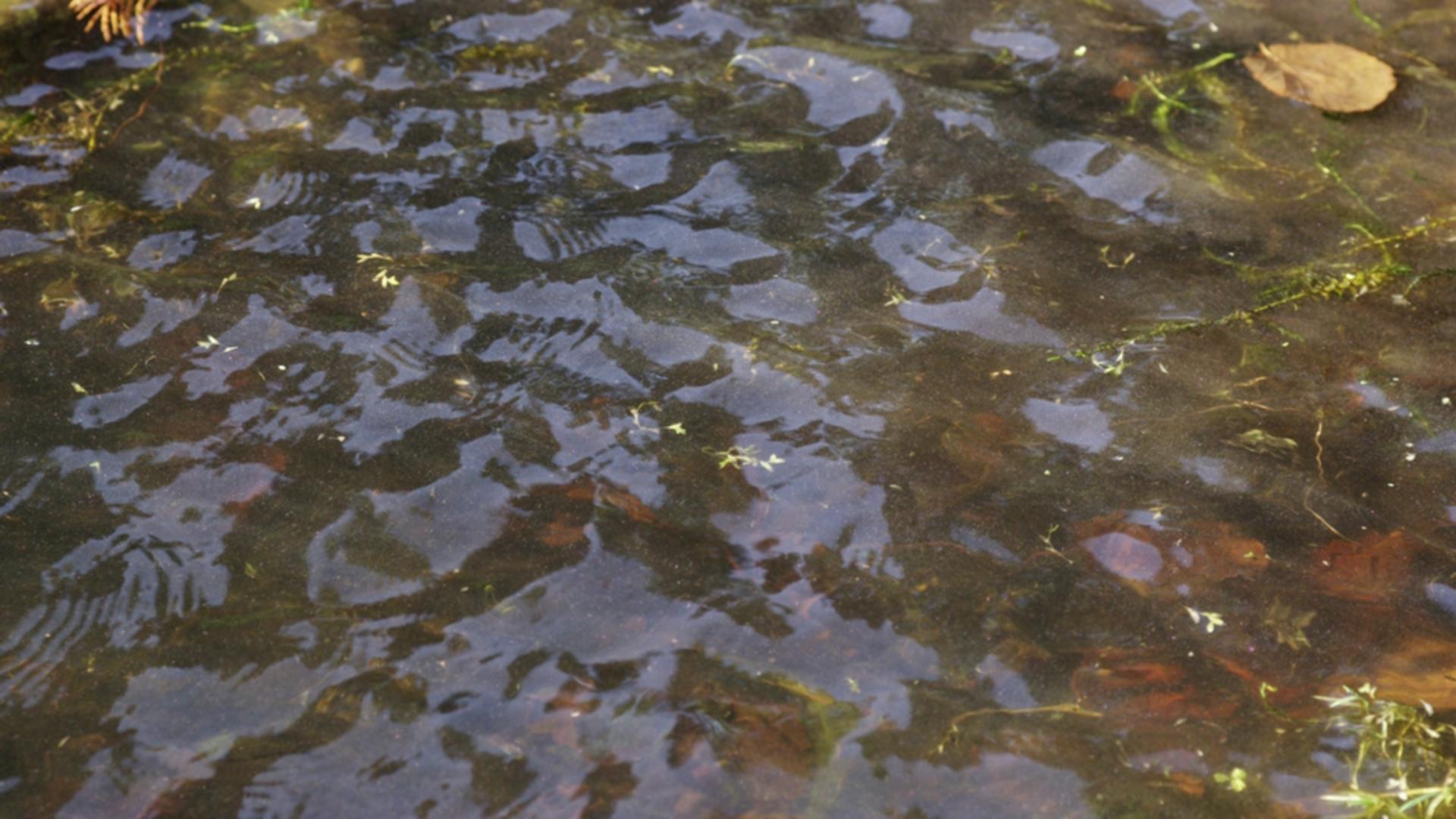}}\\
    \vspace{-3mm}
    \subfloat[]{\includegraphics[width=0.110\linewidth]{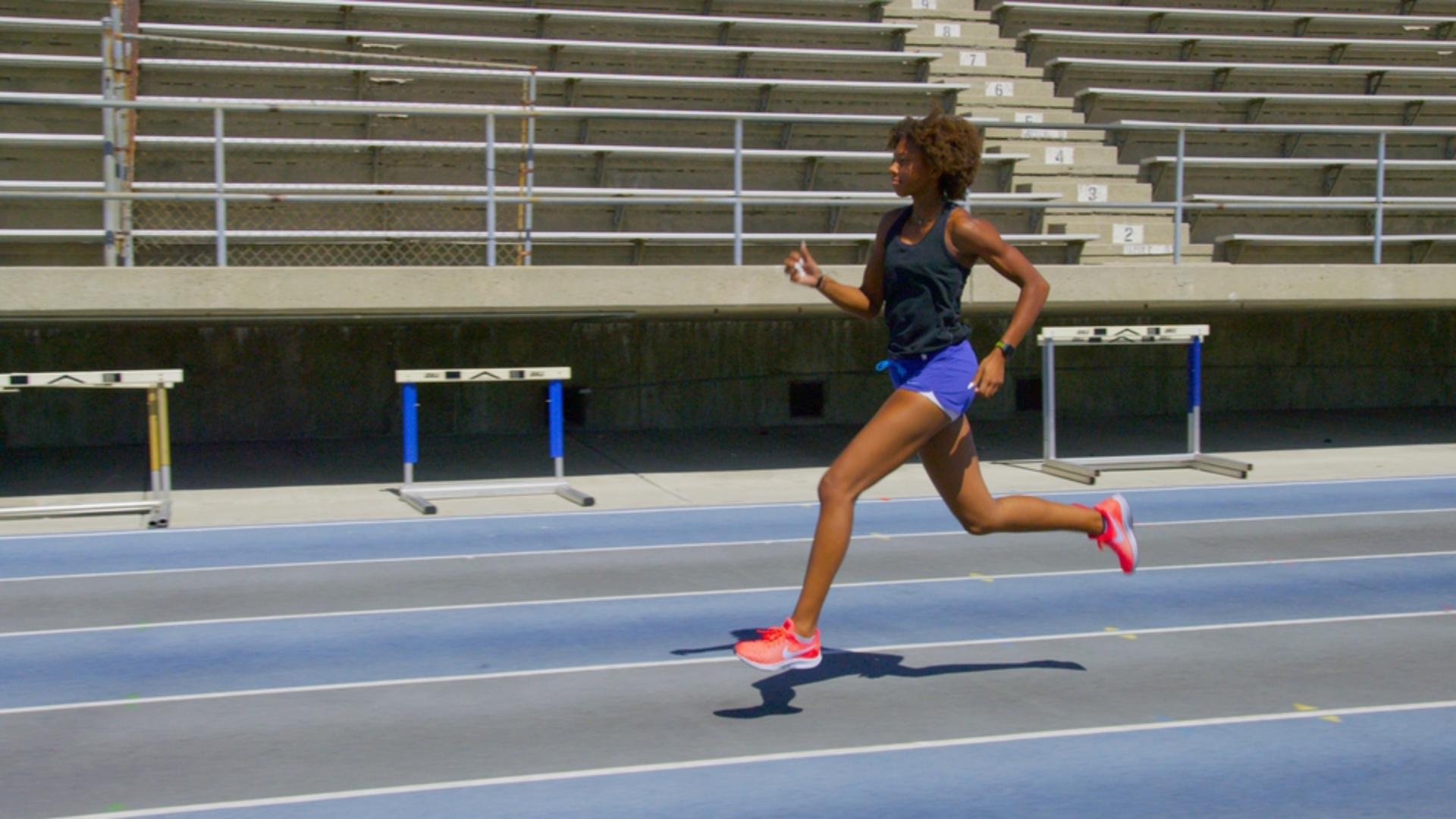}}\;\!\!\!\!
    \subfloat[]{\includegraphics[width=0.110\linewidth]{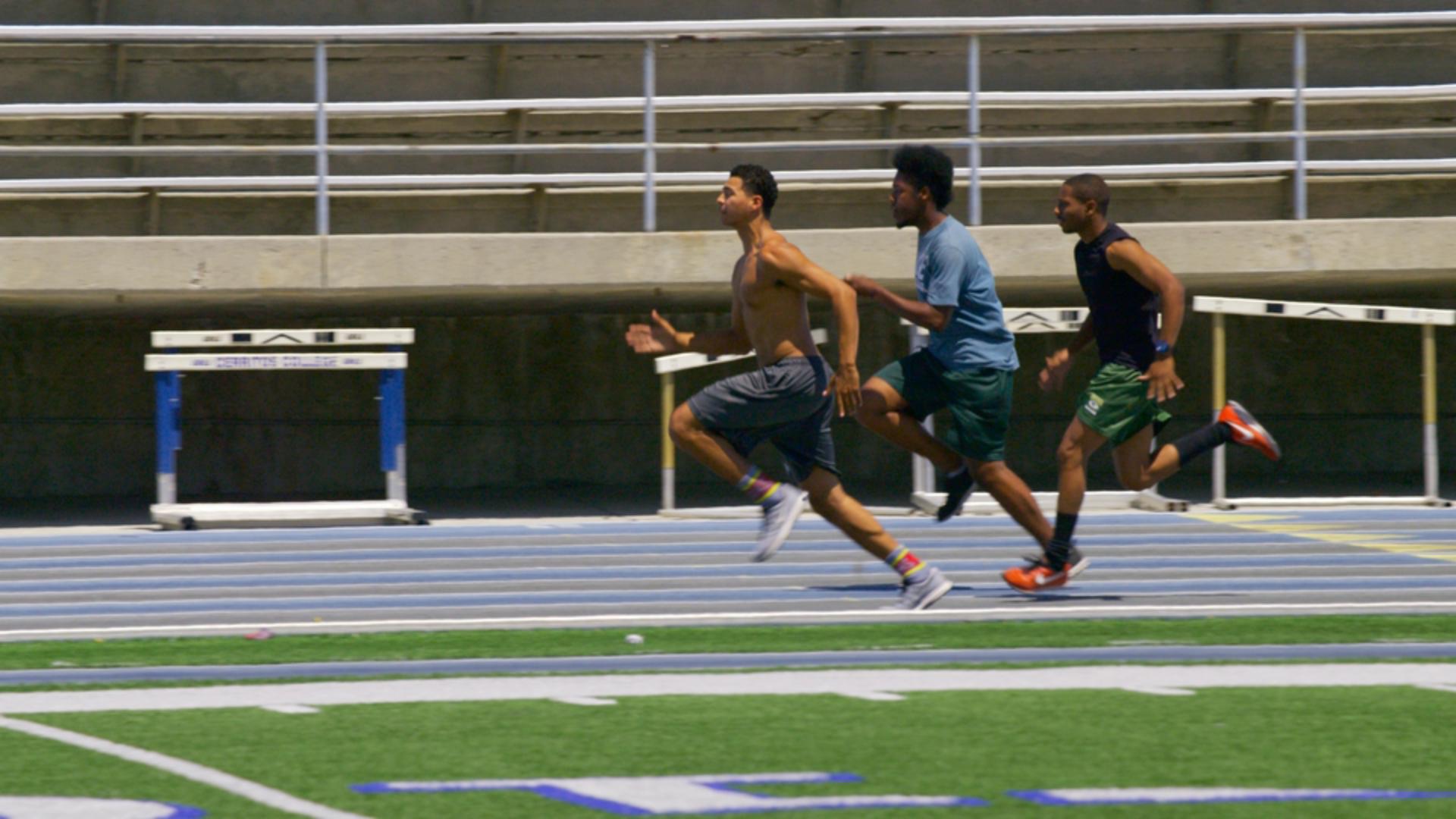}}\;\!\!\!\!
    \subfloat[]{\includegraphics[width=0.110\linewidth]{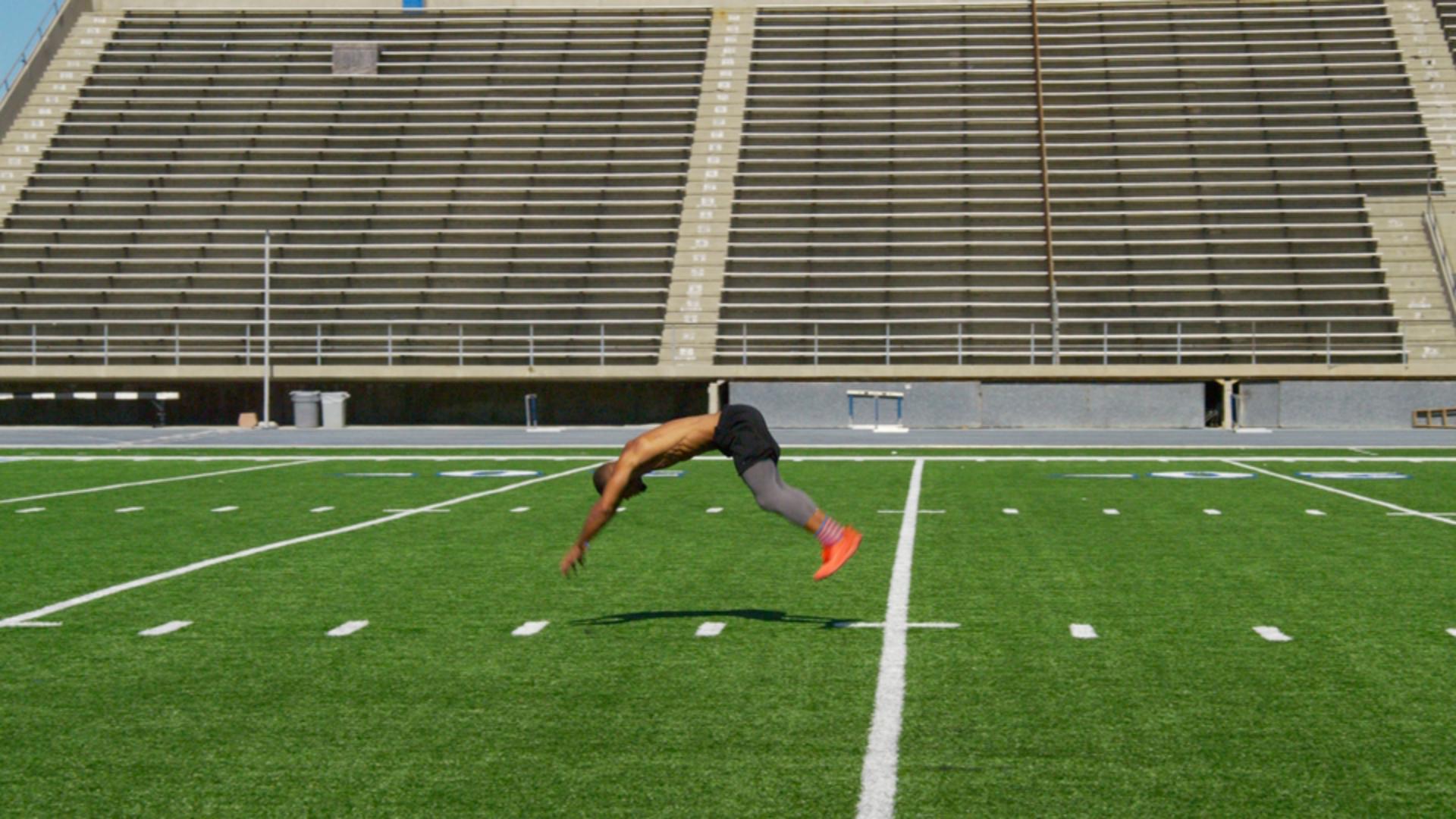}}\;\!\!\!\!
    \subfloat[]{\includegraphics[width=0.110\linewidth]{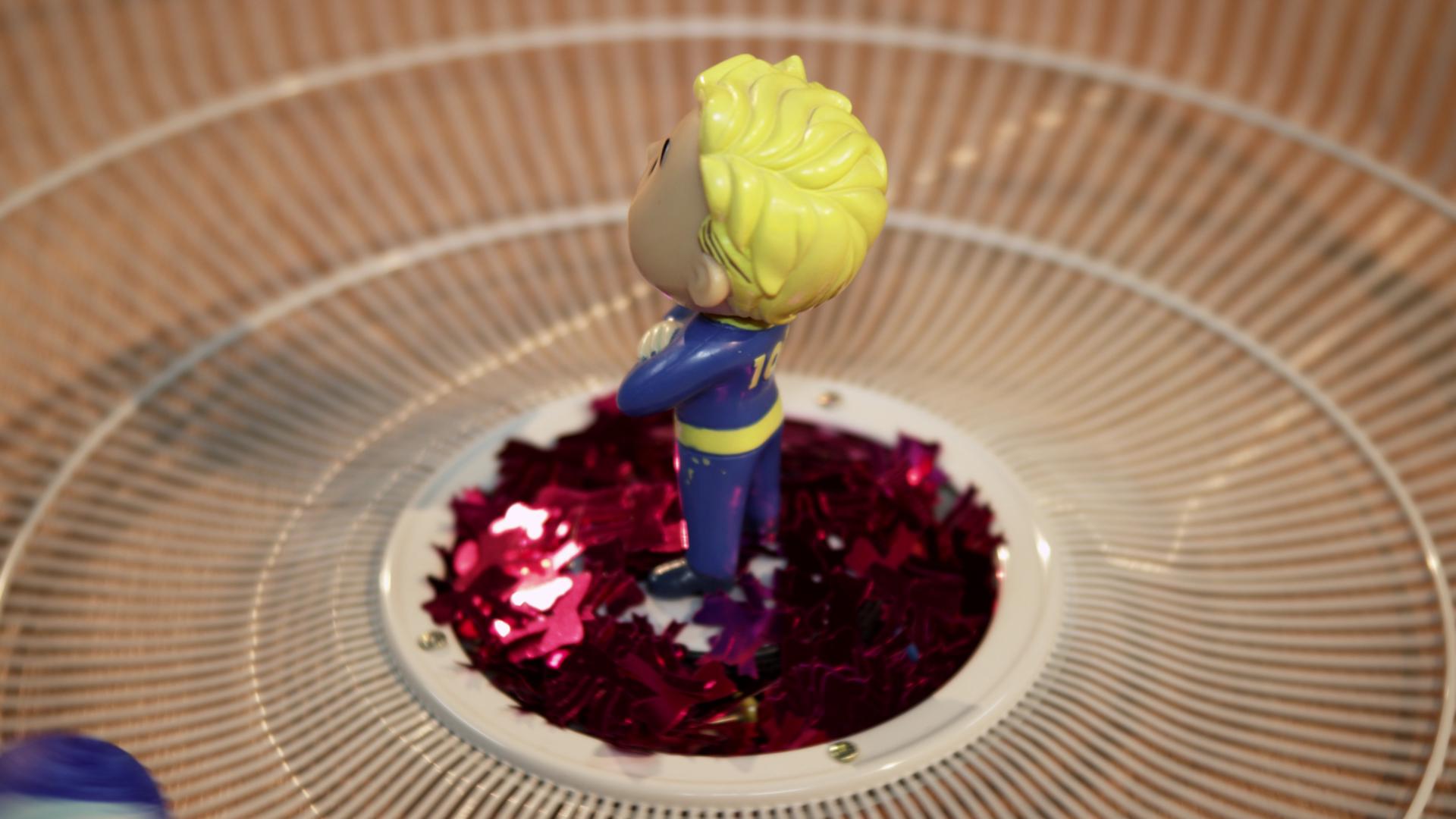}}\;\!\!\!\!
    \subfloat[]{\includegraphics[width=0.110\linewidth]{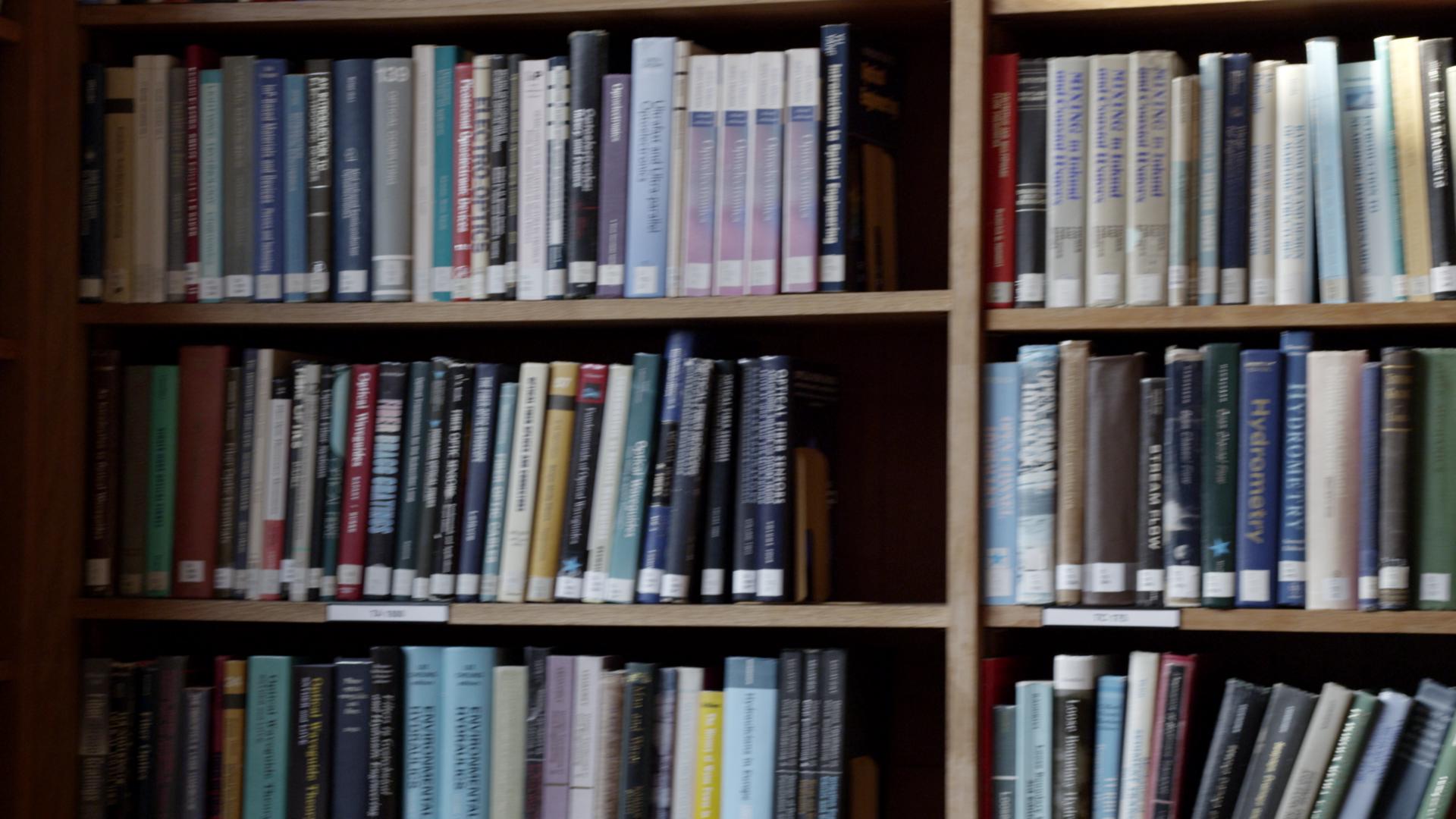}}\;\!\!\!\!
    \subfloat[]{\includegraphics[width=0.110\linewidth]{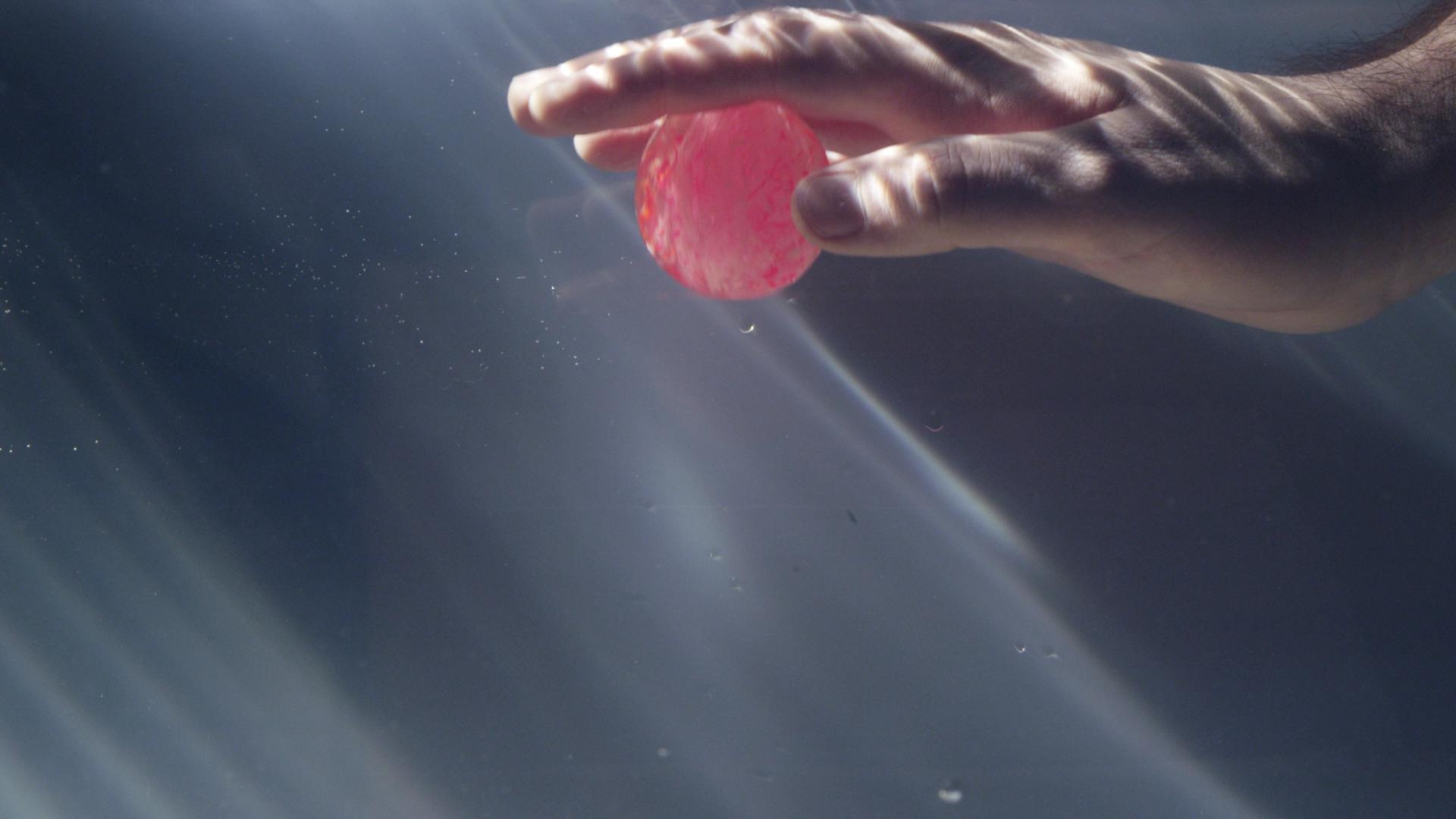}}\;\!\!\!\!
    \subfloat[]{\includegraphics[width=0.110\linewidth]{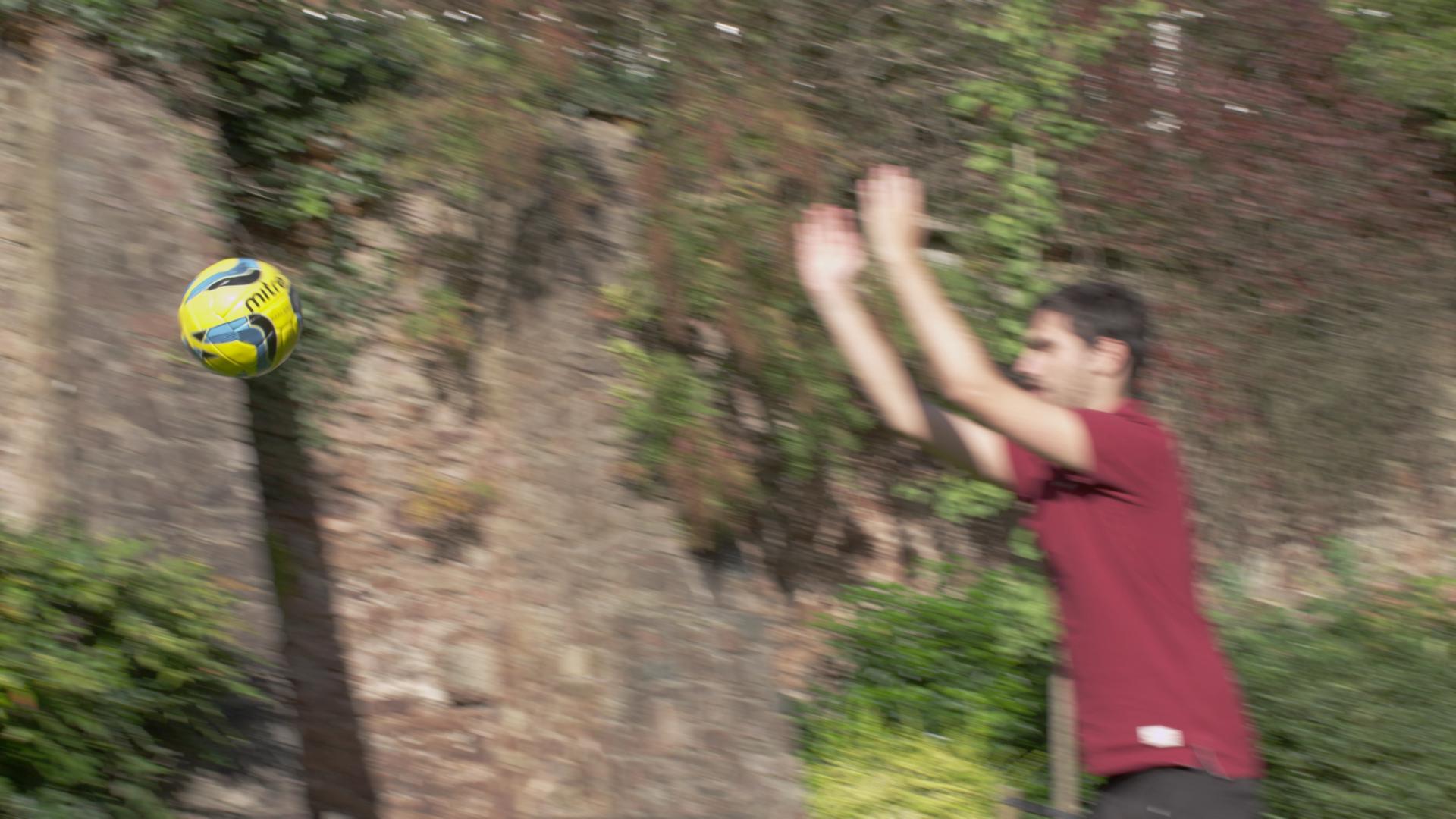}}\;\!\!\!\!
    \subfloat[]{\includegraphics[width=0.110\linewidth]{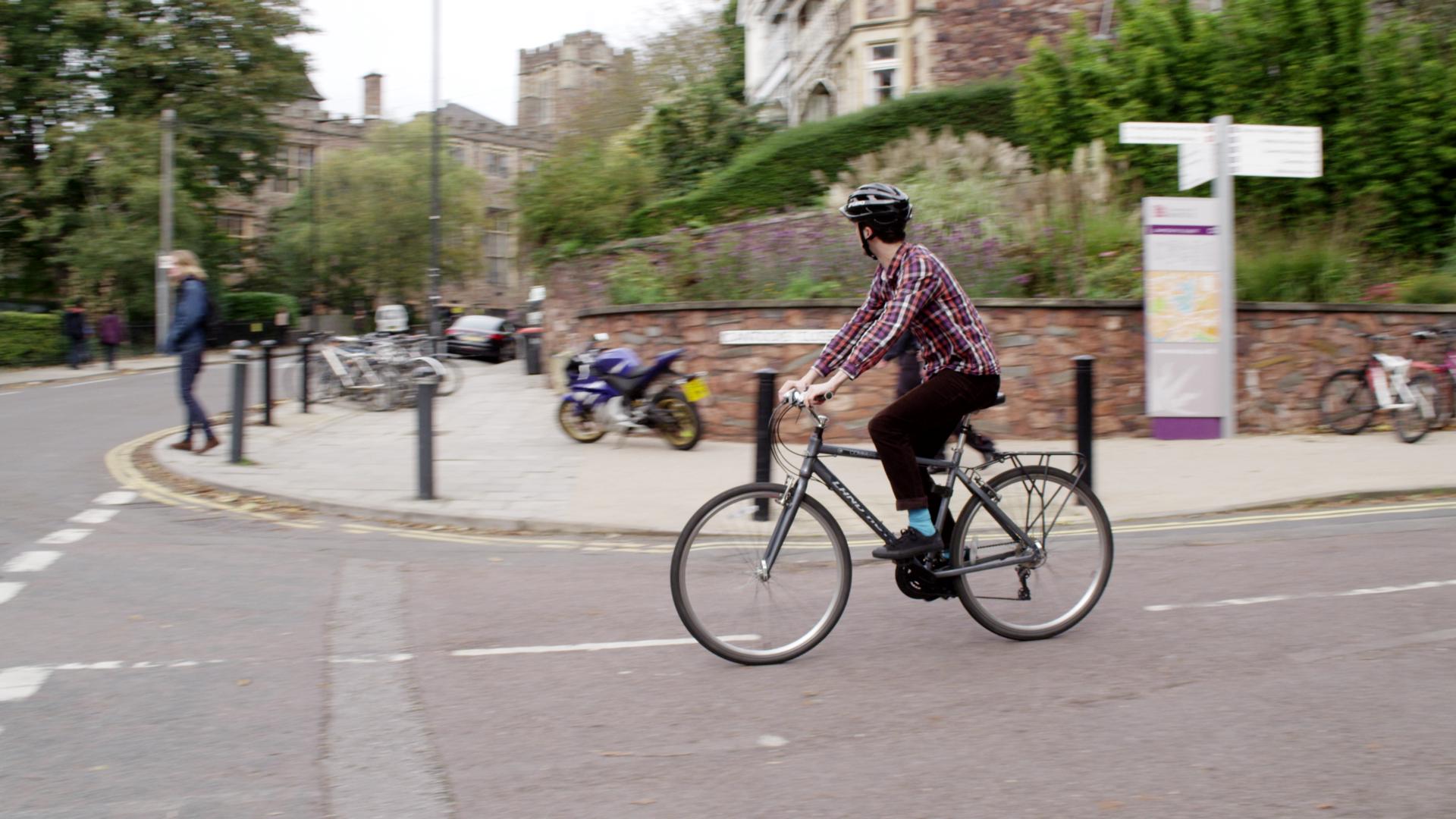}}\;\!\!\!\!
    \subfloat[]{\includegraphics[width=0.110\linewidth]{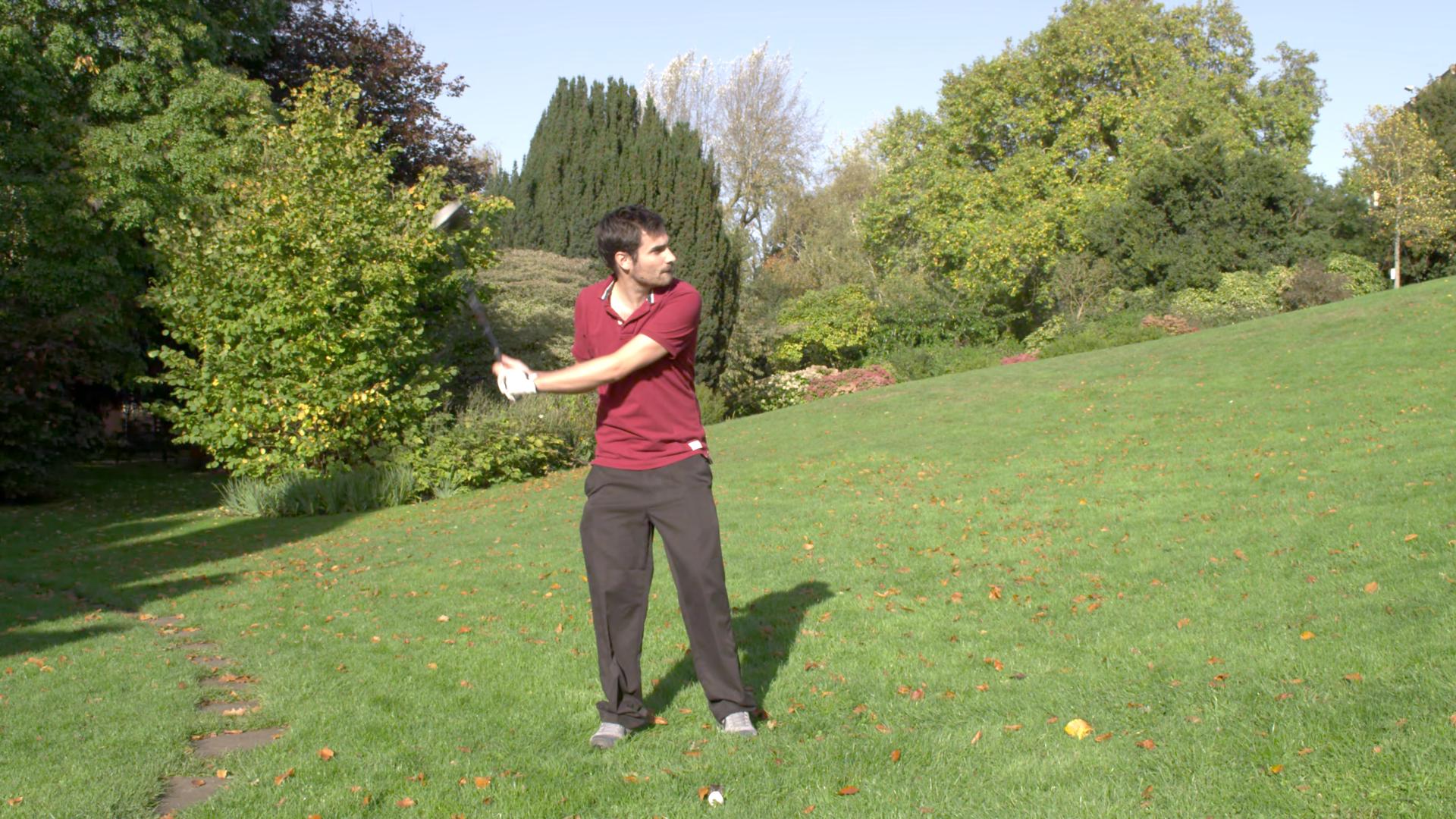}}\\
    \vspace{-3mm}
    \subfloat[]{\includegraphics[width=0.110\linewidth]{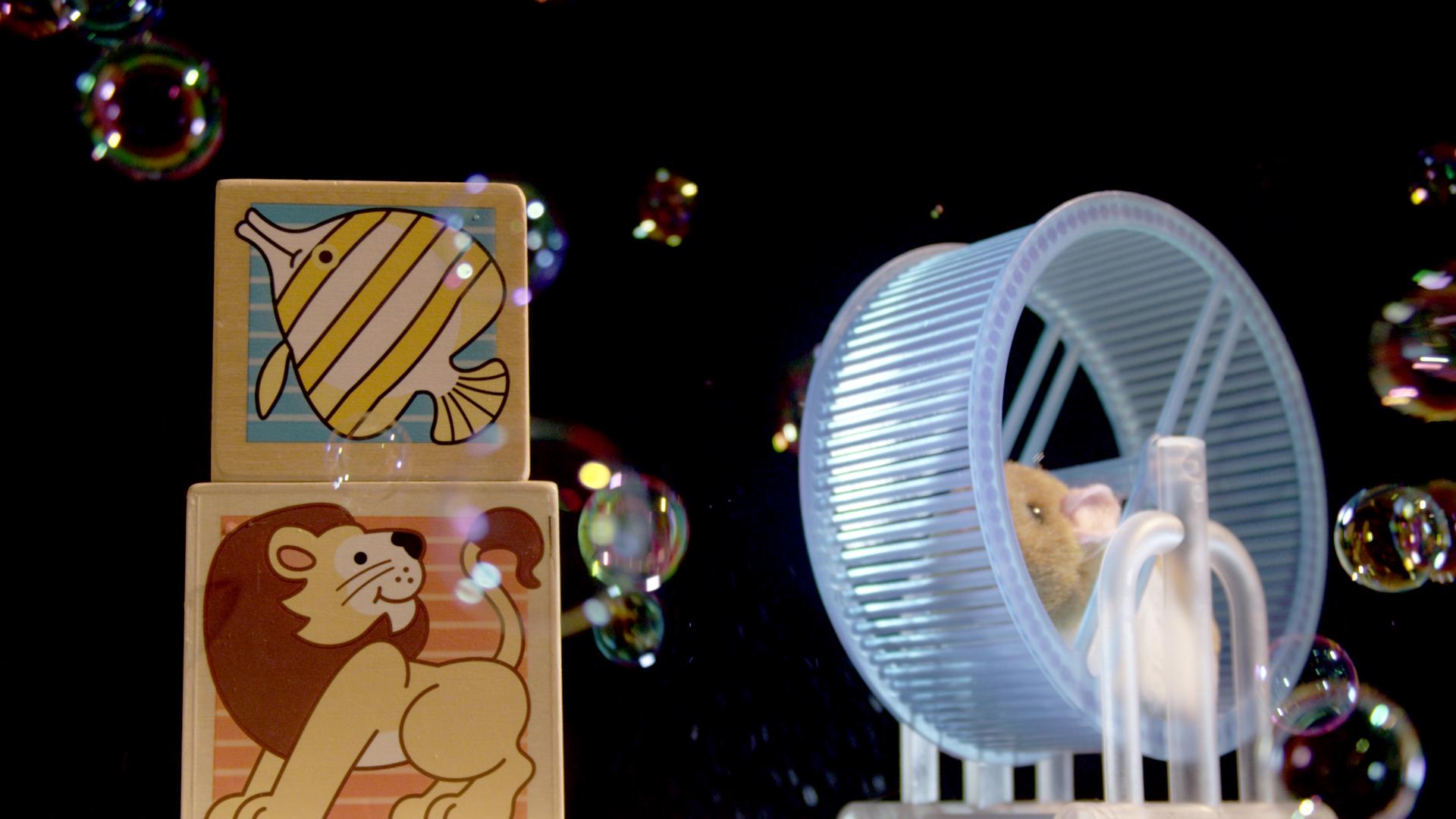}}\;\!\!\!\!
    \subfloat[]{\includegraphics[width=0.110\linewidth]{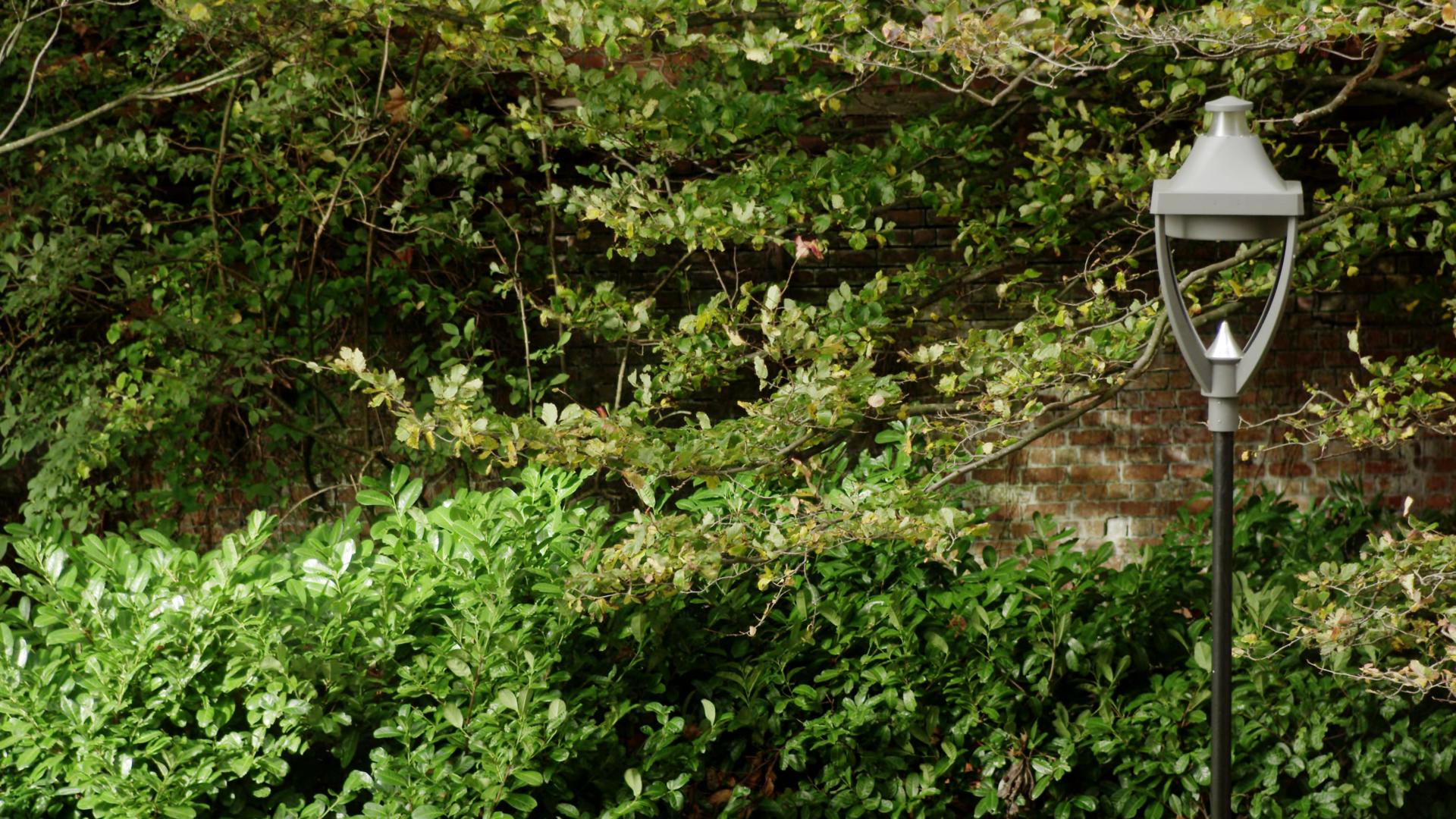}}\;\!\!\!\!
    \subfloat[]{\includegraphics[width=0.110\linewidth]{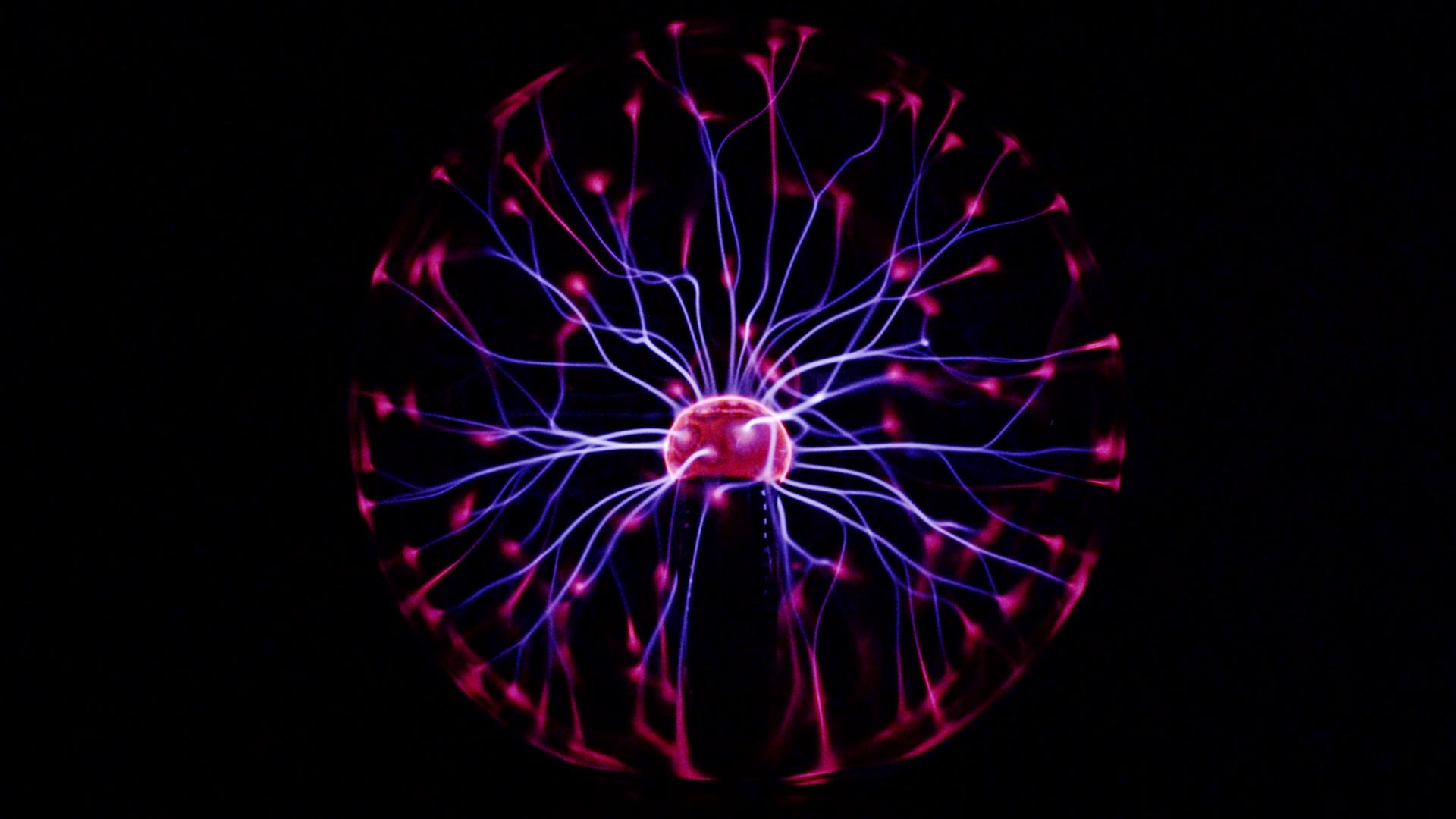}}\;\!\!\!\!
    \subfloat[]{\includegraphics[width=0.110\linewidth]{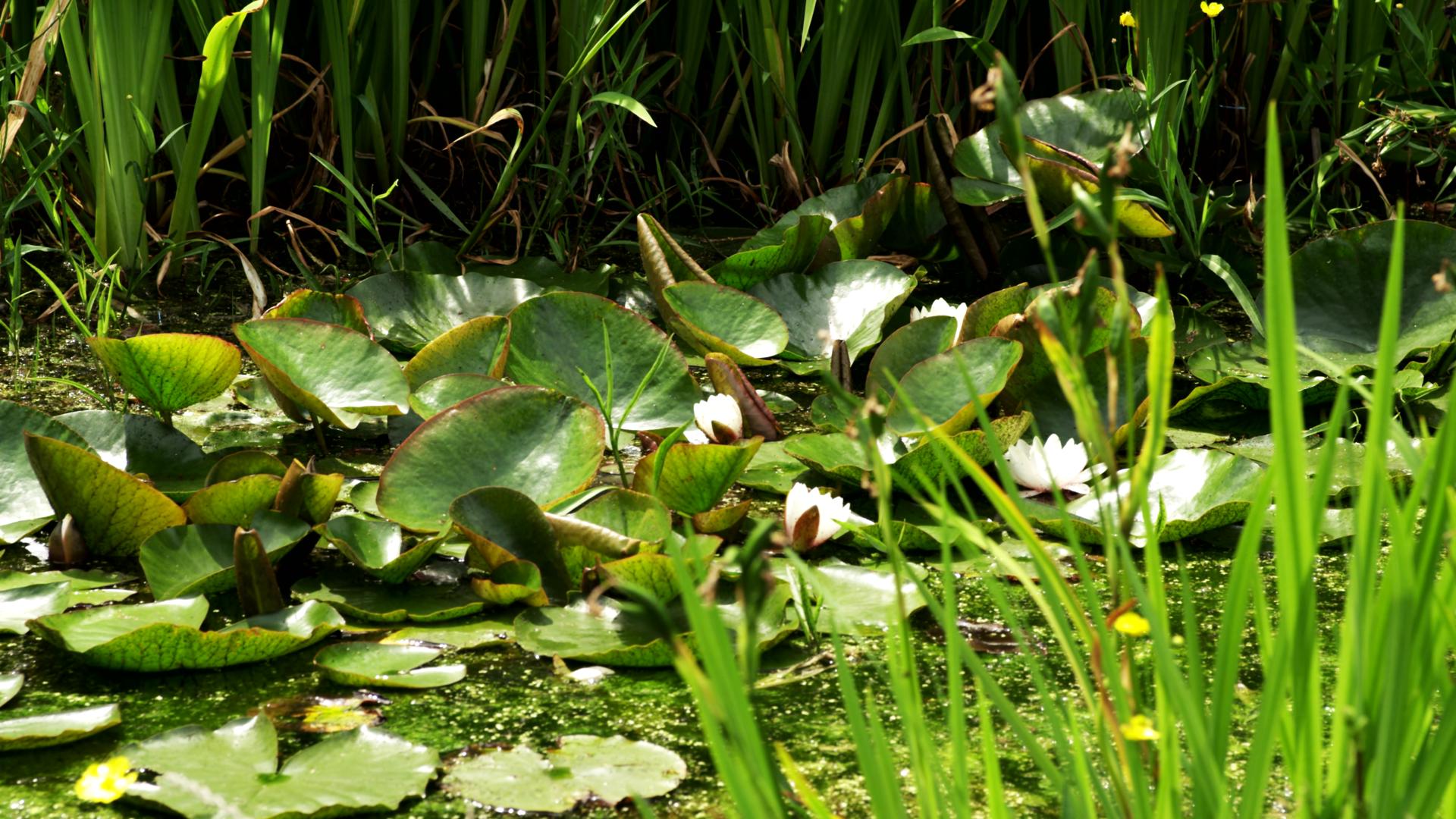}}\;\!\!\!\!
    \subfloat[]{\includegraphics[width=0.110\linewidth]{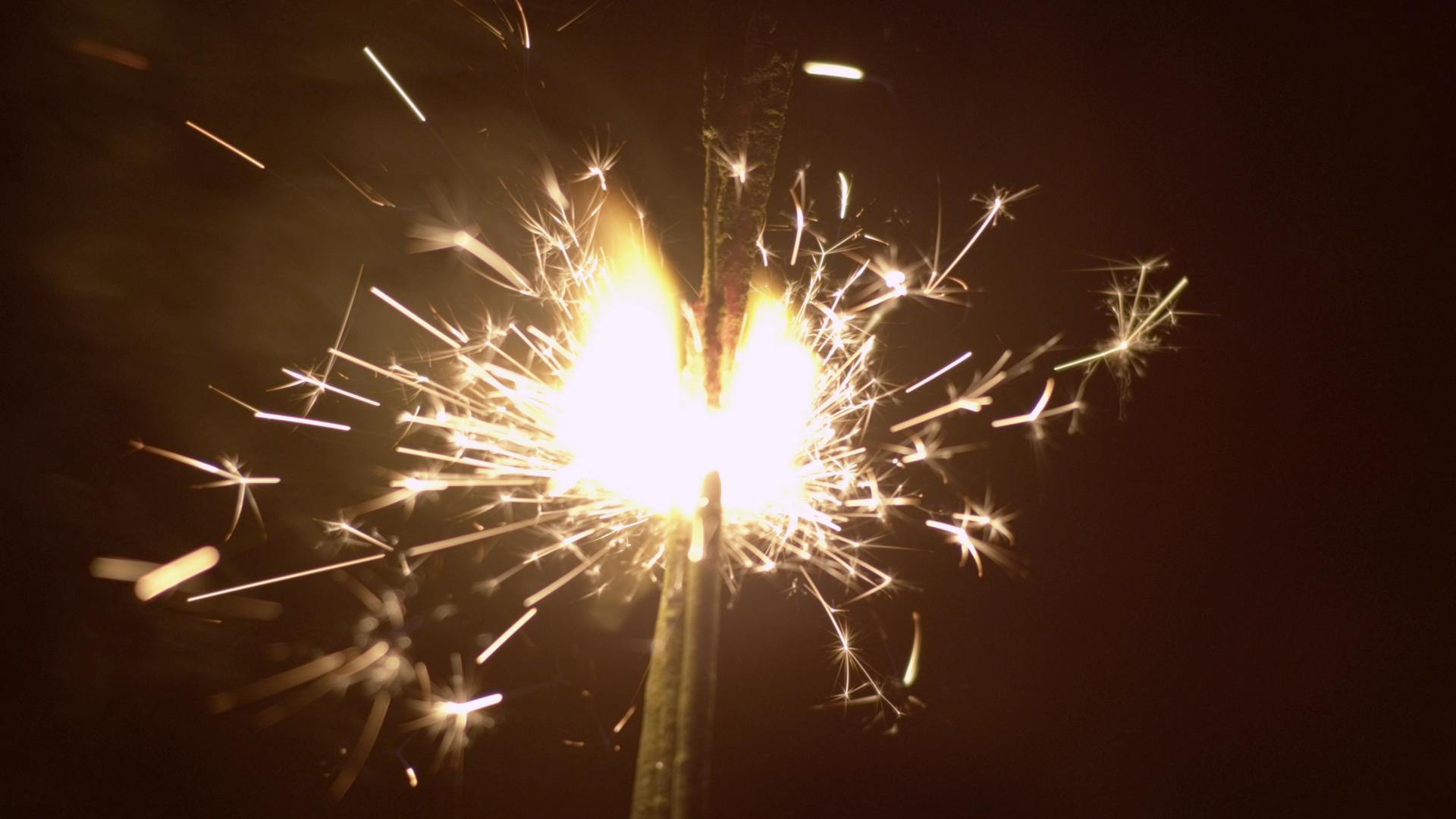}}\;\!\!\!\!
    \subfloat[]{\includegraphics[width=0.110\linewidth]{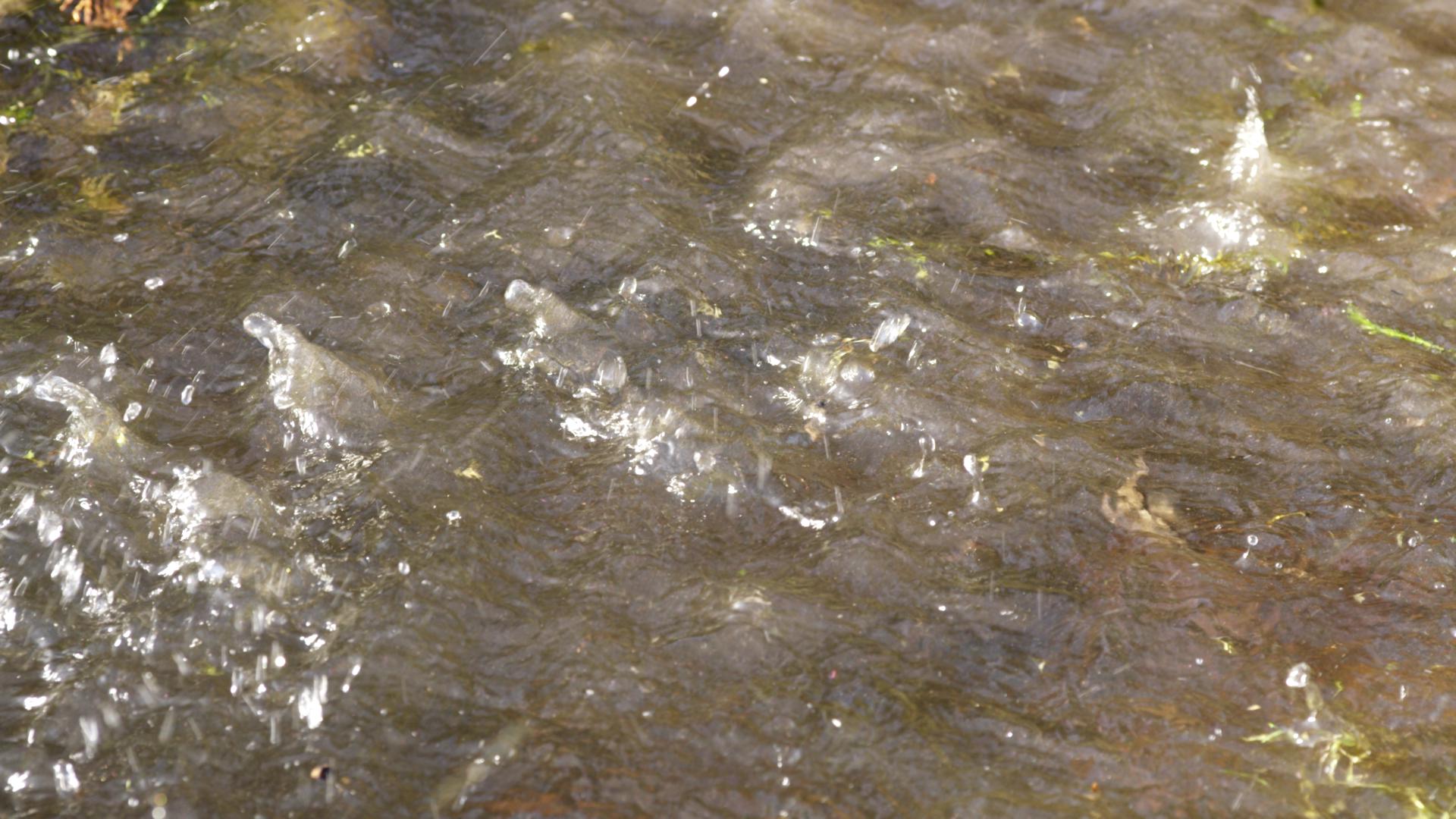}}\;\!\!\!\!
    \subfloat[]{\includegraphics[width=0.110\linewidth]{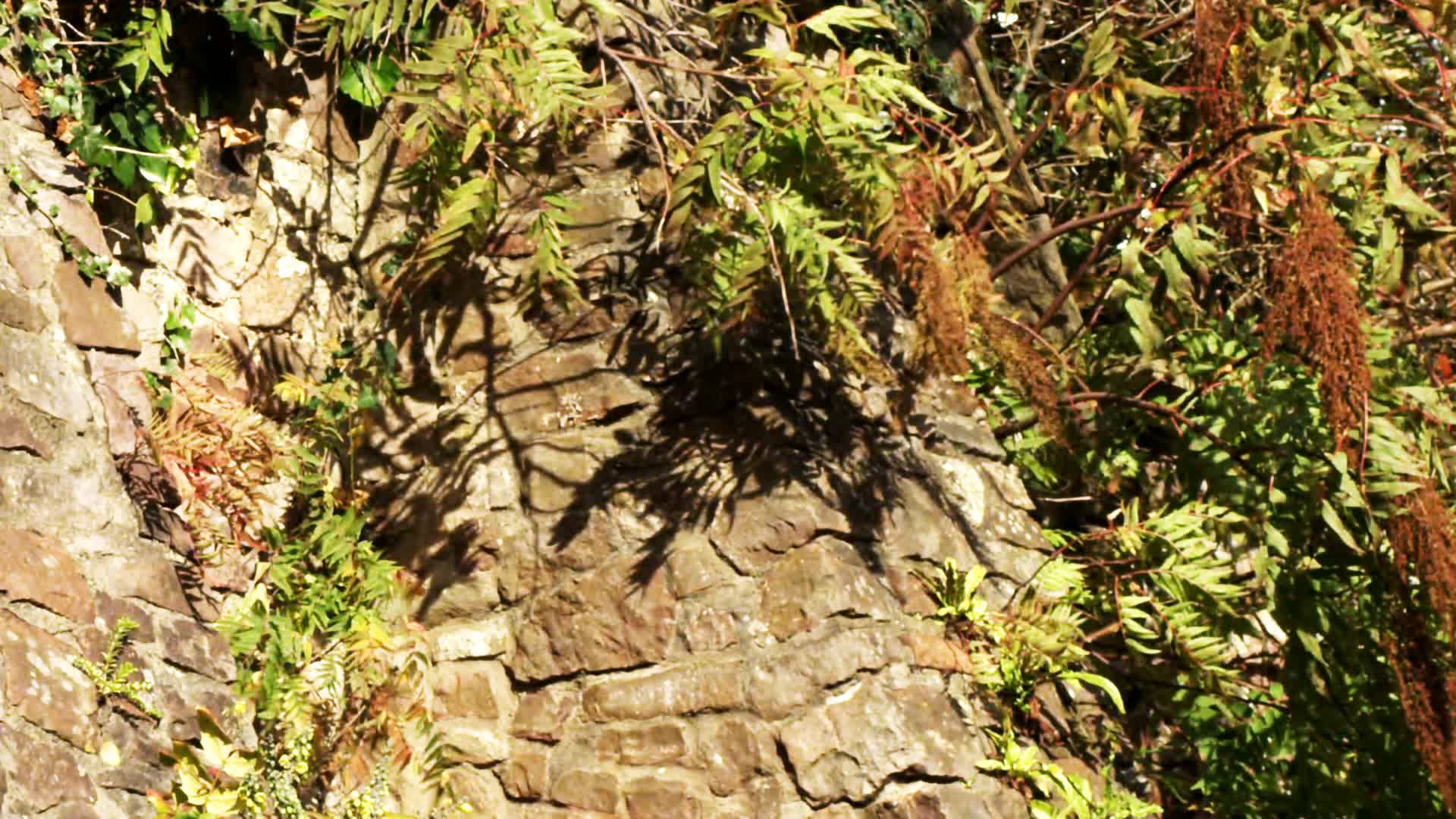}}\;\!\!\!\!
    \subfloat[]{\includegraphics[width=0.110\linewidth]{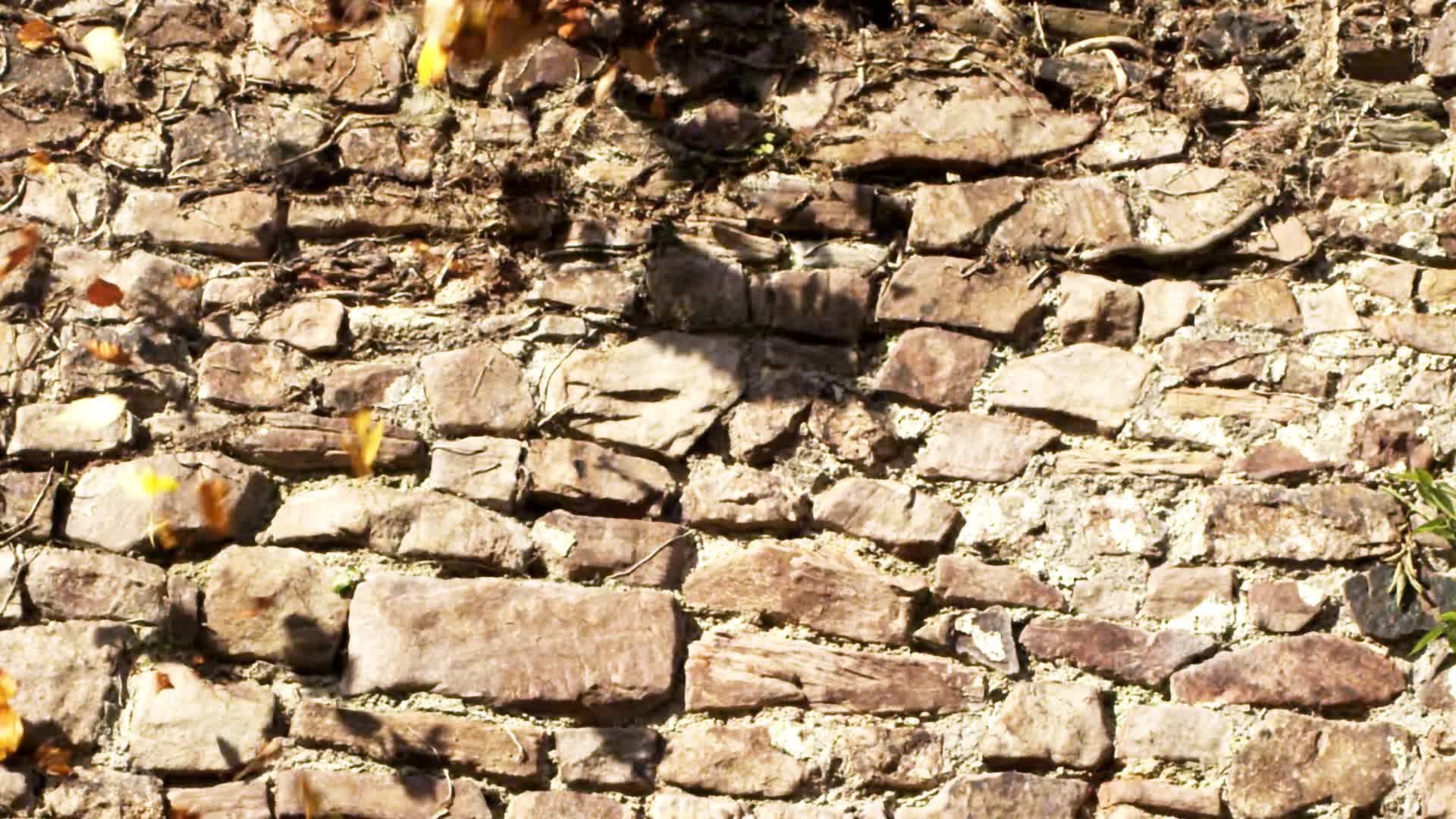}}\;\!\!\!\!
    \subfloat[]{\includegraphics[width=0.110\linewidth]{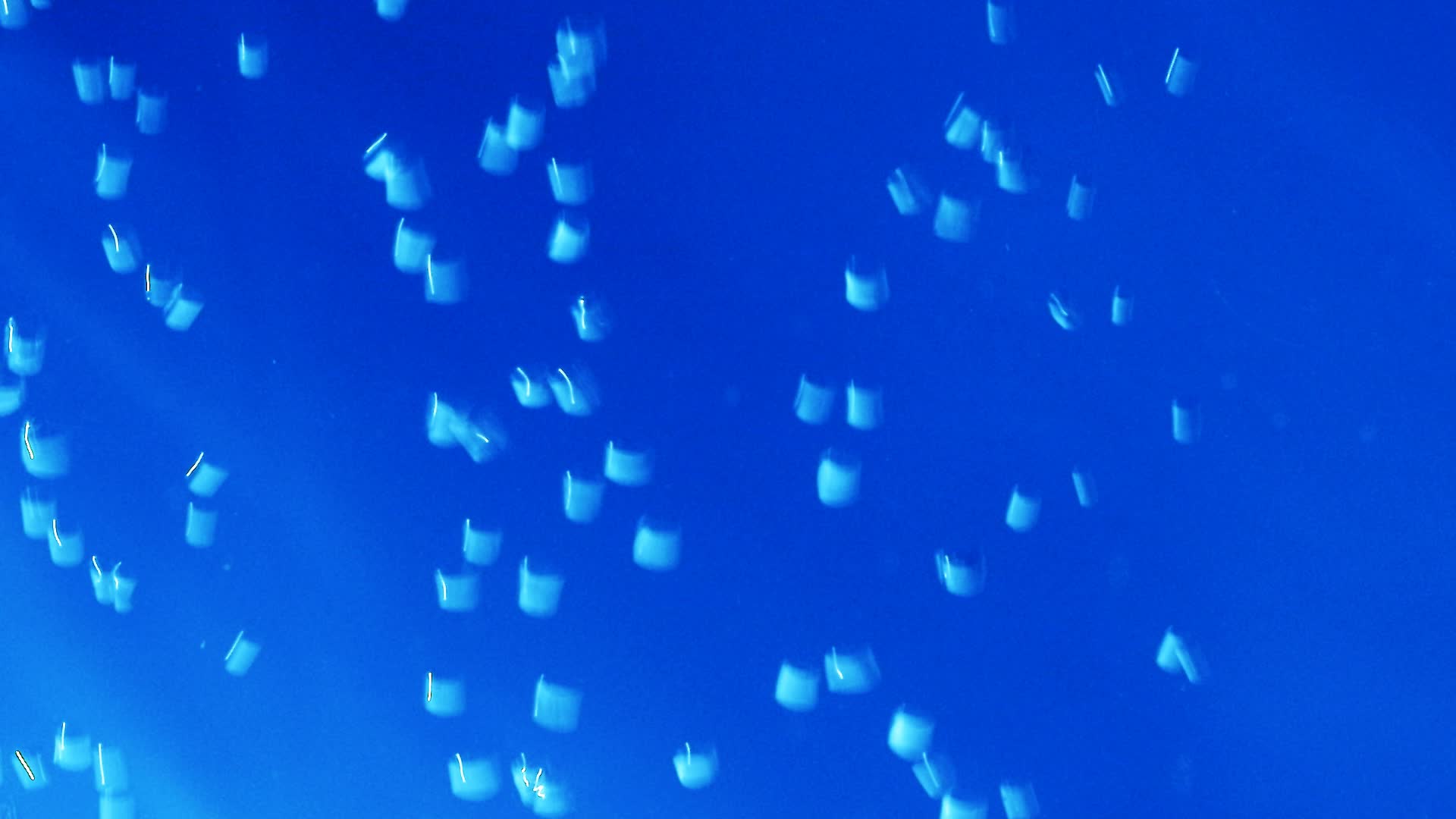}}\\
    \vspace{-3mm}
    \subfloat[]{\includegraphics[width=0.110\linewidth]{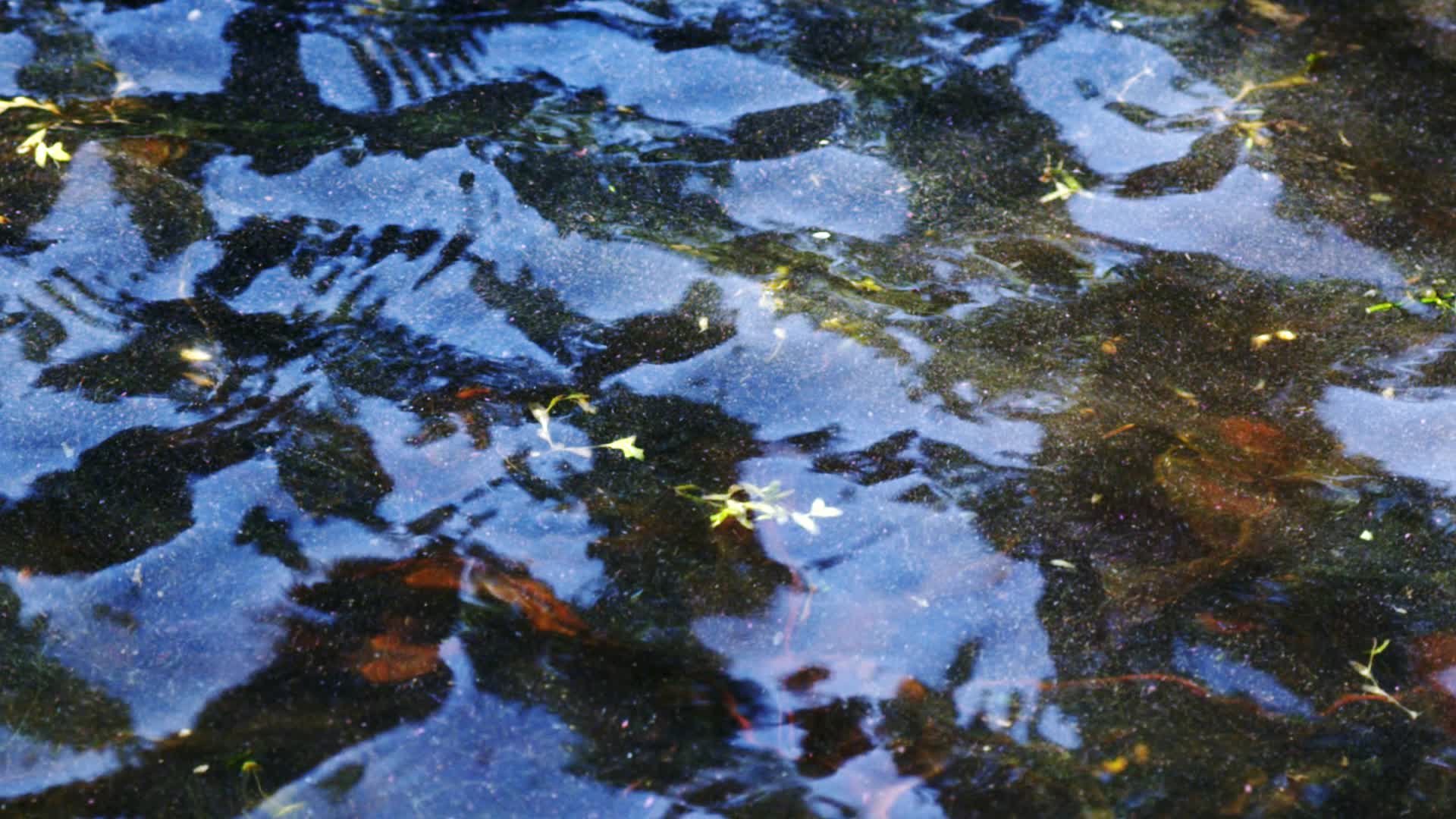}}\;\!\!\!\!
    \subfloat[]{\includegraphics[width=0.110\linewidth]{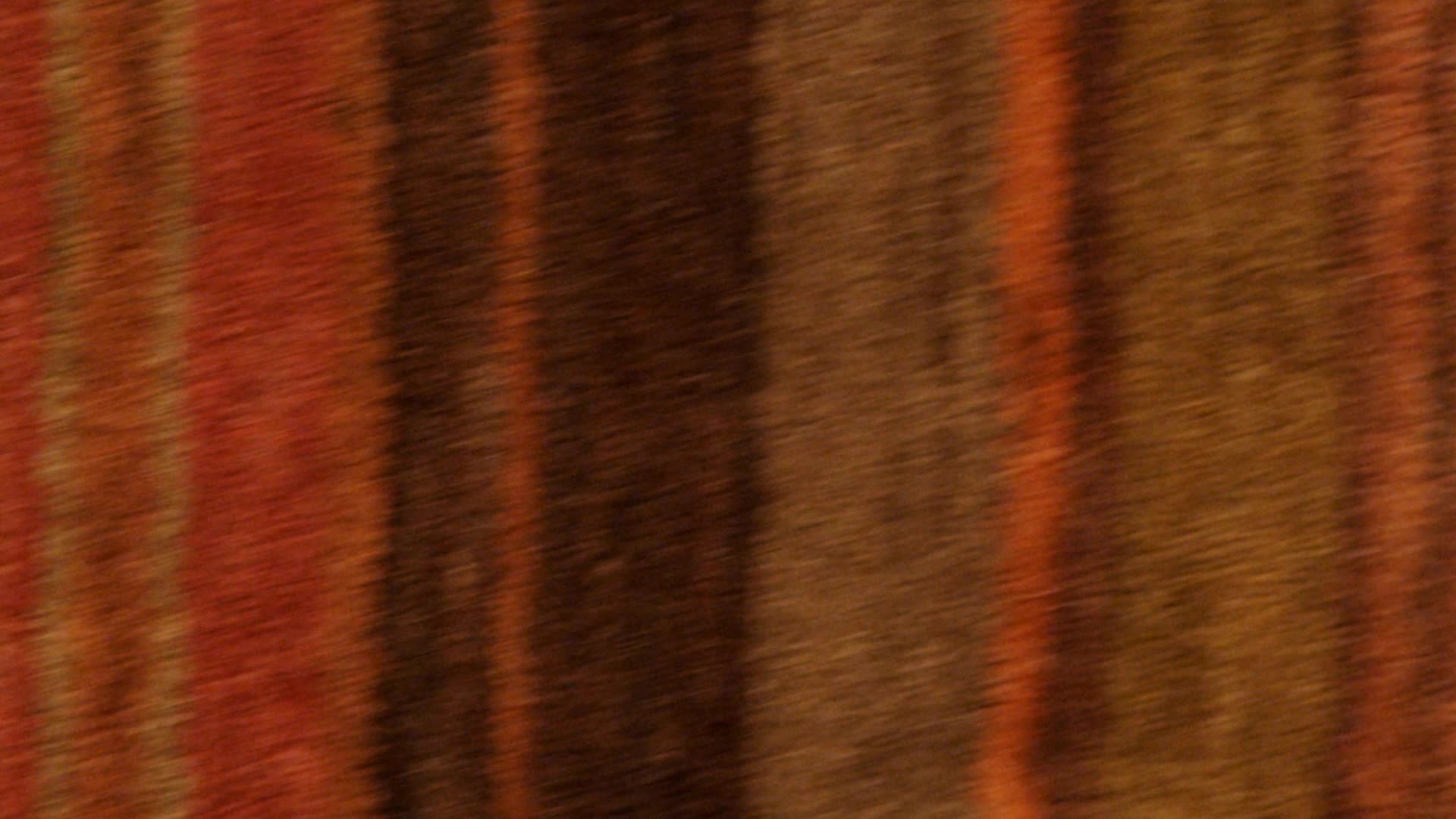}}\;\!\!\!\!
    \subfloat[]{\includegraphics[width=0.110\linewidth]{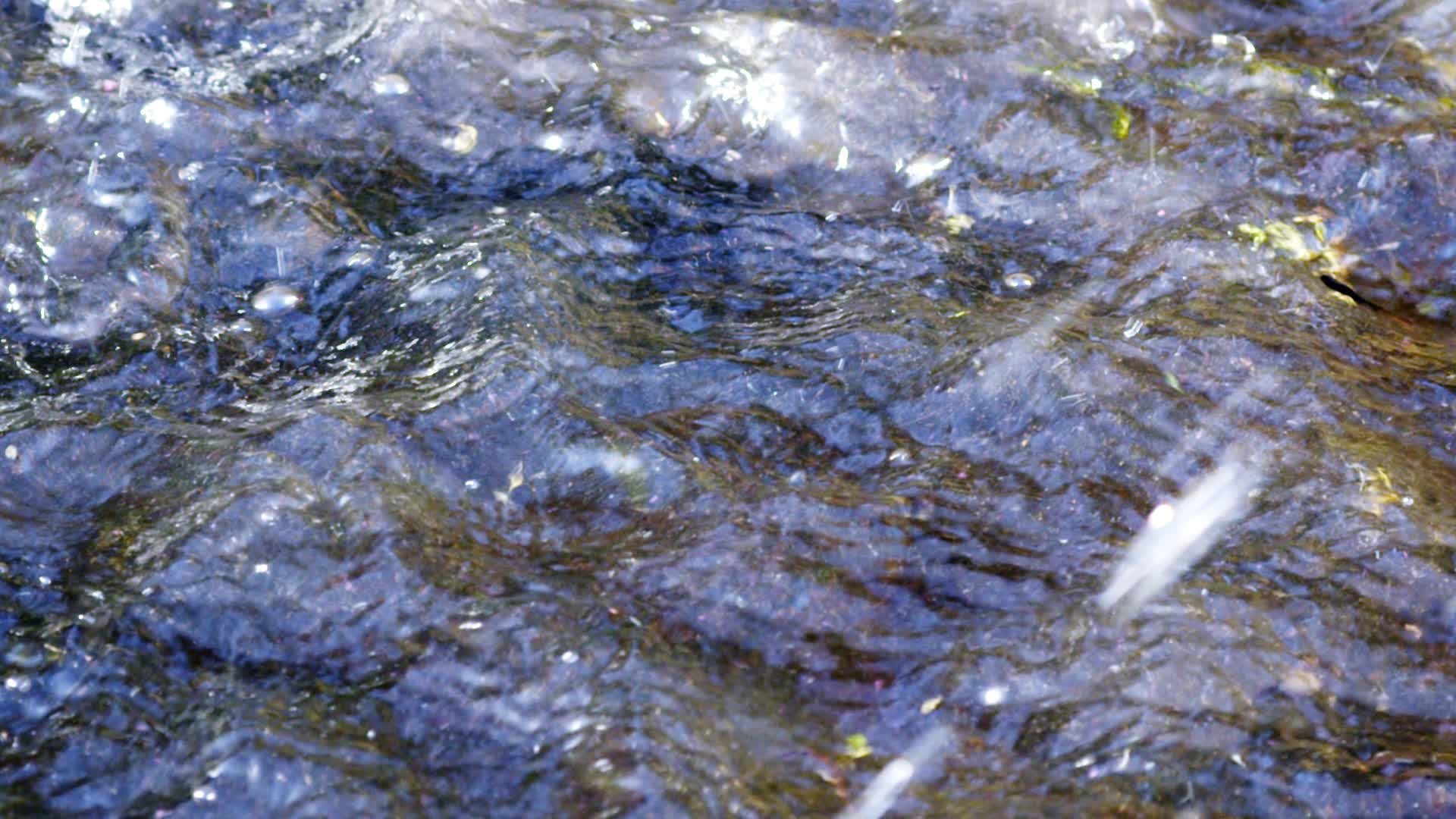}}\;\!\!\!\!
    \subfloat[]{\includegraphics[width=0.110\linewidth]{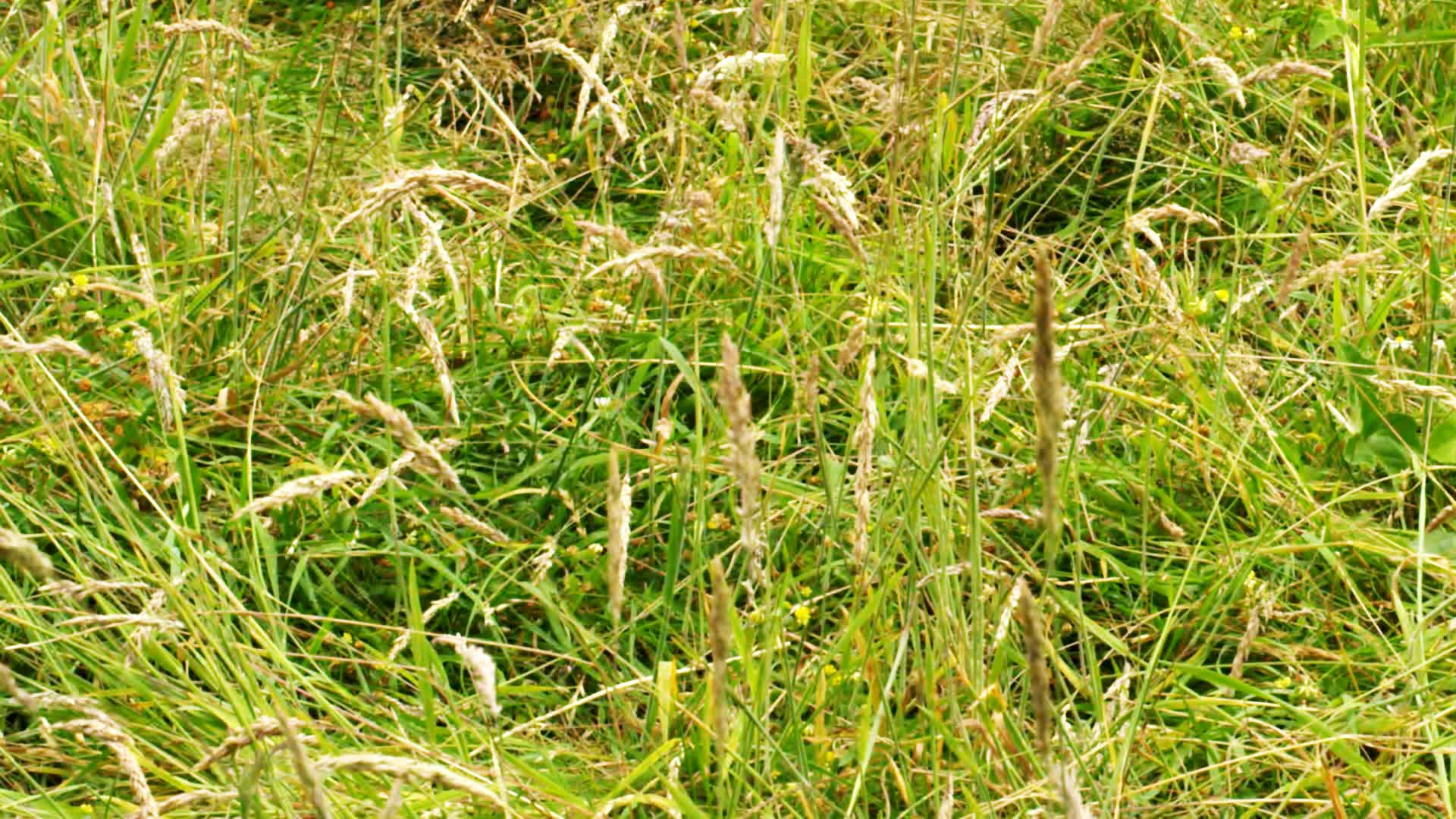}}\;\!\!\!\!
    \subfloat[]{\includegraphics[width=0.110\linewidth]{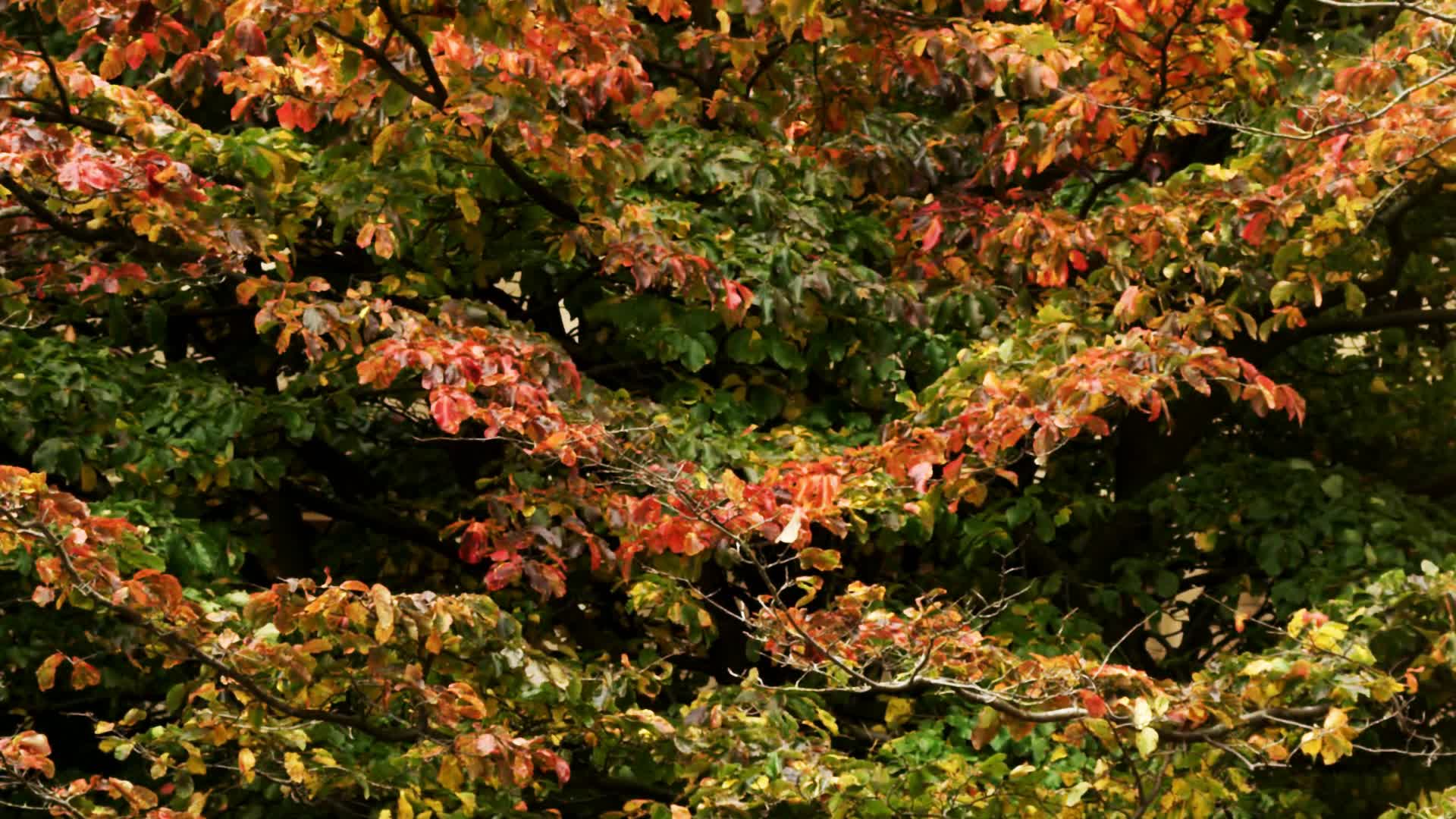}}\;\!\!\!\!
    \subfloat[]{\includegraphics[width=0.110\linewidth]{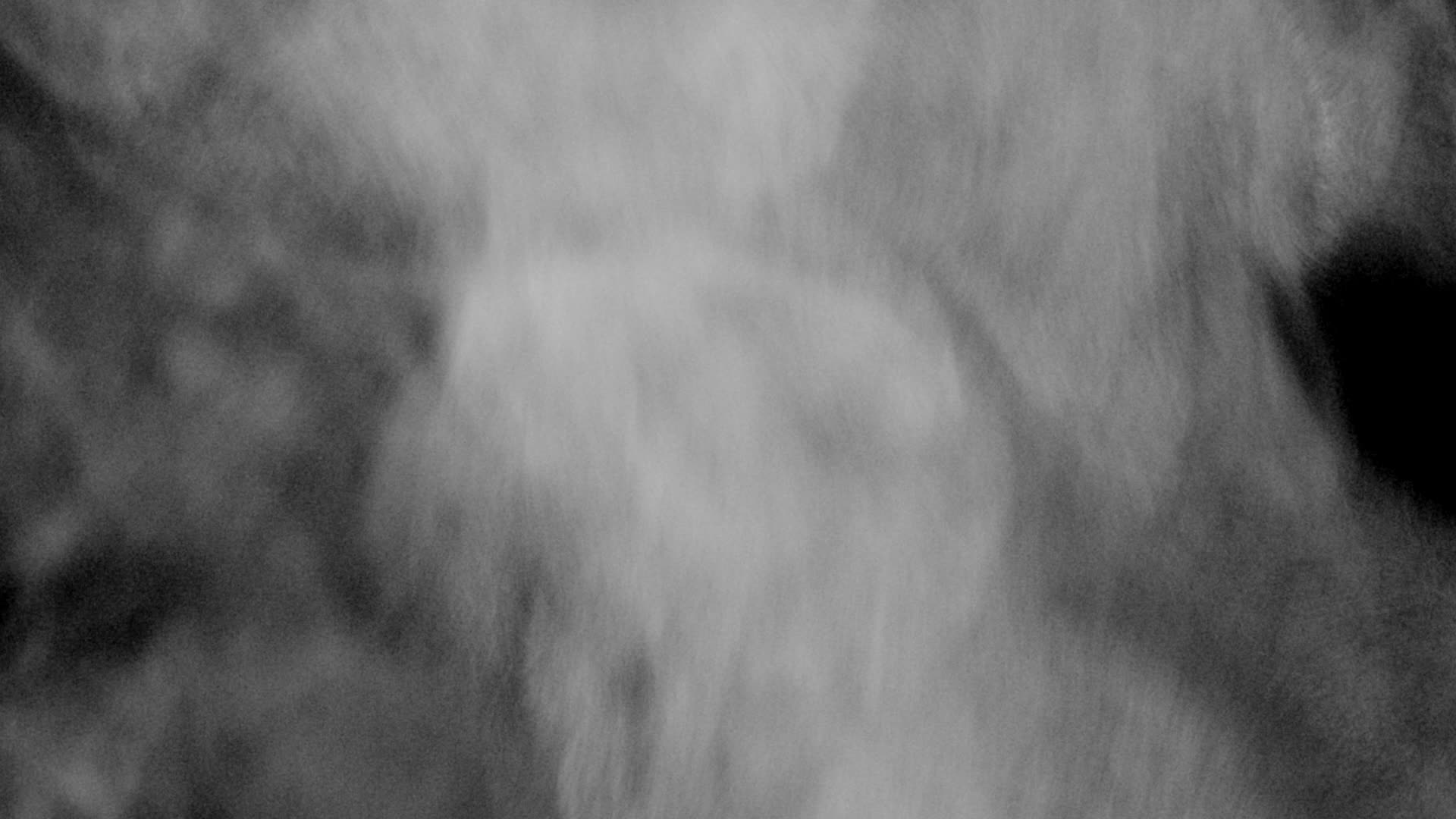}}\;\!\!\!\!
    \subfloat[]{\includegraphics[width=0.110\linewidth]{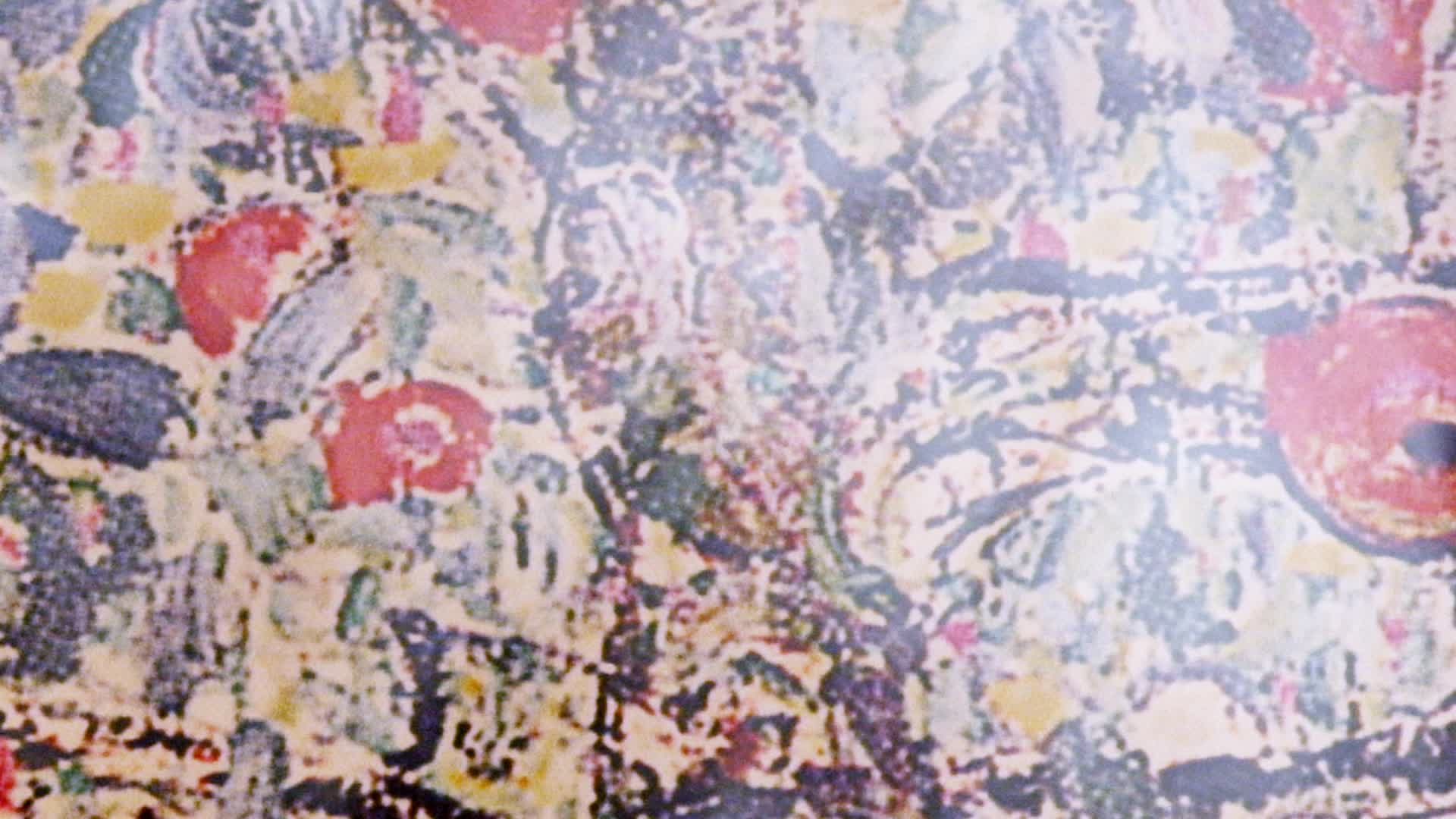}}\;\!\!\!\!
    \subfloat[]{\includegraphics[width=0.110\linewidth]{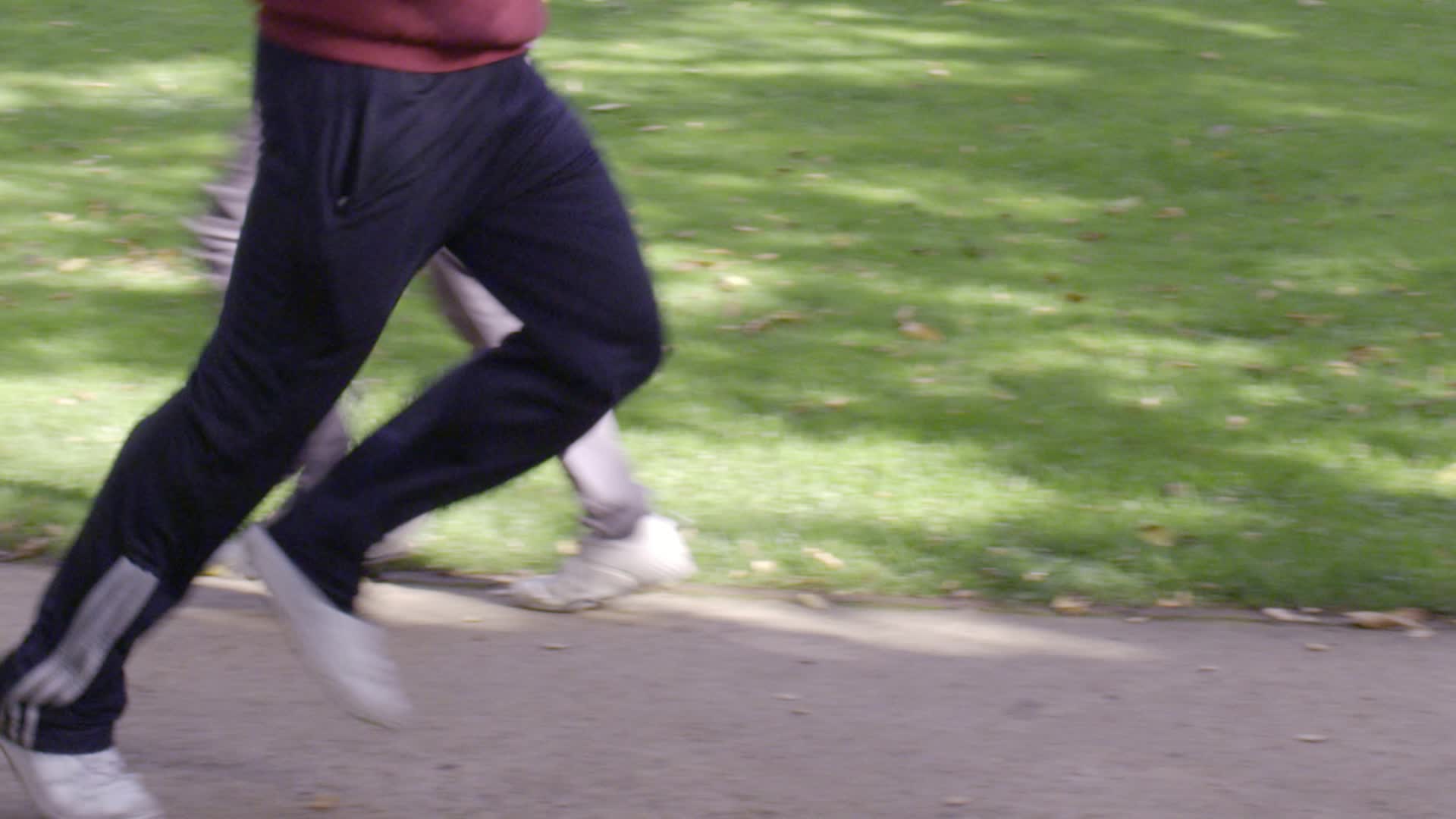}}\;\!\!\!\!
    \subfloat[]{\includegraphics[width=0.110\linewidth]{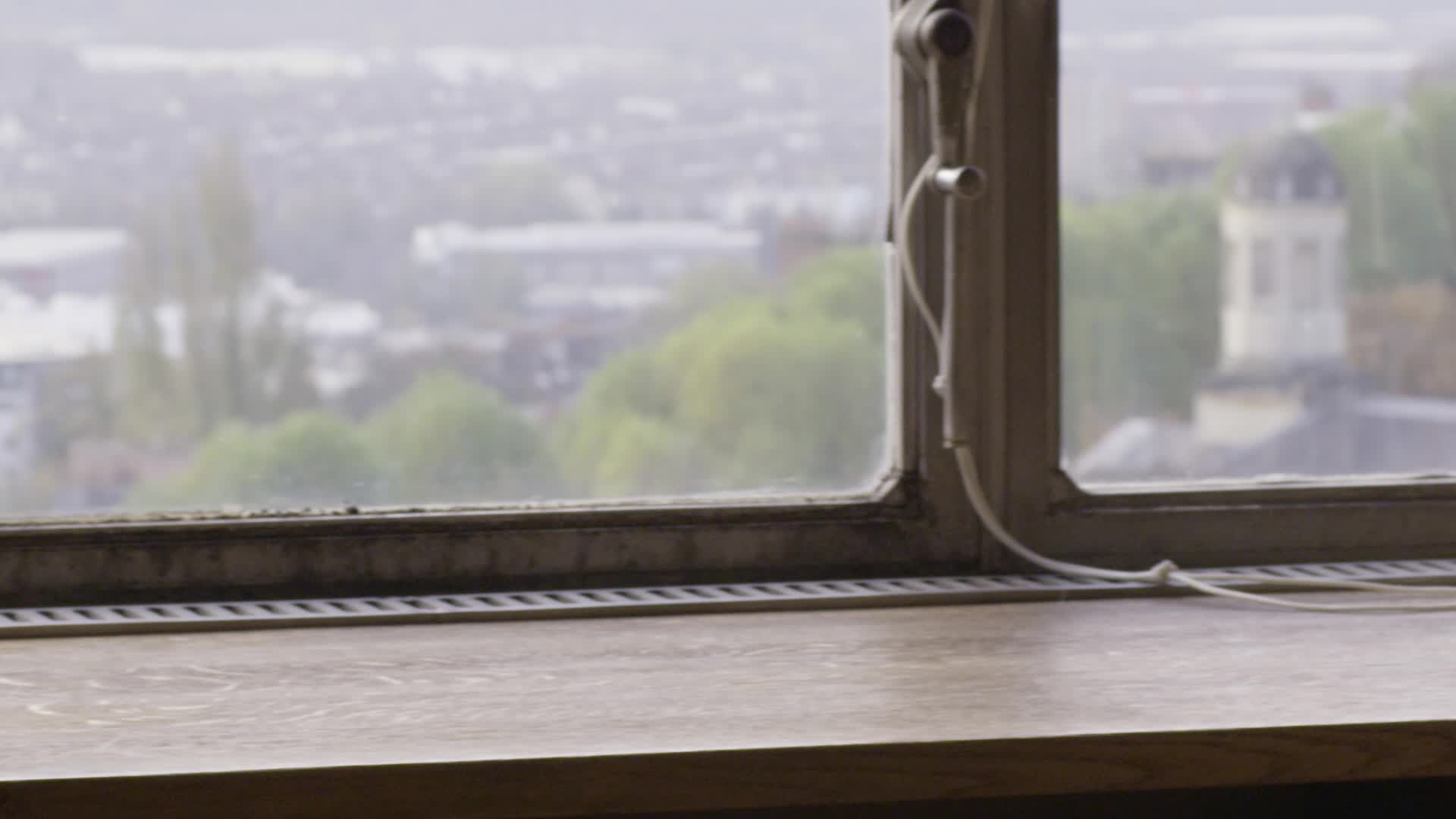}}
    \caption{\label{fig:sampleframe}Sample frames from the 36 source sequences of the BVI-VFI database. (1)-(12): sequences at 960$\times$540. (13)-(24): sequences at 1920$\times$1080. (25)-(36): sequences at 3840$\times$2160.}
\end{figure*}
\begin{table}[t]
\centering
\caption{The uniformity and range characteristics of the 36 source sequences in the BVI-VFI database.}
\label{tab:uniformity}
\begin{tabular}{c|c|c|c|c|c}
\toprule
Feature & SI & TI & CF & DTP & MV \\
\midrule
Uniformity (0-1) & 0.93 & 0.85 & 0.95 & 0.87 & 0.87 \\
Range (0-1)     & 0.88 & 0.97 & 0.74 & 0.99 & 0.99\\
\bottomrule	
\end{tabular}
\end{table}

The BVI-VFI database contains 108 reference sequences in total, which were generated from 36 different source videos captured at 120fps and resampled to 60 and 30 fps. The 36 source videos have three different spatial resolutions: 12 at 3840$\times$2160 (UHD-1), 12 at 1920$\times$1080 (HD), and 12 at 960$\times$540. For each resolution group, we: (i) first created a selection pool consisting of 120fps, YUV 4:2:0 8 bit video candidates collected from various sources; (ii) then calculated five video features: Motion Vector (MV), Dynamic Texture Parameter (DTP), Spatial Information (SI), Temporal information (TI) and Colourfulness (CF) for each candidate; (iii) used the selection algorithm described in \cite{zhang2018bvi} to determine the final source sequences in the BVI-VFI database. {This  selection procedure ensures that extreme cases of each feature are included, thus covering the most challenging VFI scenarios such as large motions and dynamic textures (see Fig.~\ref{fig:sampleframe} for examples).} Among all these five features, MV and DTP describe the overall motion magnitude and textural complexity in the video whereas the other three features are  employed to characterise the range and diversity of spatio-temporal activity and color in the video database~\cite{mackin2018study,madhusudana2021subjective,lee2021subjective}. A description of  MV, SI, TI and CF can be found in \cite{winkler2012analysis,moss2015optimal}, and DTP is described in \cite{zhang2018bvi}.

All 22 HD source candidates (in YUV 4:2:0 8 bit 120fps format) came from the BVI-HFR dataset~\cite{mackin2018study}. They were first truncated to five seconds following the research study in \cite{moss2015optimal}. {The choice of five-second duration is further justified because the source videos do not contain scene cuts and the featured motion characteristics in most videos do not vary significantly, thus allowing subjects to perceive temporal artefacts more easily. The effectiveness of the video duration has also been validated by high subject consistency in Section~\ref{sec:dataprocessing}.} Twelve\footnote{The number of source sequences for each resolution group was determined based on the trade-off between available resources for computation and subjective testing, and the optimisation of feature uniformity and range values.} sequences were selected from this pool using the algorithm in~\cite{zhang2018bvi} to ensure wide feature coverage range and uniform feature distribution~\cite{winkler2012analysis} across the whole database\footnote{The preliminary results based on HD reference and distorted sequences have been published in \cite{danier2022subjective}.}. We then followed the same procedure for the UHD-1 resolution group, for which 27 source candidates were collected from a variety of sources, including five from the LIVE-YT-HFR dataset~\cite{madhusudana2021subjective}, six from the UVG dataset~\cite{mercat2020uvg}, and 16  captured using a RED Epic-X video camera at the University of Bristol. All sequences were also trimmed to five seconds (600 frames) and converted to YUV 4:2:0 8 bit format to ensure consistency with the above-mentioned HD sources. In addition, because of the limited spatial resolution (up to 1920$\times$1080) of the high frame rate display employed in the subjective experiment, we further cropped HD representations of these UHD-1 videos. Specifically, we generated 9 cropped candidates for each UHD-1 video ($x\in\{0,960,1920\}, y\in\{0,540,1080\}$ where $(x,y)$ is the top-left coordinate of the crop), and selected the crop that has the highest content similarity to the original UHD-1 video in terms of two feature descriptors: MV and DTP. By doing so, we obtained 27 cropped videos, which retain the characteristics of UHD-1 content despite their smaller spatial extent. From these 27 sources, we selected 12 UHD-1 source videos using the same procedure as for the HD resolution group. For the resolution group of 960$\times$540, due to the lack of publicly available 120fps videos at this resolution, we generated a selection pool by spatially down-sampling (using a Lanczos3 filter) the unused 120fps videos from the previously collected HD and UHD-1 candidates. Then we applied again the same sequence selection process to obtain the 12 source sequences at 960$\times$540. 

Finally, to obtain reference sequences sampled at various frame rates, the twelve 1920$\times$1080, twelve cropped 3840$\times$2160 and the twelve 960$\times$540 source videos at 120fps were temporally sub-sampled to 60fps and 30fps by frame dropping, resulting in 108 references. An alternative sub-sampling method, frame averaging, can create visible ghosting artefacts in cases with small shutter angles and introduce additional motion blur. Both of these artefacts can seriously deteriorate the frame interpolation performance~\cite{shen2020blurry, zhang2020video}. In contrast, although frame dropping reduces the original shutter angle which may introduce motion judder, the resulting frames provide a superior basis for VFI applications. Sample frames of the all the final source sequences are shown in Fig.~\ref{fig:sampleframe}.
Table~\ref{tab:uniformity} reports the range and uniformity characteristics of the source sequences in BVI-VFI. It can be observed that the selected source sequences offer a wide and uniform coverage for all the spatial and temporal features measured.

\begin{figure*}[t]
    \centering
    \subfloat{\includegraphics[width=0.2475\linewidth]{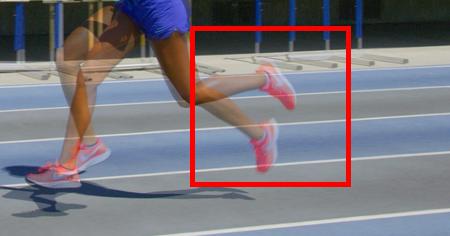}}\;\!\!\!
    \subfloat{\includegraphics[width=0.122\linewidth]{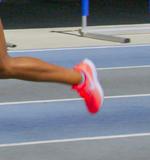}}\;\!\!\!
    \subfloat{\includegraphics[width=0.122\linewidth]{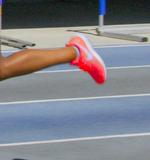}}\;\!\!\!
    \subfloat{\includegraphics[width=0.122\linewidth]{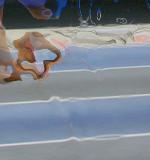}}\;\!\!\!
    \subfloat{\includegraphics[width=0.122\linewidth]{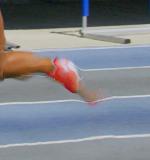}}\;\!\!\!
    \subfloat{\includegraphics[width=0.122\linewidth]{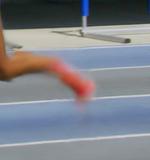}}\;\!\!\!
    \subfloat{\includegraphics[width=0.122\linewidth]{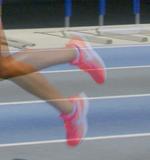}}\\
    \vspace{-3mm}
    \subfloat{\includegraphics[width=0.2475\linewidth]{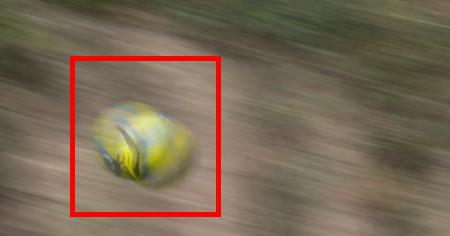}}\;\!\!\!
    \subfloat{\includegraphics[width=0.122\linewidth]{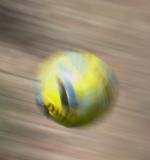}}\;\!\!\!
    \subfloat{\includegraphics[width=0.122\linewidth]{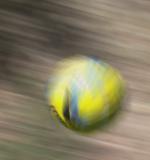}}\;\!\!\!
    \subfloat{\includegraphics[width=0.122\linewidth]{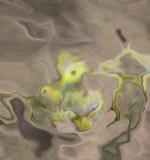}}\;\!\!\!
    \subfloat{\includegraphics[width=0.122\linewidth]{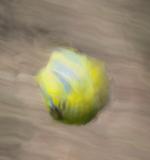}}\;\!\!\!
    \subfloat{\includegraphics[width=0.122\linewidth]{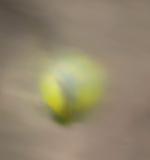}}\;\!\!\!
    \subfloat{\includegraphics[width=0.122\linewidth]{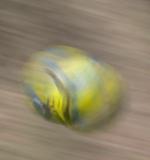}}
    \vspace{-3mm}
    \setcounter{subfigure}{0}
    \subfloat[Overlay]{\includegraphics[width=0.2475\linewidth]{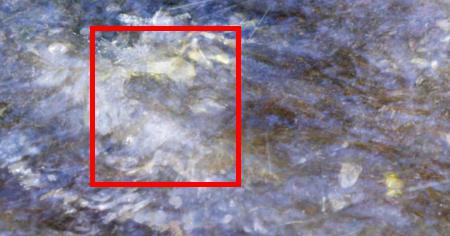}}\;\!\!\!
    \subfloat[Original]{\includegraphics[width=0.122\linewidth]{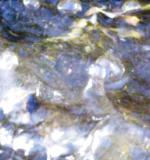}}\;\!\!\!
    \subfloat[Repeat]{\includegraphics[width=0.122\linewidth]{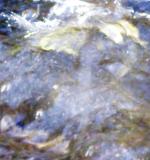}}\;\!\!\!
    \subfloat[DVF]{\includegraphics[width=0.122\linewidth]{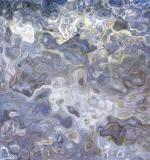}}\;\!\!\!
    \subfloat[QVI]{\includegraphics[width=0.122\linewidth]{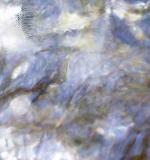}}\;\!\!\!
    \subfloat[ST-MFNet]{\includegraphics[width=0.122\linewidth]{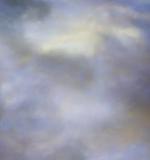}}\;\!\!\!
    \subfloat[Average]{\includegraphics[width=0.122\linewidth]{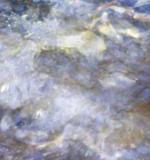}}
    \caption{Example frame blocks generated by five VFI algorithms. It is noted that for frame repeating, although the result appears less distorted, the video exhibits motion juddering.}
	\label{fig:egdistortion}
\end{figure*}

\subsection{Distorted Sequence Generation}

To generate different distorted versions of the 108 reference sequences, we first halved their frame rates by dropping every second frame, and then reconstructed the dropped frames using five VFI algorithms: frame repeating, frame averaging (where the middle frame is generated by averaging every two frames), DVF~\cite{liu2017video}, QVI~\cite{xu2019quadratic} and ST-MFNet~\cite{danier2022spatio}. The first two methods were included because they have very low computational complexity and produce unique artefact types, motion judder and motion blur respectively. The other three algorithms are all based on deep learning. {and were chosen for the following reasons.}
{
\begin{itemize}
    \item \textbf{DVF} is one of the earliest and most well-known deep learning-based VFI method that is based on flow estimation. It inspired a series of new VFI methods that improved upon DVF, and is representative of the class of VFI methods that rely on a linear motion assumption to perform frame warping.
    \item \textbf{QVI} is the first flow-based VFI method that explicitly models higher-order motion (second order). Many following works~\cite{liu2022jnmr, park2021asymmetric, liu2020enhanced} have drawn from the components of QVI to develop new methods. Therefore, QVI is a representative flow-based VFI method that assume non-linear motions.
    \item \textbf{ST-MFNet} is a state-of-the-art VFI method that is representative of kernel-based VFI approaches, where interpolation kernels are predicted and used for frame synthesis. Equipped with kernel-based warping and deep CNNs for frame processing, ST-MFNet can generate similar artefacts to many other kernel-based~\cite{adaconv, lee2020adacof, ding2021cdfi} and end-to-end methods~\cite{choi2020channel,kalluri2020flavr}.
\end{itemize}
}
For the deep learning-based models, we employ the model parameters pre-trained as in \cite{danier2022spatio} due to their proven performance on various challenging content. As a result, a total of 540 (108$\times$5) distorted videos were obtained. Fig.~\ref{fig:egdistortion} shows example frames interpolated by all five VFI methods, where it can be seen that diverse artefacts types have been generated. It should be noted that we did not employ video compression during the test sequence generation process, hence our content is free from compression artefacts, ensuring a focus solely on the perceptual quality of VFI-generated content.

\subsection{Summary}
{To summarise, in total the BVI-VFI database contains 108 reference videos and 540 distorted (interpolated) videos generated by applying five VFI algorithms to each reference video. The 108 reference sequences correspond to 36 source sequences at three frame rates and three spatial resolutions. The number of sequences employed for subjective testing was necessarily constrained due to the laboratory-based experiment conducted. Nevertheless, Table~\ref{tab:uniformity} indicates that the sequences employed provide a satisfactory coverage of video features.}

\section{Subjective Experiments}\label{sec:subjective}
This section describes the large-scale subjective test to collect quality scores for the sequences in BVI-VFI.

\subsection{Experimental Setup}
The psychophysical experiment was conducted in a darkened laboratory-based environment set up according to \cite{itu2002500}. The sequences were displayed on a BENQ XL2720Z HFR monitor with a screen size of 598$\times$336mm. The spatial and temporal resolutions of the display were set to 1920$\times$1080 and 120fps. {The peak luminance level of the display is 200lux, and there was no default motion blur/deblur settings enabled.} Note that all sequences were played back at their native spatial and temporal resolutions. The viewing distance was set to 1008mm, which is three times the picture height for HD and cropped UHD-1 sequences and six times the picture height for 540p sequences{, resulting in an effective resolution of 56 pixels per degree}. These are all compliant with ITU-R BT.500-14~\cite{itu2002500}. The monitor was connected to a Windows PC, with Matlab Psychtoolbox 3.0 software~\cite{pychotoolbox} used to control the experiment and collect user input from a wireless mouse.

\subsection{Experimental Procedure}
We employed the Double Stimulus Continuous Quality Scale (DSCQS) methodology~\cite{itu2002500} for this subjective experiment. {Compared to the single-stimulus (SSCQS) alternative which has been frequently used in previous work~\cite{madhusudana2021subjective, lee2021subjective, hosu2017konstanz, sinno2018large}, DSCQS better supports the detection of more subtle artefacts, as typically introduced in VFI. It thus provides a more accurate measure of the perceptual degradation~\cite{bull2021intelligent}, and reduces context effects~\cite{pinson2003comparing}.} In each trial of a session, the subject is presented with two sequences, A and B, one of which is the reference sequence and the other is a distorted version generated by one of the VFI algorithms. The display order of the reference and distorted videos is randomised, and the subject is unaware of which one is the reference. Sequences A and B are shown to the subject twice in an alternating fashion (i.e., A, B, A, B) interleaved with grey screens containing text informing the subject which video (A or B) will be played next. After videos are played, the subject is presented with a grey screen showing the question: ``Please rate the quality of video A'' and another such question for video B. The user is then asked to provide their answers using two continuous sliders (for A and B) with five evenly spaced ticks labelled \textit{Bad}, \textit{Poor}, \textit{Fair}, \textit{Good}, \textit{Excellent}, which correspond to 0, 25, 50, 75, 100 respectively. There are two-second grey-screen intervals between video presentations, and the rating process of the viewers are not timed.

A total of 189 subjects were paid to participate in this experiment. In order to reduce viewer fatigue caused by excessively long experiment sessions while ensuring sufficient raw subjective scores (at least 20) for each distorted video, we divided the participants into nine groups and each group of subjects were presented with sequences associated with only four source content types, which include 60 (4 content $\times$ 3 frame rates $\times$ 5 VFI methods) distorted sequences and their reference counterparts. In the test session, each participant was presented with 60 different trials (in randomised order), and each distorted video was rated by at least 20 subjects. At the beginning of each session, the participant was assessed for visual acuity with a Snellen chart and colour blindness with an Ishihara chart. The subject was then briefed about the experiment and given detailed instructions. To further familiarise him/her with the experiment, four practice trials were then performed, which contain different sequences from those in BVI-VFI. Finally, each formal test session took approximately 30 minutes on average.

\begin{figure}[t]
    \centering
    {
    \includegraphics[width=0.7\linewidth]{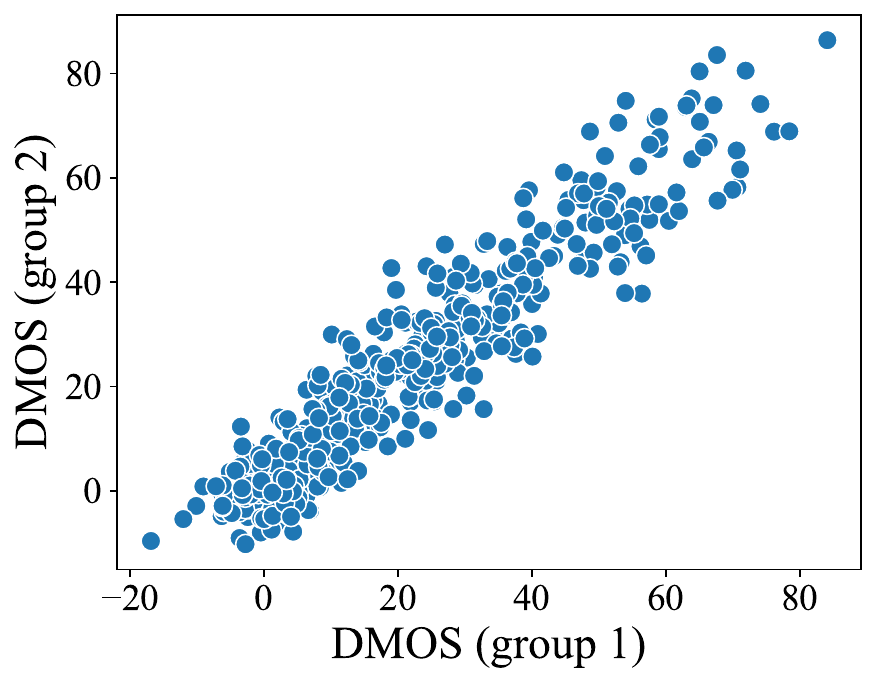}
    \caption{Scatter plot of DMOS values given by two randomly separated equal-size groups of subjects. Note that lower DMOS indicates higher perceived quality. }
	\label{fig:intersubject}
    }
\end{figure}

\begin{figure}[t]
    \centering
    {
    \subfloat[]{\includegraphics[width=0.49\linewidth]{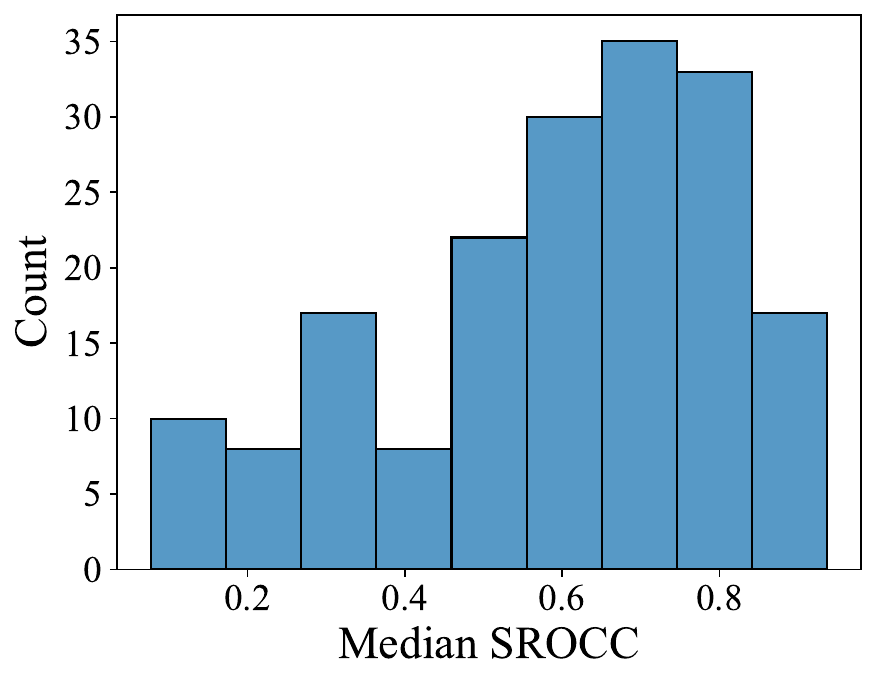}}
    \subfloat[]{\includegraphics[width=0.49\linewidth]{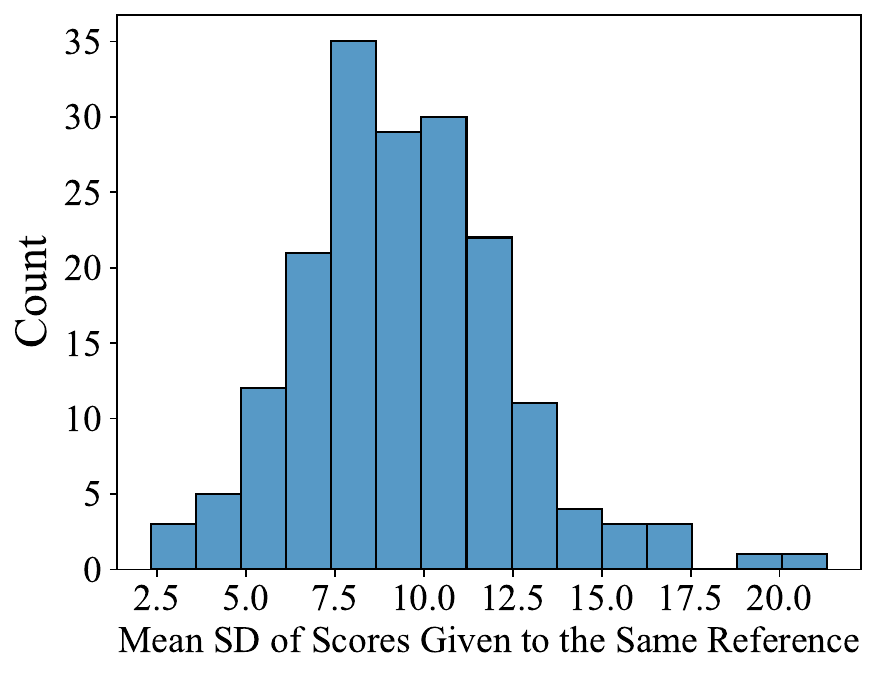}}
    \caption{(a) Histogram of median SRCC achieved by each user in self-consistency examination based MOS on reference videos. (b) Histogram of the average standard deviation of multiple scores given to the same reference video by each subject.}
	\label{fig:selfconsistency}
    }
\end{figure}

\subsection{Data Processing and Validation} \label{sec:dataprocessing}

As mentioned above, each of the 189 participants took part in only one session consisting of 60 trials. This resulted in 60 pairs of scores given by each subject for the 60 reference-distortion tuples, and each trial (reference-distortion pair) was rated by at least 20 subjects. {In order to obtain the quality scores for all the distorted videos, we first compute the differential opinion score $d_{ij}$ for each subject $i$ and for each distorted video $j$. This  computes the difference between the subject's score $s_{ij}^{\mathrm{ref}}$ for the reference version of that video and their score $s_{ij}$ for the distorted video. Namely,
\begin{equation}
    d_{ij} = s_{ij}^{\mathrm{ref}} - s_{ij}.
\end{equation}
The Differential Mean Opinion Score $\mathrm{DMOS}_{j}$ for each distorted video $j$ is obtained by averaging over its $N_j$ viewers,
\begin{equation}
    \mathrm{DMOS}_j = \frac{1}{N_j} \sum_{i=1}^{N_j} d_{ij}.\label{eqn:dmos}
\end{equation}
}

To investigate the credibility of the subjective data collected, we follow previous works~\cite{hossfeld2013best,madhusudana2021subjective} to measure the inter- and intra-subject consistencies using the DMOS values. Firstly, The inter-subject consistency~\cite{hossfeld2013best} is assessed by measuring the correlation between the average differential scores (DMOS) given by two randomly separated equal-size groups of subjects. Due to the subject grouping described, the random split was performed within each group. We performed 1000 times of such random split and at each time computed the Spearman's Ranked Order Correlation Coefficient (SRCC) between the DMOS of the two groups. The median SRCC value for the 1000 splits was found to be 0.910, and the standard deviation was 0.006. This indicates a high level of consistency among subjects. Fig.~\ref{fig:intersubject} shows the distribution of the average differential values for one such split.

\begin{figure}[t]
    \centering
    \subfloat[]{\includegraphics[width=0.49\linewidth]{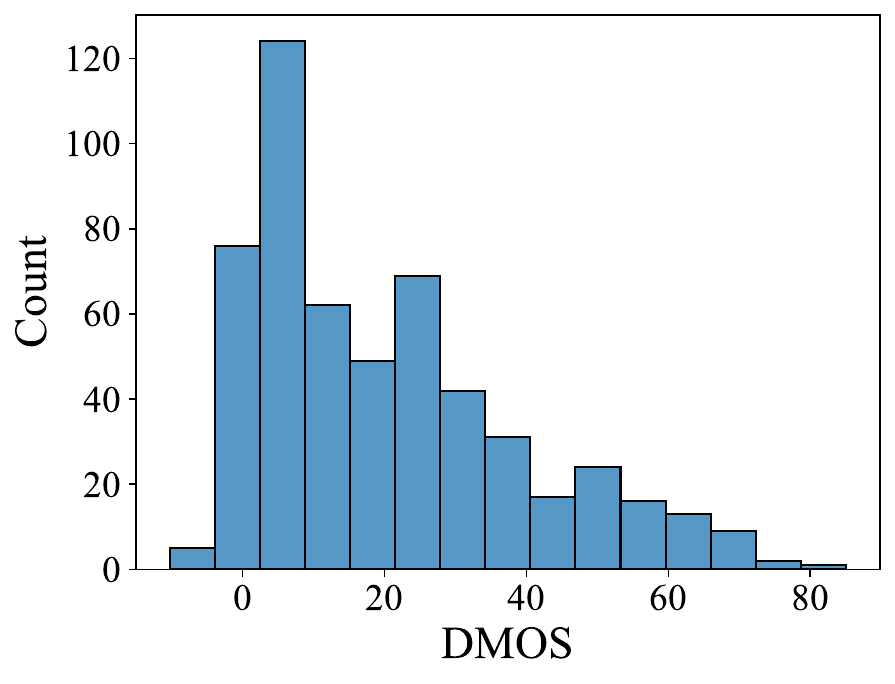}}
    \subfloat[]{\includegraphics[width=0.49\linewidth]{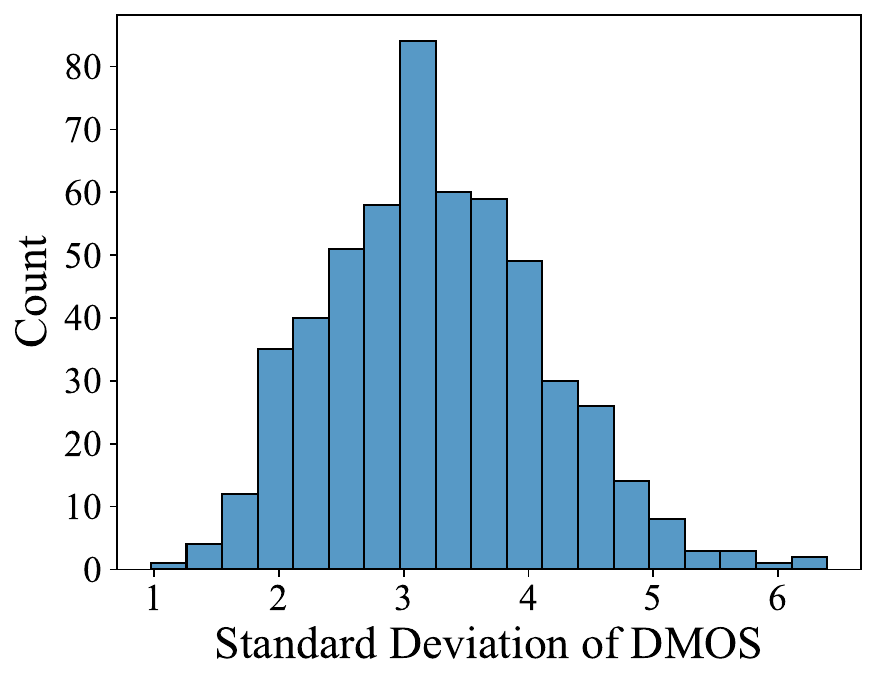}}\\
    \subfloat[]{\includegraphics[width=0.49\linewidth]{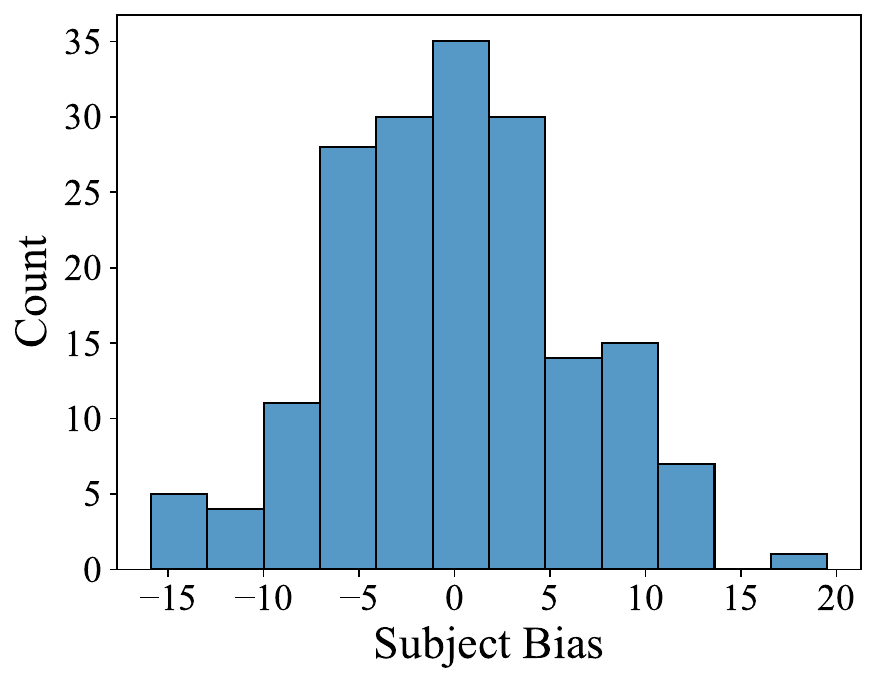}}
    \subfloat[]{\includegraphics[width=0.49\linewidth]{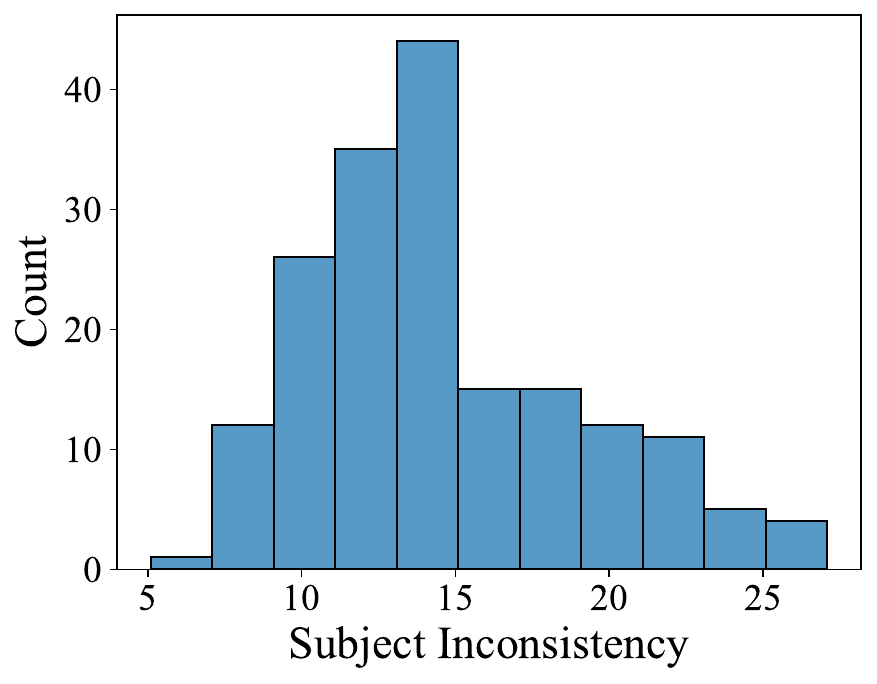}}
    \caption{{(a) Final DMOS distribution. (b) Standard deviation of DMOS over subjects. (c) Subject bias distribution. (d) Subject inconsistency distribution.}}
	\label{fig:p910}
\end{figure}

Furthermore, we also checked the intra-subject consistency~\cite{hossfeld2013best}, which is a measurement of how well the judgement of each subject agrees with all the viewers who viewed the same videos. This is evaluated by calculating the SRCC between the differential scores given by each subject to his/her allocated videos and the DMOS values given by the group for the same videos. Performing this process for every subject, we obtained a median SRCC value of 0.712 with a standard deviation of 0.119, which are reasonable values that indicate the reliability of the collected opinion scores~\cite{madhusudana2021subjective}.

\begin{figure*}[t]
    \centering
    {
    \includegraphics[width=1\linewidth]{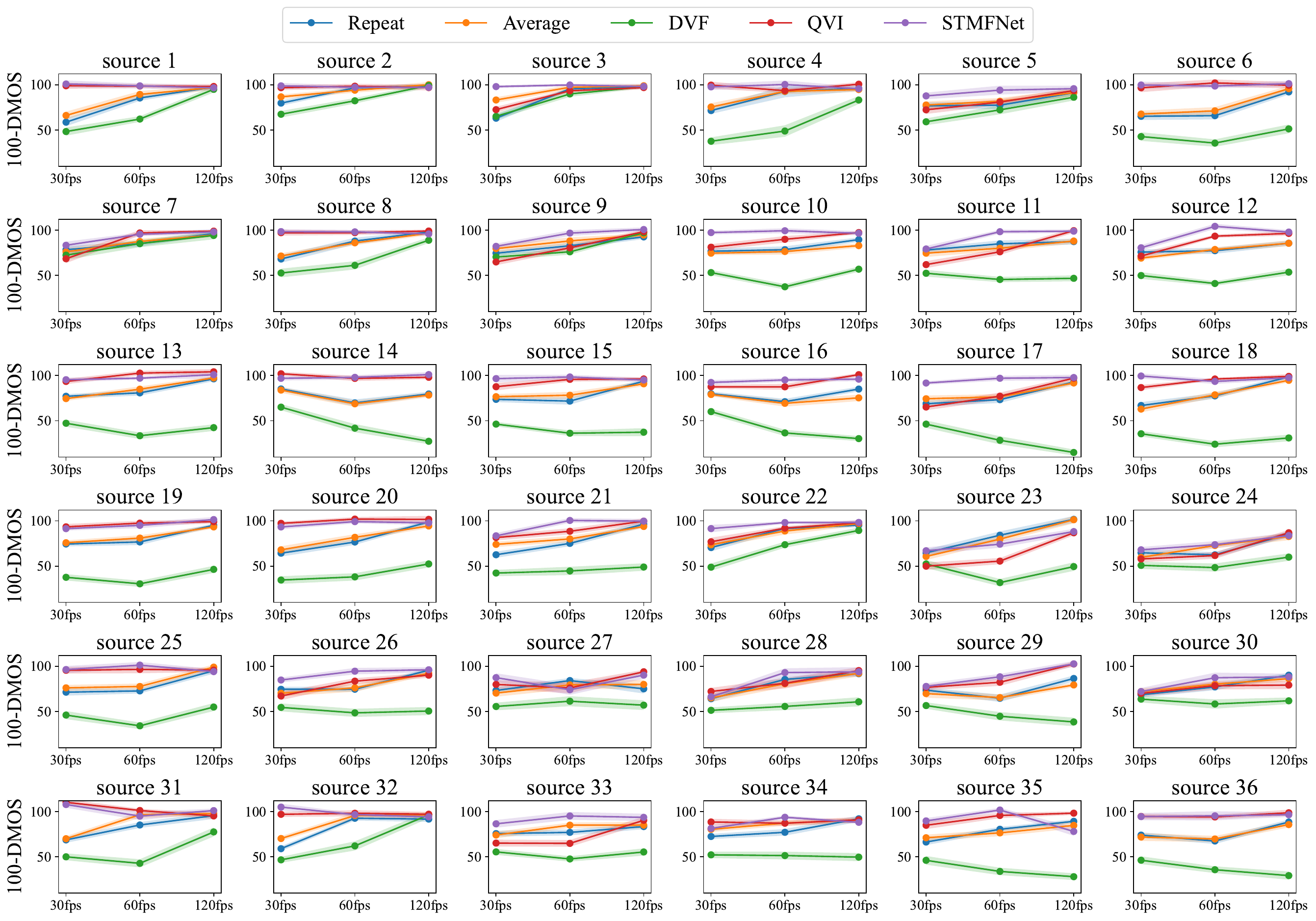}
        \vspace{-1em}
    \caption{Plots of 100-DMOS against frame rate for each VFI algorithm, for each content. Note higher DMOS means worse quality, so higher values of (100-DMOS) denote better quality. The shaded areas indicate the standard error of the subjective data.}
	\label{fig:dmosvsfpsforeachcontent}
    }
    \vspace{-1em}
\end{figure*}

{It is noted that, in each test session of our DSCQS experiment, each subject viewed and rated the same reference video multiple times. Specifically, each of the 12 reference videos in the session (see Sec. IV-B) are rated by the subject five times. We exploit this property to further measure the self-consistency of the subjects. Concretely, for each subject, we randomly split the five opinion scores given to each of the 12 reference videos into two groups (i.e. one group contains two scores for each reference and the other contains three). Within each group, we then average all the scores given to each reference sequence to obtain 12 mean opinion scores (MOS) for the references. The SRCC value is then calculated between the two sets of MOS values obtained from each group, which indicates the consistency of the subject when repeatedly rating the same videos. Such split was performed 1000 times and we recorded the median SRCC value for each subject. The histogram of the median SRCC values for all users are shown in Fig.~\ref{fig:selfconsistency} (a), where it is observed that most of the users achieved median SRCC values greater than 0.7, while there are also non-negligible number of subjects whose median SRCCs were below 0.5. The main reason for these lower SRCC values can be that subjects tend to provide relative scores instead of absolute ones. That is, the scores given to the same reference can vary significantly depending on the distorted content in the comparison. For example, when a reference video is viewed in comparison to a video interpolated by DVF~\cite{liu2017video}, it tends to receive very high scores because the subjects often try to give extreme scores to reflect the perceived quality difference. Therefore, it should be noted that the intra-/inter-subject consistency values measured on DMOS data above can be more appropriate indicators of subject consistency.} {For reference, we also calculated the standard deviation (SD) of the five scores given by each subject for each reference video (resulting in 12 SDs for each subject). The distribution of these mean SDs is shown in Fig.~\ref{fig:selfconsistency} (b).}

{Finally, to obtain more reliable DMOS scores, we follow ITU-T Recommendation P.910~\cite{p910} to perform subject screening. Specifically, the DMOS values obtained in Eqn.~\ref{eqn:dmos} are processed by the Alternating Projection (AP) algorithm specified in \cite{p910}, which alternates between (i) estimating the bias and inconsistency of all subjects using current DMOS data, and (ii) estimating new DMOS values by considering the subject bias and inconsistency.  As the algorithm terminates, it outputs final DMOS values for all the distorted videos where the subject bias is removed; the contribution of the subjects are then re-weighted according to their consistency. As side information, this algorithm also outputs the bias and inconsistency measures of all subjects. Figure~\ref{fig:p910} shows the histograms of (a) the final DMOS values, (b) the standard deviaions of DMOS, (c) subject bias, and (d) subject inconsistency. All the subsequent analysis and experiments are performed on the post-screening DMOS values.}

\subsection{Analysis of Subjective Data}
\begin{figure}[t]
    \centering
    {
    \includegraphics[width=0.8\linewidth]{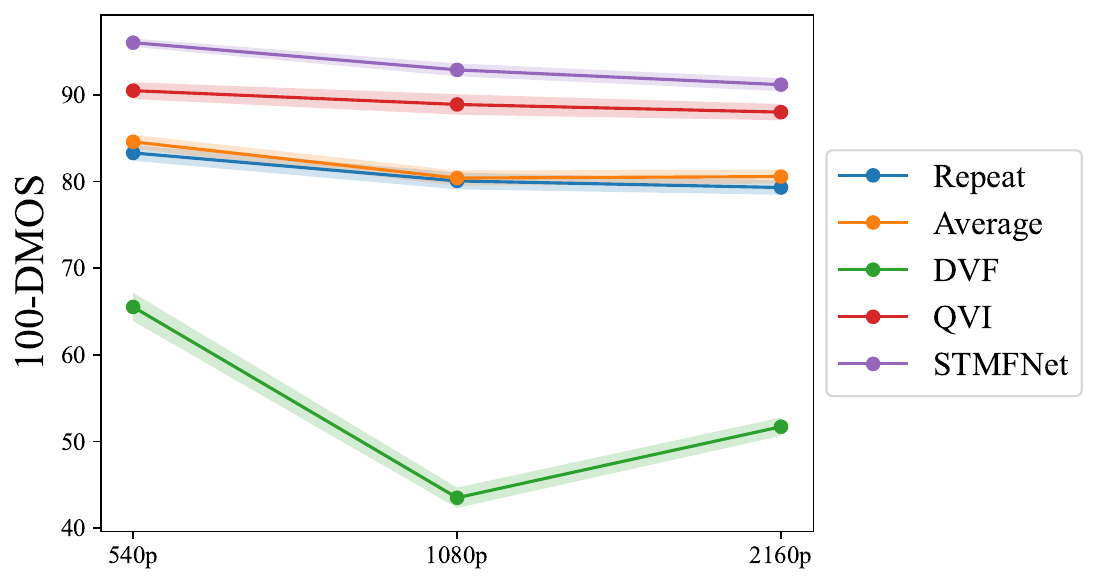}
    \caption{The average (100-DMOS) values for each VFI method at various resolutions. The shaded areas indicate standard deviations over sequences.}
	\label{fig:dmosvssr}
    }
\end{figure}

{To investigate the performance of the five VFI algorithms on different source content at different frame rates, we plot the DMOS achieved by each method against the frame rate for all the source sequences in Fig~\ref{fig:dmosvsfpsforeachcontent}. Note that higher values of 100-DMOS indicate better perceptual quality.} We can observe that in general, the more recent deep learning (DL)-based VFI methods, ST-MFNet and QVI, achieved the lower DMOS values, with ST-MFNet performing the best in most cases. The other DL-based method, DVF, received the highest overall DMOS values, which may be due to the assumption of simple linear motion between adjacent frames, i.e. the optical flows from the interpolated middle frame to both of its neighbouring frames are constrained to be symmetric. Such an assumption clearly fails in scenarios with non-linear motions, which occur in many sequences in the BVI-VFI database. {The two non-DL methods, frame repeating and averaging, achieved similarly mediocre DMOS scores. It is noted that the variation of their DMOS values w.r.t. both frame rate (Fig.~\ref{fig:dmosvsfpsforeachcontent}) and spatial resolution (Fig.~\ref{fig:dmosvssr}) follow a similar trend, while previous studies~\cite{mackin2016visibility,denes2020perceptual,jindal2021perceptual} have found different quality degradation patterns caused by motion judder (frame repeating) and blur (frame averaging). A possible reason for this is insufficient precision of the scoring scale.}

\begin{figure*}[t]
    \centering
    \subfloat {\includegraphics[width=0.195\linewidth]{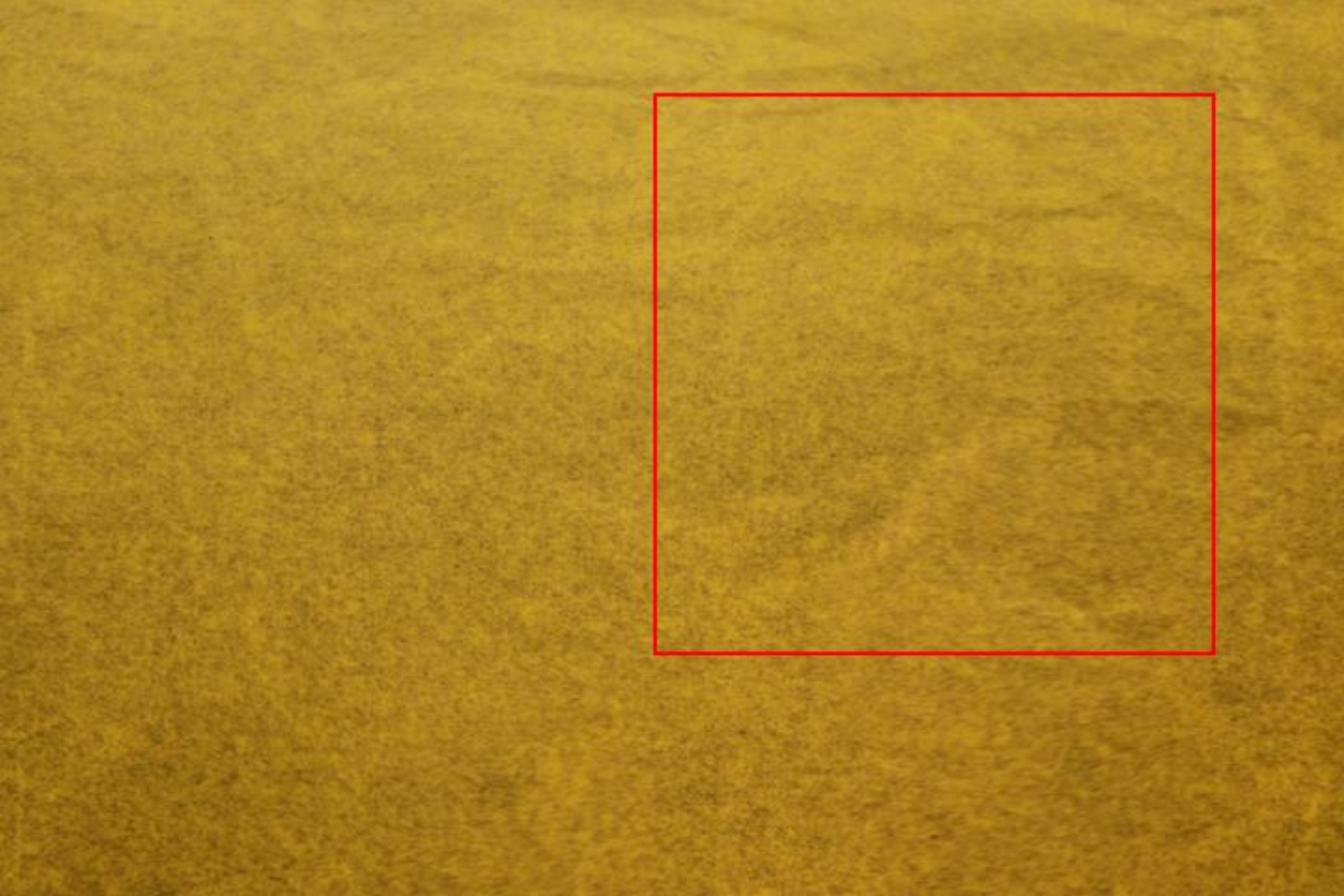}}\;\!\!\!
	\subfloat {\includegraphics[width=0.130\linewidth]{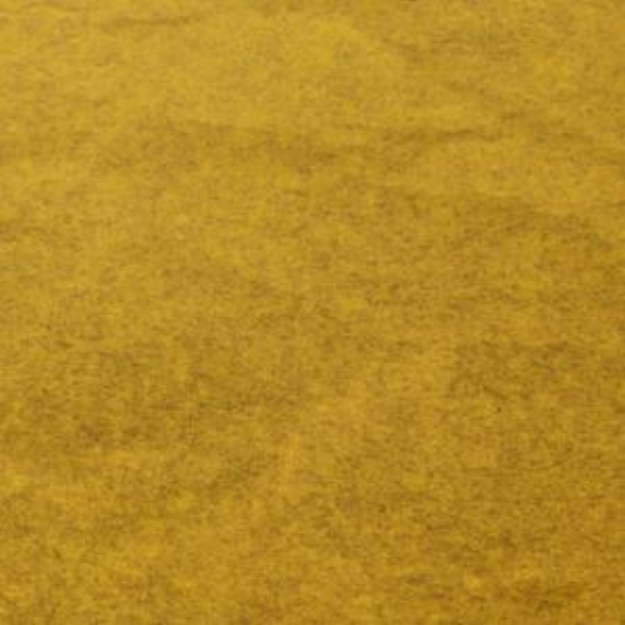}}\;\!\!\!
	\subfloat {\includegraphics[width=0.130\linewidth]{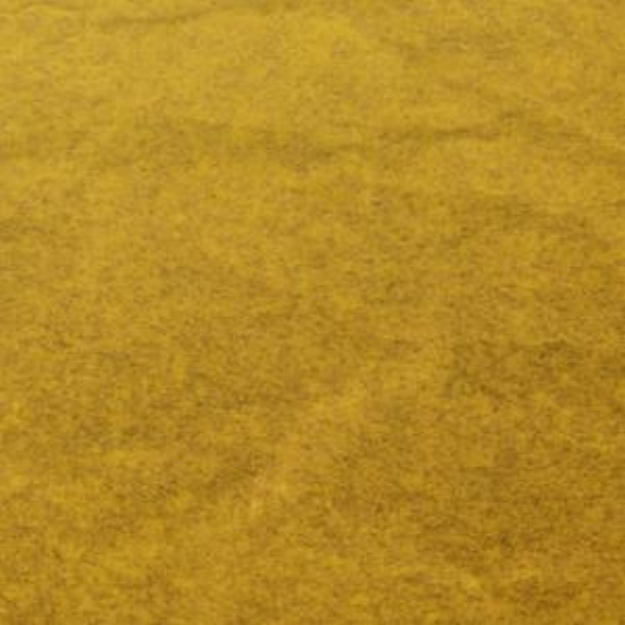}}\;\!\!\!
    \subfloat {\includegraphics[width=0.130\linewidth]{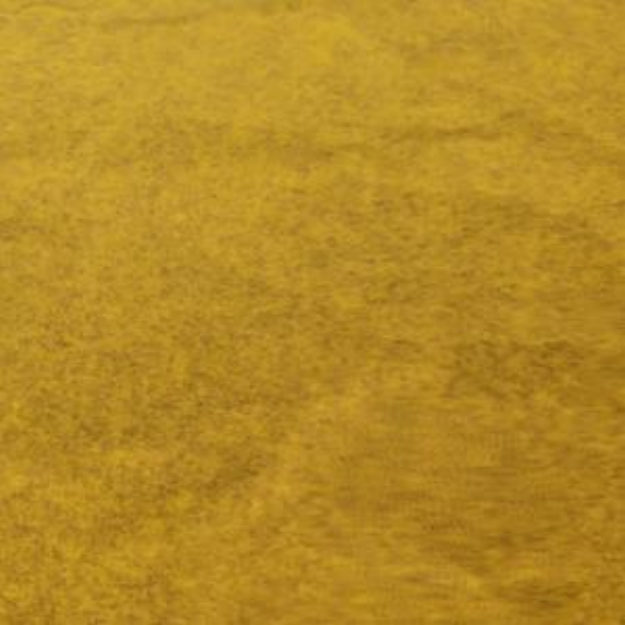}}\;\!\!\!
    \subfloat {\includegraphics[width=0.130\linewidth]{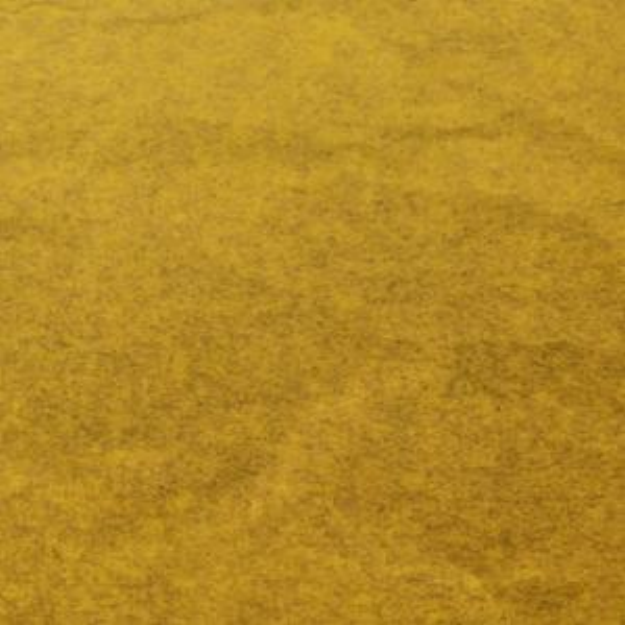}}\;\!\!\!
    \subfloat {\includegraphics[width=0.130\linewidth]{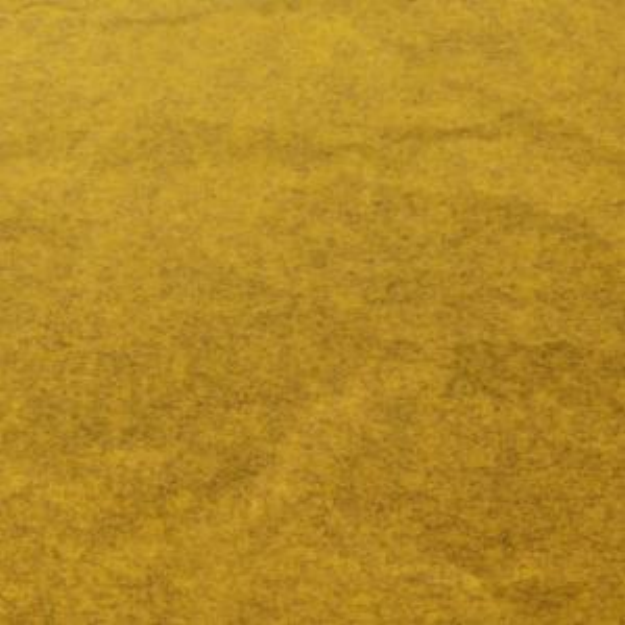}}\;\!\!\!
    \subfloat {\includegraphics[width=0.130\linewidth]{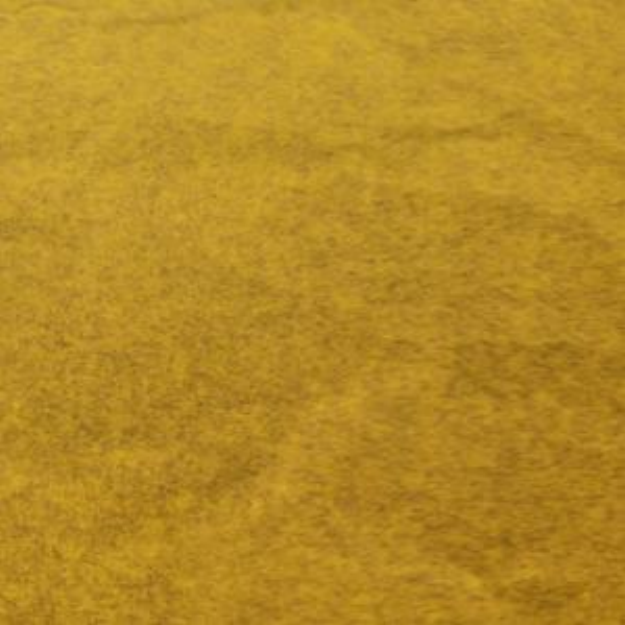}}\\
    \vspace{-3mm}
    \subfloat {\includegraphics[width=0.195\linewidth]{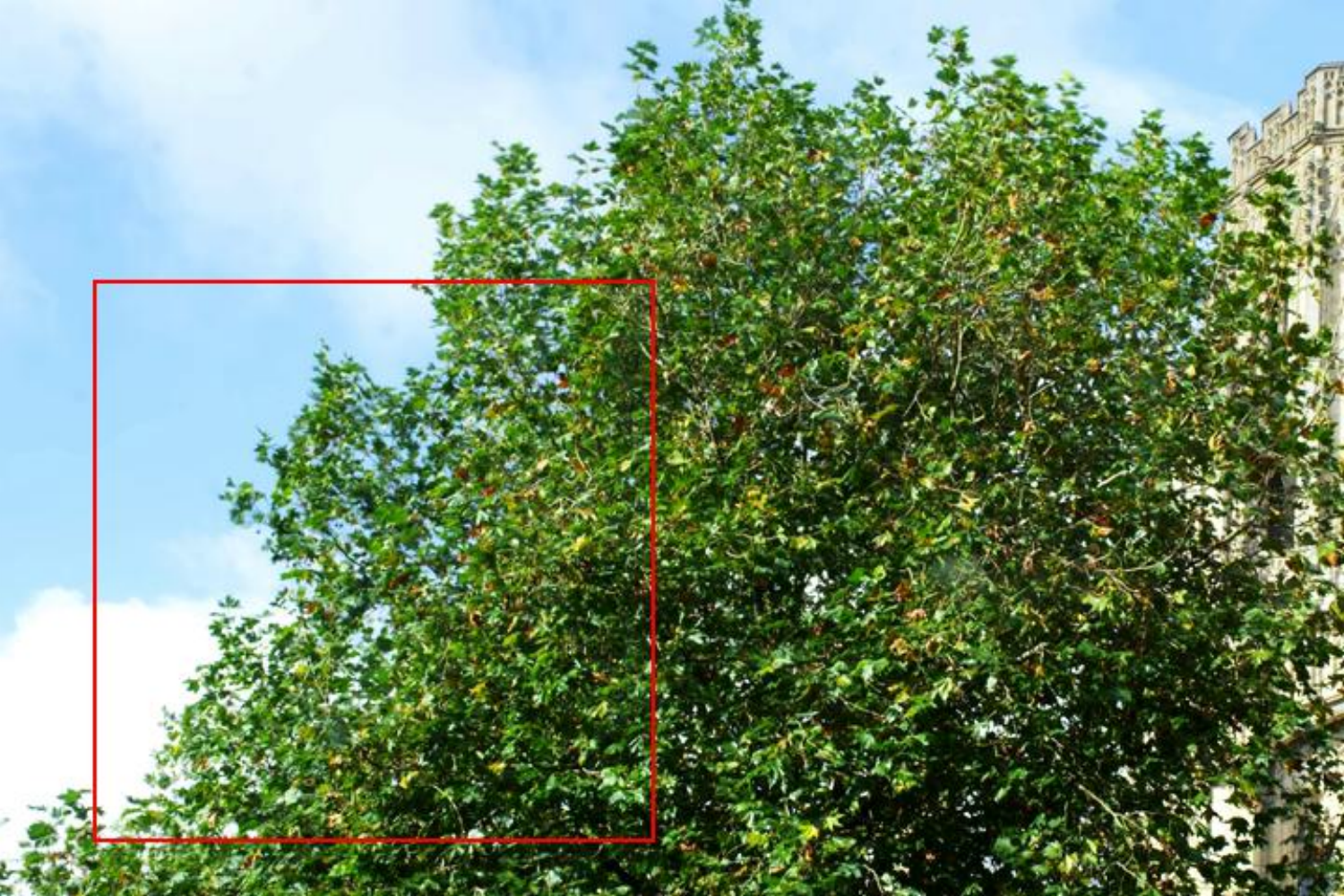}}\;\!\!\!
	\subfloat {\includegraphics[width=0.130\linewidth]{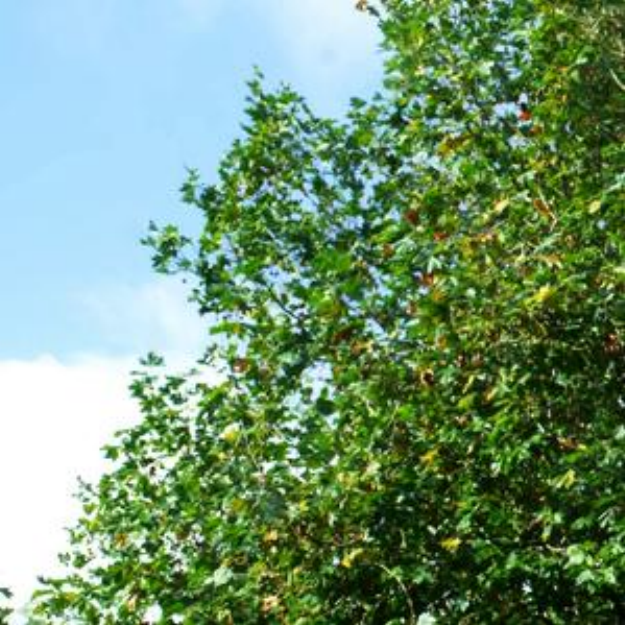}}\;\!\!\!
	\subfloat {\includegraphics[width=0.130\linewidth]{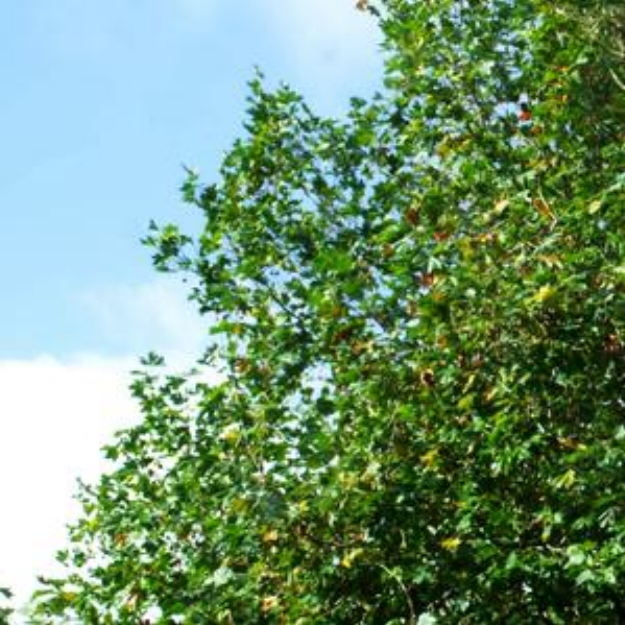}}\;\!\!\!
    \subfloat {\includegraphics[width=0.130\linewidth]{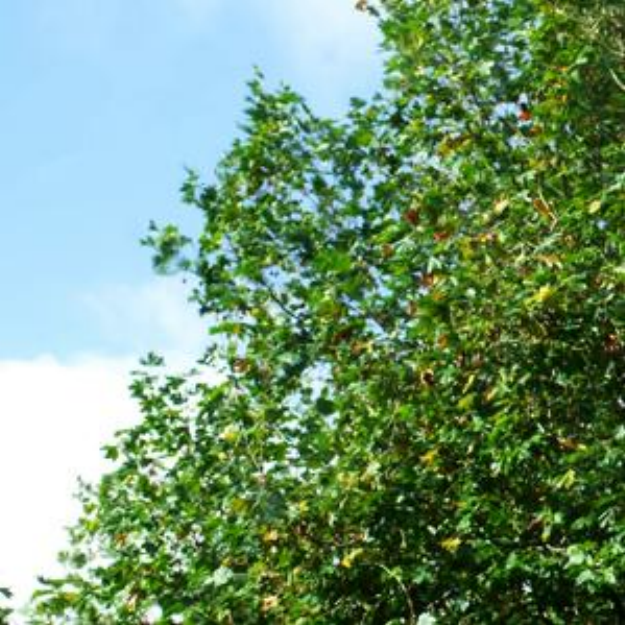}}\;\!\!\!
    \subfloat {\includegraphics[width=0.130\linewidth]{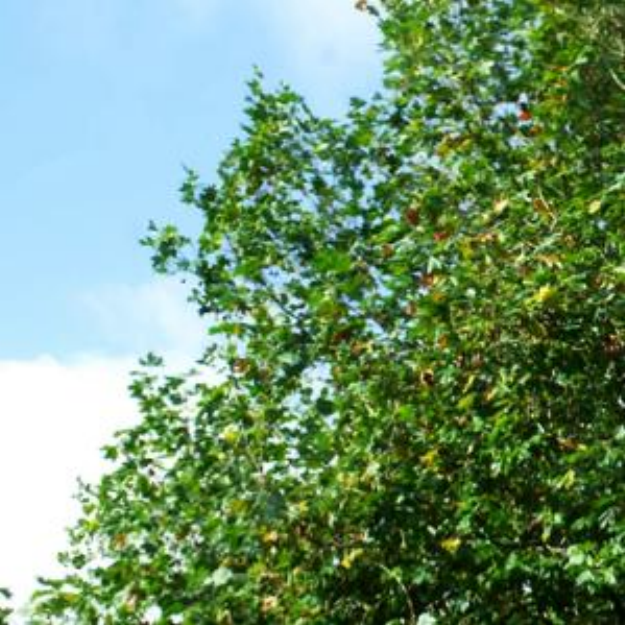}}\;\!\!\!
    \subfloat {\includegraphics[width=0.130\linewidth]{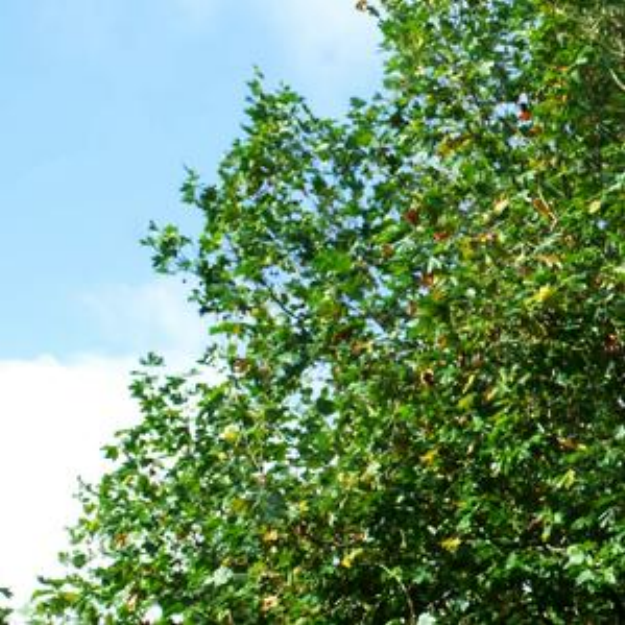}}\;\!\!\!
    \subfloat {\includegraphics[width=0.130\linewidth]{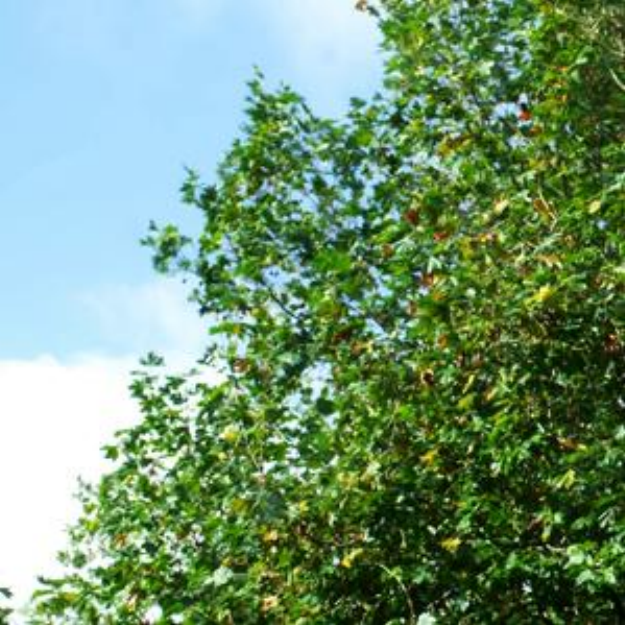}}\\
    \vspace{-3mm}
    \subfloat {\includegraphics[width=0.195\linewidth]{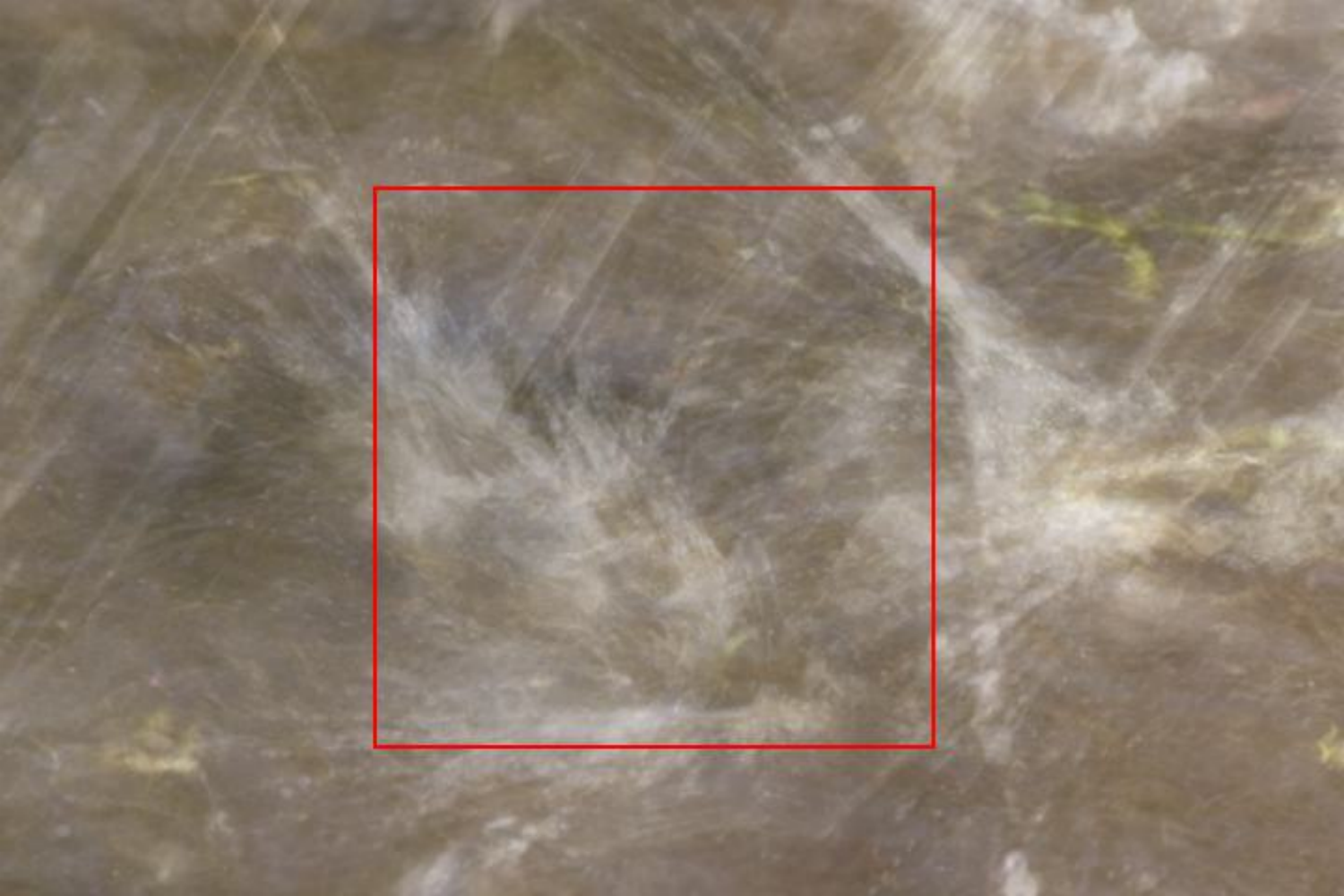}}\;\!\!\!
	\subfloat {\includegraphics[width=0.130\linewidth]{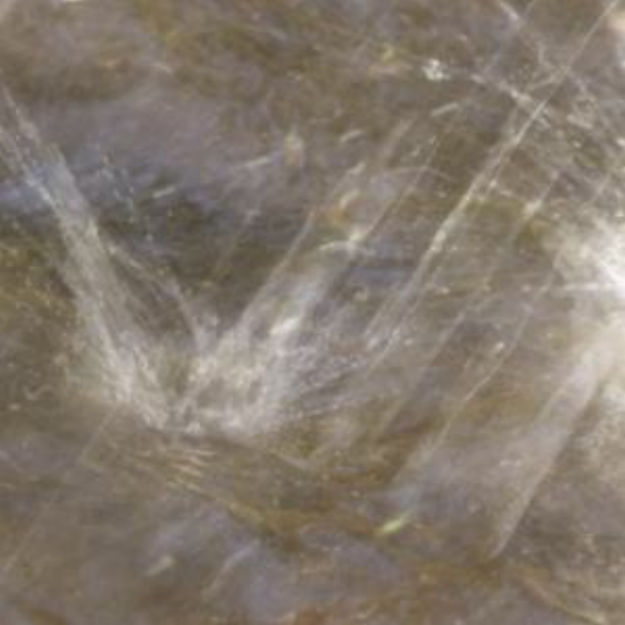}}\;\!\!\!
	\subfloat {\includegraphics[width=0.130\linewidth]{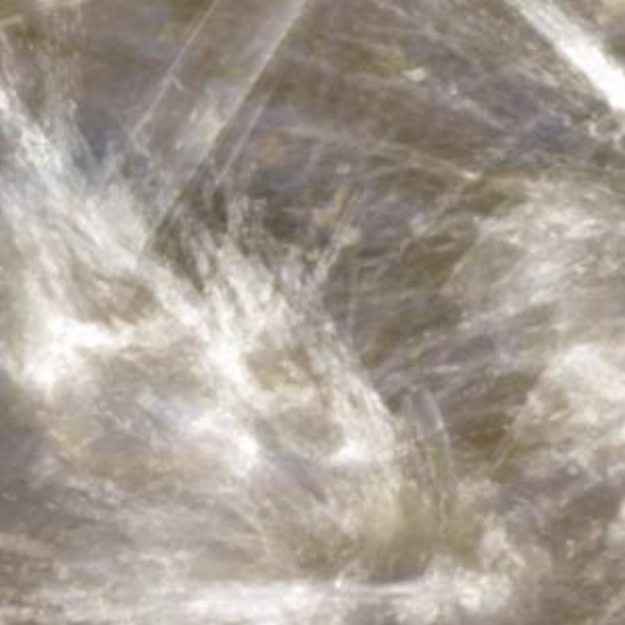}}\;\!\!\!
    \subfloat {\includegraphics[width=0.130\linewidth]{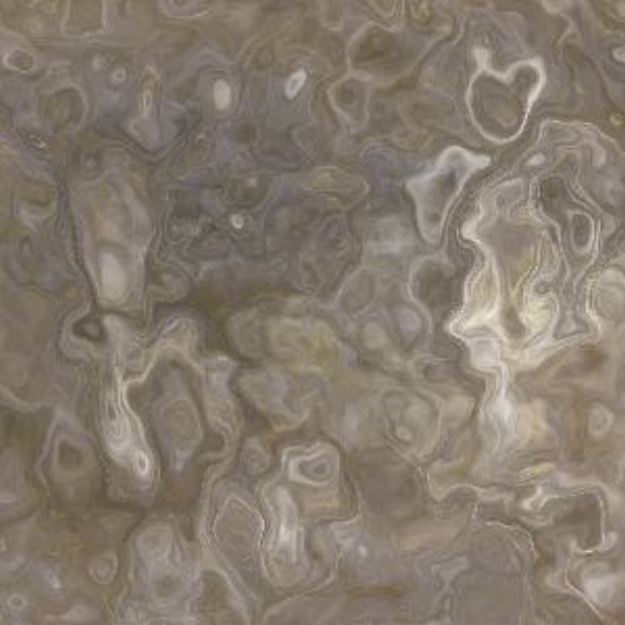}}\;\!\!\!
    \subfloat {\includegraphics[width=0.130\linewidth]{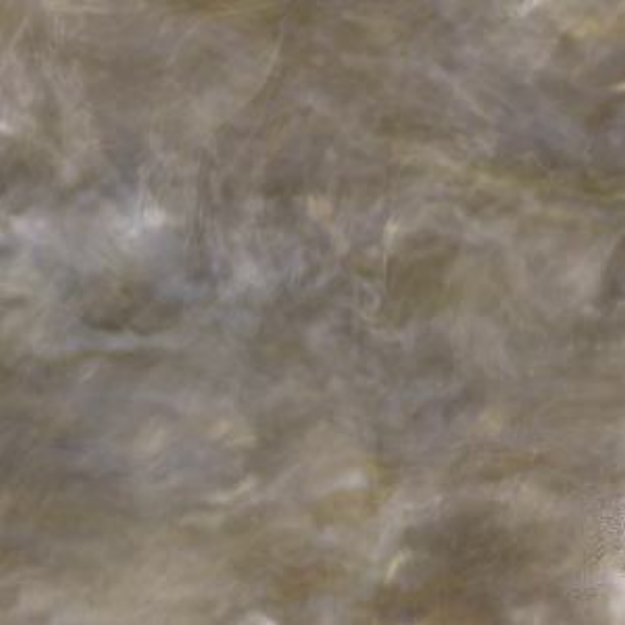}}\;\!\!\!
    \subfloat {\includegraphics[width=0.130\linewidth]{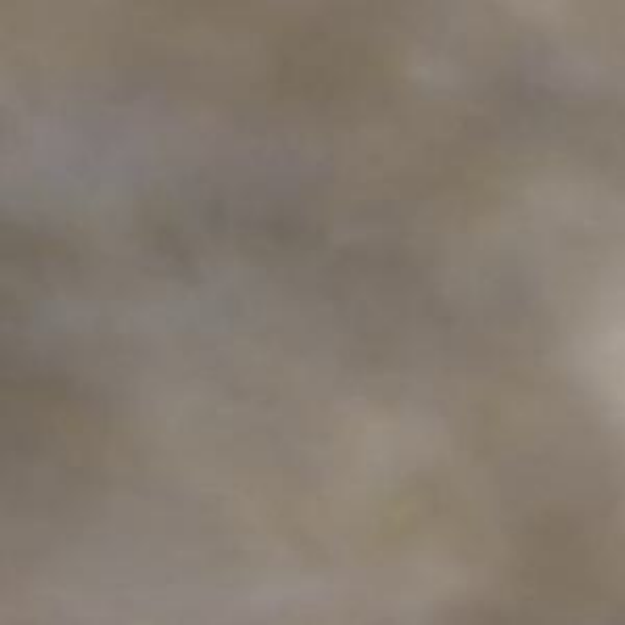}}\;\!\!\!
    \subfloat {\includegraphics[width=0.130\linewidth]{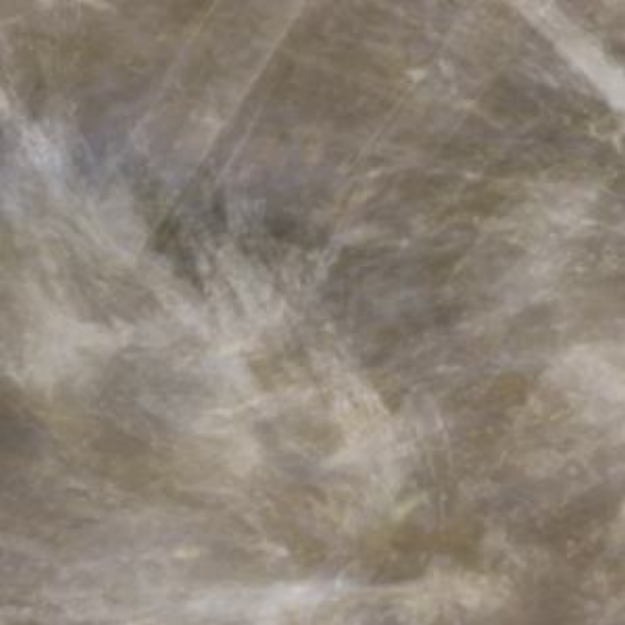}}\\
    \vspace{-3mm}
    \setcounter{subfigure}{0}
    \subfloat[Overlaid inputs] {\includegraphics[width=0.195\linewidth]{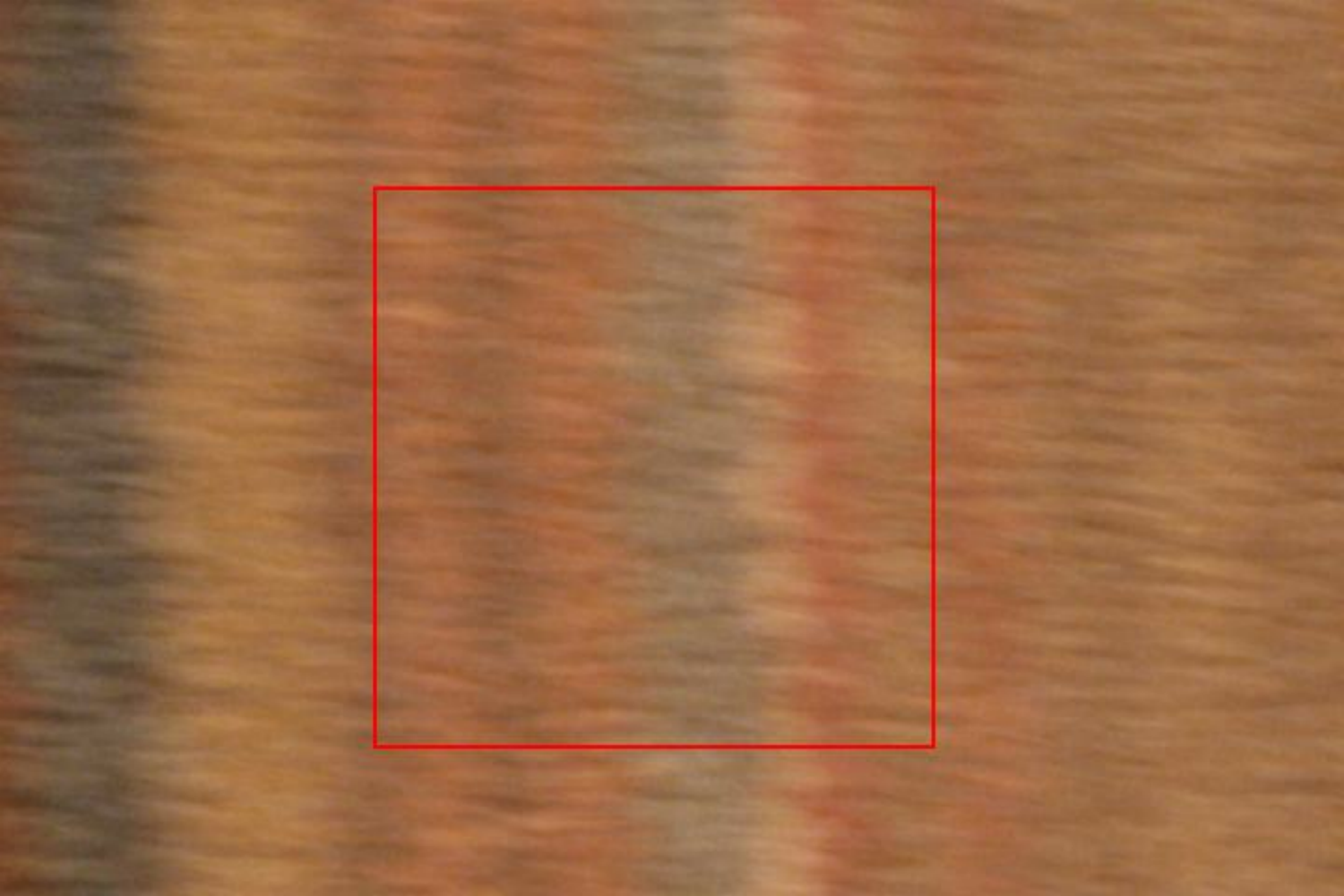}}\;\!\!\!
	\subfloat[Original] {\includegraphics[width=0.130\linewidth]{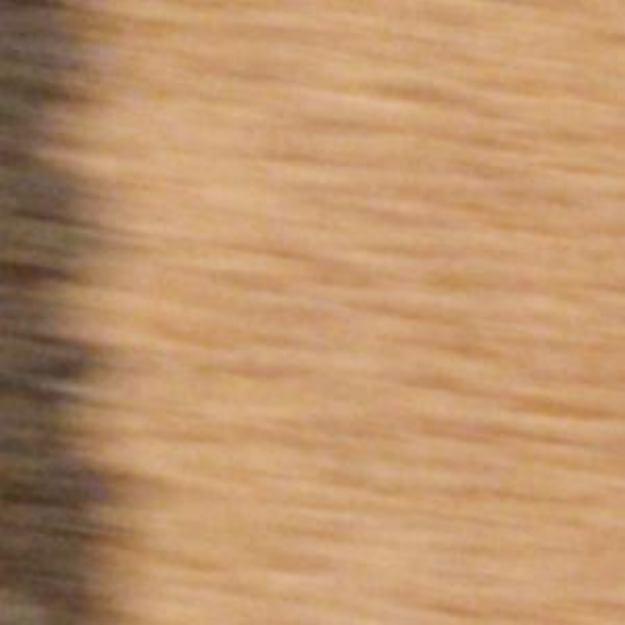}}\;\!\!\!
	\subfloat[Repeat] {\includegraphics[width=0.130\linewidth]{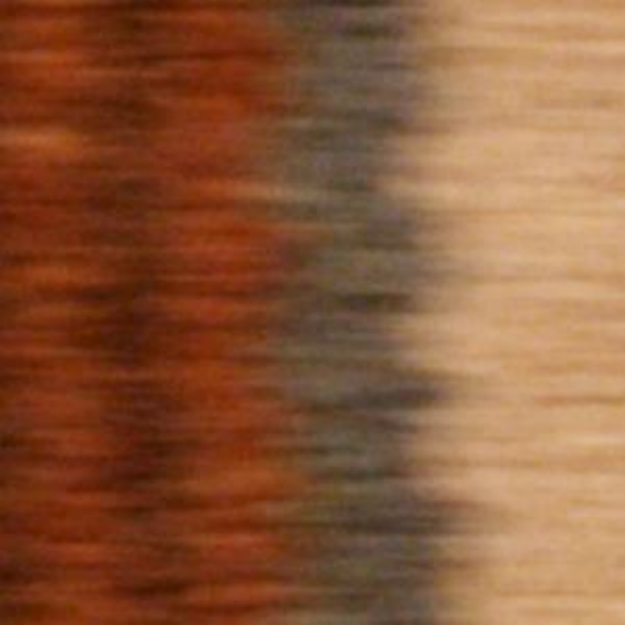}}\;\!\!\!
    \subfloat[DVF] {\includegraphics[width=0.130\linewidth]{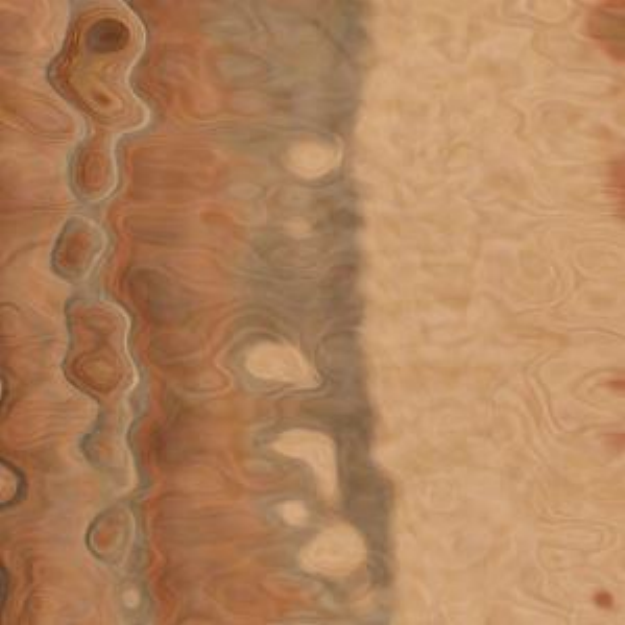}}\;\!\!\!
    \subfloat[QVI] {\includegraphics[width=0.130\linewidth]{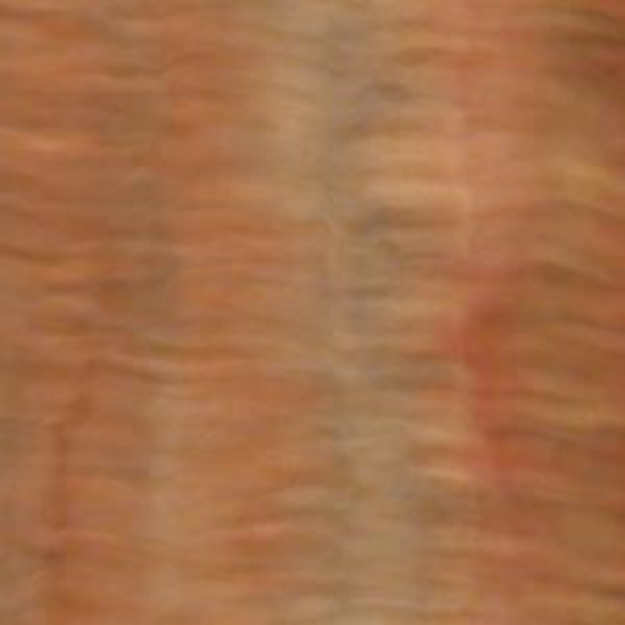}}\;\!\!\!
    \subfloat[ST-MFNet] {\includegraphics[width=0.130\linewidth]{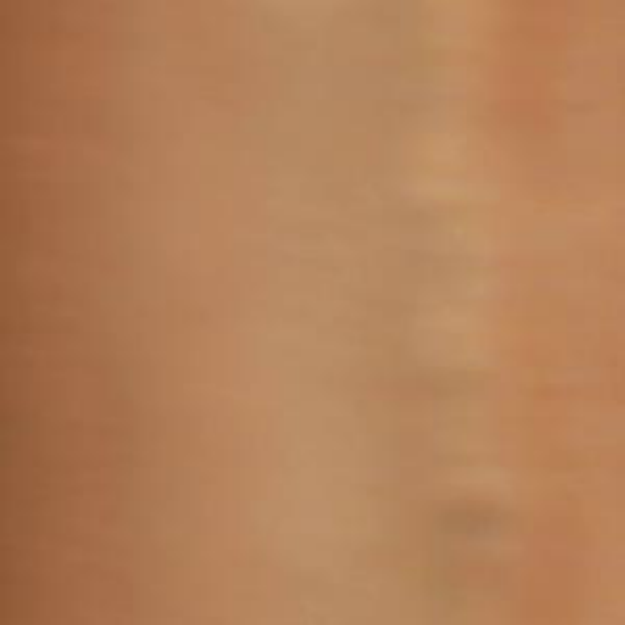}}\;\!\!\!
    \subfloat[Average] {\includegraphics[width=0.130\linewidth]{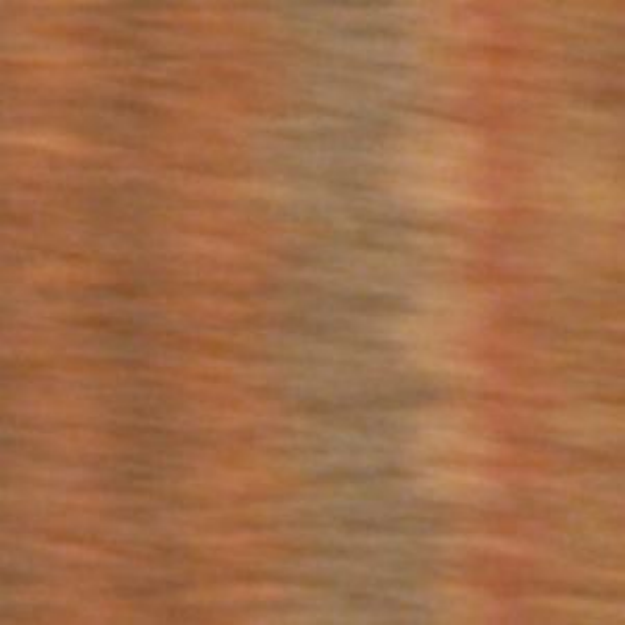}}
    \caption{Visual examples of sequences where the VFI methods performed well and poorly. The first two rows show sequences that achieved the lowest overall DMOS {(the best quality)}, while the last two rows correspond to sequences with highest overall DMOS {(the worst quality)}.}
	\label{fig:goodpoorexamples}
\end{figure*}

It can also be observed that there are several scenes where the simple non-DL methods achieve similar or even better perceived quality than the two best-performing DL-based algorithms. Most of these scenarios involve dynamic textures~\cite{peteri2010dyntex}, which exhibit rapid and irregular motions of unstructured entities, such as leaves (sources 3, 5),  water (sources 7, 9, 24, 28, 30), fire (source 23) and smoke (sources 33 and 34). There are two reasons for this. Firstly, such complex motions, sometimes combined with large motion magnitude, are too challenging for these two advanced DL-based algorithms QVI and ST-MFNet, causing them to produce frames with unacceptable quality and obvious interpolation artefacts. Secondly, the texture masking effect introduced by such dynamic textures decreases the sensitivity of the human visual system, thus making it difficult to distinguish between different distortions and resulting in similar levels of perceived quality. The images on the third row of Fig.~\ref{fig:egdistortion} (which correspond to source 30) illustrate both scenarios, where the complex scene of splashing water exhibits texture masking and poses significant challenges to DL-based VFI methods.

{In terms of the effect of frame rate on the interpolation quality, Fig.~\ref{fig:dmosvsfpsforeachcontent} shows that the difference in the DMOS values achieved by the state-of-the-art DL-based methods and those obtained by the simple non-DL methods becomes smaller at higher frame rates. It is reasonable to conjecture that, at even higher frame rates, this difference can become negligible and the computationally heavy DL-based methods can be replaced by simple frame repeating or averaging. However, this needs to be confirmed by further study.}

\begin{figure*}[t]
    \centering
    {
    \includegraphics[width=\linewidth]{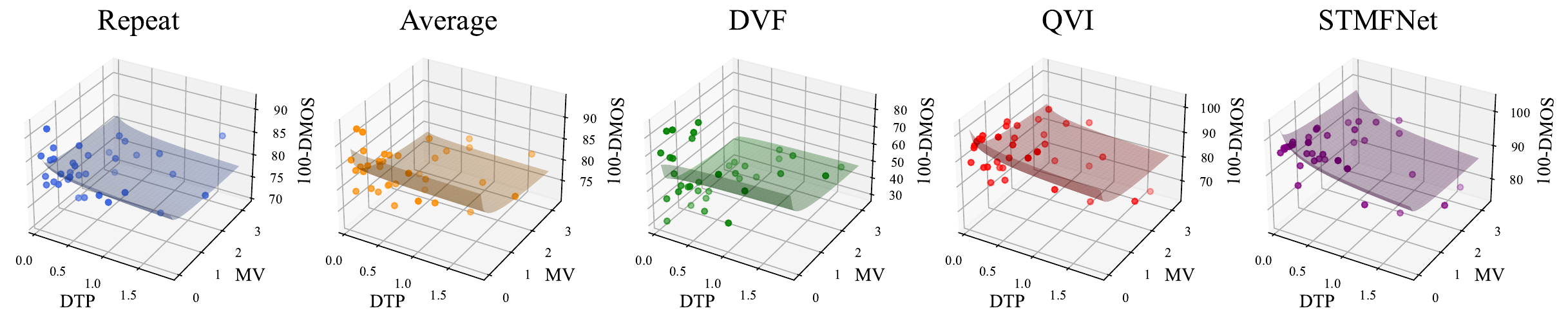}
    \includegraphics[width=\linewidth]{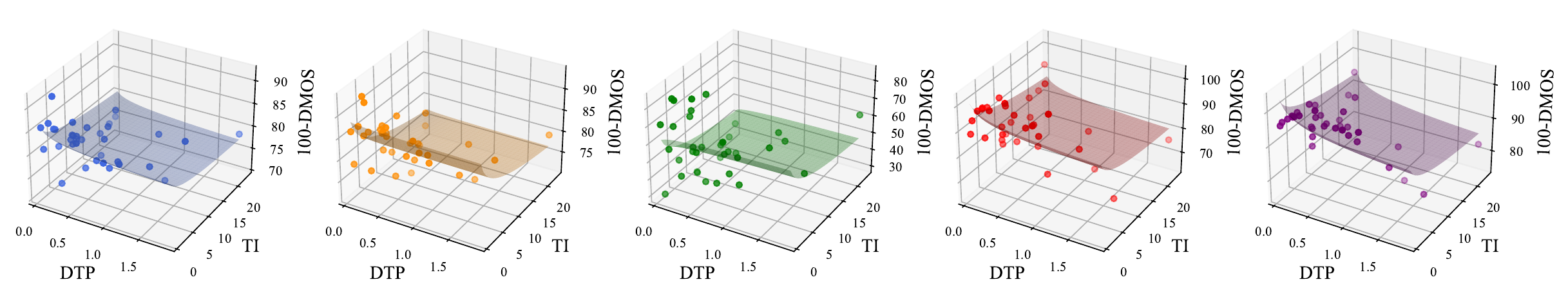}
    \caption{DMOS values for the interpolated sequences plotted against DTP and MV (first row)/TI (second row).}
	\label{fig:dmosvsdtpvsmv}
    }
\end{figure*}

\begin{figure*}[t]
    \centering
    {
    \includegraphics[width=\linewidth]{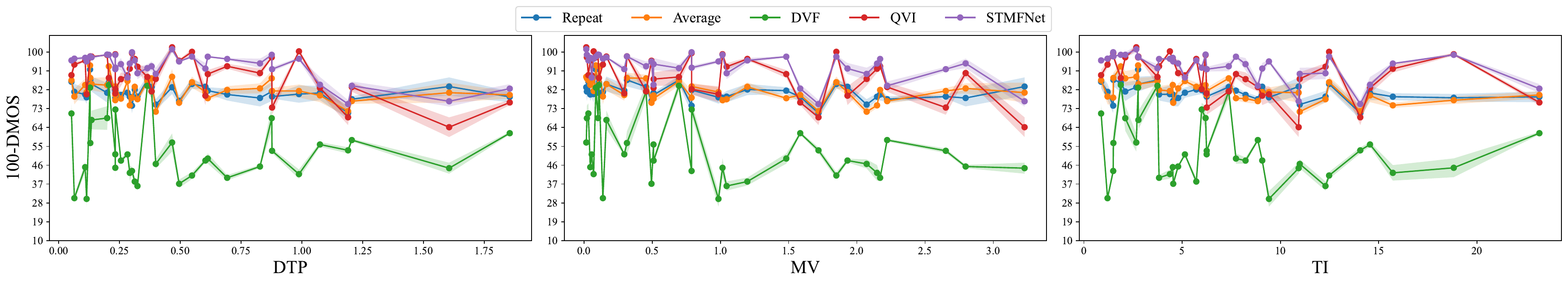}
    \vspace{-2em}
    \caption{Two-dimensional plots of 100-DMOS against different video features. The shaded area denotes the standard error over sequences.}
	\label{fig:dmosvsfeatures_2d}
    }
\end{figure*}

{Fig.~\ref{fig:goodpoorexamples} shows sampled blocks from the test sequences that received the lowest (the first two rows) and the highest overall DMOS (the last two rows), corresponding to the best and worst overall quality respectively. The first example corresponds to a $960\times 540$ video recording a relatively static scene - the yellow paper being blown slightly, which results in low values of motion vector (MV=0.08) and dynamic texture parameter (DTP=0.01) features. The low scene dynamics could be the reason for the better performance of other VFI algorithms. The second example corresponds to a scene of foliage with random movements, but the MV (0.11) and DTP (0.59) values are higher than in the first example. The reason for the higher perceived quality in this case could be the texture masking effect of the human vision system~\cite{bull2021intelligent}, which is able to ``hide'' artefacts to some extent. Regarding the poor-quality examples, the first one (the third row) is a $1920\times 1080$ video showing water splash, with high MV (1.72) and DTP (1.19) values that made the VFI particularly challenging. Although the texture masking still exists, it appears that the effect was not strong enough to hide the evident artefacts produced by the VFI methods. Lastly, the second example where VFI performed poorly involves horizontal camera movement in a scene that contains vertically aligned textures, resulting in a reasonable DTP of 0.40 but a high MV of 2.07. As shown in Fig.~\ref{fig:goodpoorexamples}, all methods showed strong artefacts that are easily perceived.}

{To further study the effect of video features (MV, DTP and TI) on VFI performance, we plot the DMOS values (100-DMOS) of the distorted videos against these features for each algorithm in Fig.~\ref{fig:dmosvsdtpvsmv}, where a surface function is fitted on the scatter points. It is observed from Fig.~\ref{fig:dmosvsdtpvsmv} that in general, the combination of higher DTP and MV/TI values lead to degraded VFI quality, implying that the combination of these feature directly affects the VFI difficulty. However, looking at the effect of individual video features on the perceived VFI quality in Fig.~\ref{fig:dmosvsfeatures_2d}, the trend is less clear, implying that a single feature can be less indicative of the VFI difficulty.}

{Fig.~\ref{fig:dmosvssr} shows the overall perceived quality at different spatial resolutions for each VFI algorithm. It is observed that for all algorithms except DVF, the perceived quality of their interpolation results decreases as the spatial resolution increases. However, it should be noted that at various resolutions, the source content is different, which indicates that the results are also affected by the video content.}

\section{Evaluation of Objective Quality Metrics}\label{sec:objective}
One of the key motivations for developing the new BVI-VFI database was to provide a rigorous framework for evaluating the performance of quality assessment metrics on frame interpolated content. In this section, we perform a comprehensive evaluation with 33 conventional and state-of-the-art image and video quality assessment methods. While the majority of the tested metrics are full/reduced-reference models (which make use of certain information from the pristine reference frames), several no-reference ones (which predict the quality solely based on the distorted frames) are also included.

\begin{table*}[t]
\centering
\caption{{The performance of the tested quality metrics on the DMOS values in BVI-VFI dataset across different frame rates and interpolation methods. In each column, the best and the second best models are \textbf{boldfaced} and \underline{underlined}.}}
\label{tab:overallfpsmethod}
\scriptsize
\begin{tabular}{c||c|c||c|c||c|c||c|c||c|c||c|c|c|c}
\toprule
  & \multicolumn{2}{c||}{30fps} & \multicolumn{2}{c||}{60fps} & \multicolumn{2}{c||}{120fps} & \multicolumn{2}{c||}{non-DL} & \multicolumn{2}{c||}{DL} & \multicolumn{4}{c}{Overall}\\
\midrule
Model & SRCC & PLCC & SRCC & PLCC & SRCC & PLCC & SRCC & PLCC & SRCC & PLCC & SRCC & KRCC & PLCC & RMSE \\
\midrule 
ST-GREED       & 0.11          & 0.16          & 0.14          & 0.33          & 0.03          & 0.27          & 0.06          & 0.11          & 0.08          & 0.30          & 0.06          & 0.04          & 0.26          & 18.04          \\
CONTRIQUE   & 0.17          & 0.33          & 0.19          & 0.39          & 0.27          & 0.45          & 0.13          & 0.17          & 0.42          & 0.44          & 0.25          & 0.17          & 0.39          & 17.23          \\
deepIQA\_FR & 0.26          & 0.36          & 0.43          & 0.43          & 0.44          & 0.40          & 0.36          & 0.43          & 0.48          & 0.47          & 0.45          & 0.30          & 0.43          & 16.83          \\
PieAPP      & 0.40          & 0.42          & 0.47          & 0.45          & 0.47          & 0.41          & 0.30          & 0.34          & 0.55          & 0.50          & 0.48          & 0.34          & 0.45          & 16.66          \\
VIF         & 0.34          & 0.36          & 0.53          & 0.44          & 0.53          & 0.44          & 0.36          & 0.37          & 0.53          & 0.50          & 0.51          & 0.35          & 0.43          & 16.84          \\
C3DVQA      & 0.25          & 0.34          & 0.57          & 0.45          & {\ul 0.66}    & 0.42          & 0.37          & 0.41          & 0.60          & 0.49          & 0.54          & 0.37          & 0.43          & 16.89          \\
LPIPS       & 0.39          & 0.40          & 0.56          & 0.52          & 0.59          & 0.54          & 0.40          & 0.40          & 0.61          & 0.58          & 0.56          & 0.39          & 0.52          & 16.00          \\
DISTS       & 0.41          & 0.44          & 0.54          & 0.55          & 0.59          & \textbf{0.61} & 0.38          & 0.40          & 0.63          & 0.60          & 0.57          & 0.40          & 0.55          & 15.54          \\
VMAF        & 0.42          & 0.43          & 0.62          & 0.54          & 0.61          & 0.50          & 0.38          & 0.42          & 0.63          & 0.60          & 0.58          & 0.40          & 0.52          & 15.97          \\
FRQM        & 0.44          & 0.50          & 0.64          & {\ul 0.60}    & 0.62          & 0.57          & \textbf{0.80} & \textbf{0.82} & 0.49          & 0.47          & 0.58          & 0.41          & 0.50          & 16.14          \\
STRRED      & 0.45          & 0.35          & 0.63          & 0.46          & 0.58          & 0.45          & 0.40          & 0.41          & 0.66          & 0.47          & 0.60          & 0.42          & 0.53          & 15.86          \\
SSIM        & 0.47          & 0.49          & 0.64          & 0.56          & 0.59          & 0.49          & 0.39          & 0.42          & 0.66          & 0.64          & 0.60          & 0.42          & 0.54          & 15.74          \\
FloLPIPS    & 0.47          & 0.49          & 0.59          & 0.57          & 0.61          & {\ul 0.58}    & 0.43          & 0.46          & 0.67          & 0.63          & 0.61          & 0.43          & {\ul 0.58}    & 15.26          \\
MSSSIM      & 0.43          & 0.46          & 0.64          & 0.56          & 0.61          & 0.52          & 0.42          & 0.44          & 0.67          & 0.63          & 0.62          & 0.43          & 0.54          & 15.67          \\
FSIM        & 0.44          & 0.48          & 0.63          & 0.55          & 0.62          & 0.54          & 0.43          & 0.46          & 0.66          & 0.63          & 0.62          & 0.42          & 0.55          & 15.64          \\
GMSD        & \textbf{0.49} & {\ul 0.52}    & 0.65          & 0.58          & 0.63          & 0.53          & 0.40          & 0.47          & 0.68          & 0.66          & 0.63          & 0.43          & 0.57          & 15.36          \\
VSI         & 0.45          & 0.49          & 0.65          & 0.58          & 0.64          & 0.54          & 0.43          & 0.47          & 0.68          & 0.65          & 0.63          & 0.44          & 0.57          & 15.38          \\
FovVideoVDP & 0.42          & 0.45          & 0.64          & 0.55          & 0.61          & 0.51          & {\ul 0.54}    & {\ul 0.58}    & 0.66          & 0.61          & 0.64          & 0.44          & 0.56          & 15.44          \\
SpEED       & {\ul 0.48}    & 0.40          & {\ul 0.67}    & 0.51          & 0.63          & 0.53          & 0.43          & 0.32          & {\ul 0.70}    & 0.59          & 0.64          & {\ul 0.45}    & 0.57          & 15.37          \\
PSNR        & 0.47          & {\ul 0.52}    & {\ul 0.67}    & {\ul 0.60}    & 0.63          & 0.55          & 0.44          & 0.49          & {\ul 0.70}    & {\ul 0.68}    & {\ul 0.65}          & {\ul 0.45}    & {\ul 0.58}    & {\ul 15.16}    \\
FAST        & \textbf{0.49} & \textbf{0.54} & \textbf{0.73} & \textbf{0.65} & \textbf{0.72} & \textbf{0.61} & 0.45          & 0.47          & \textbf{0.77} & \textbf{0.76} & \textbf{0.70} & \textbf{0.50} & \textbf{0.63} & \textbf{14.54} \\
\midrule
ChipQA      & 0.02          & 0.14          & 0.06          & 0.11          & 0.04          & 0.20          & 0.01          & 0.07          & 0.06          & 0.14          & 0.01          & 0.00          & 0.12          & 18.55          \\
BRISQUE     & 0.14          & 0.24          & 0.03          & 0.15          & 0.08          & 0.10          & 0.18          & 0.22          & 0.07          & 0.19          & 0.01          & 0.01          & 0.13          & 18.52          \\
NIQE        & 0.15          & 0.25          & 0.06          & 0.15          & 0.16          & 0.20          & 0.20          & 0.20          & 0.01          & 0.24          & 0.04          & 0.02          & 0.14          & 18.49          \\
deepIQA\_NR & 0.10          & 0.16          & 0.08          & 0.10          & 0.01          & 0.11          & 0.05          & 0.08          & 0.07          & 0.12          & 0.05          & 0.04          & 0.12          & 18.54          \\
VIDEVAL     & 0.26          & 0.31          & 0.13          & 0.12          & 0.03          & 0.13          & 0.01          & 0.14          & 0.12          & 0.15          & 0.08          & 0.06          & 0.14          & 18.50          \\
VIIDEO      & 0.01          & 0.07          & 0.05          & 0.26          & 0.17          & 0.31          & 0.01          & 0.11          & 0.25          & 0.26          & 0.10          & 0.07          & 0.19          & 18.32          \\
FastVQA     & 0.10          & 0.17          & 0.18          & 0.15          & 0.22          & 0.27          & 0.28          & 0.29          & 0.11          & 0.13          & 0.16          & 0.11          & 0.14          & 18.49          \\
\bottomrule	
\end{tabular}
\end{table*}

\begin{table*}[t]
\centering
\caption{{The performance of the tested no-reference quality metrics on the MOS values in BVI-VFI dataset across different frame rates and interpolation methods. In each column, the best and the second best models are \textbf{boldfaced} and \underline{underlined}.}}
\label{tab:overallfpsmethod_nr}
\scriptsize
\begin{tabular}{c||c|c||c|c||c|c||c|c||c|c||c|c|c|c}
\toprule
  & \multicolumn{2}{c||}{30fps} & \multicolumn{2}{c||}{60fps} & \multicolumn{2}{c||}{120fps} & \multicolumn{2}{c||}{non-DL} & \multicolumn{2}{c||}{DL} & \multicolumn{4}{c}{Overall}\\
\midrule
Model & SRCC & PLCC & SRCC & PLCC & SRCC & PLCC & SRCC & PLCC & SRCC & PLCC & SRCC & KRCC & PLCC & RMSE \\ \midrule
BRISQUE     & 0.01 & 0.16 & 0.00 & 0.15 & 0.08 & 0.15 & 0.13 & 0.14 & 0.06 & \underline{0.17} & 0.00 & 0.00 & 0.12 & 22.16 \\
ChipQA      & 0.03 & 0.11 & 0.03 & \underline{0.26} & 0.06 & 0.16 & 0.01 & 0.05 & 0.05 & 0.12 & 0.03 & 0.02 & 0.12 & 22.16 \\
VIDEVAL     & 0.05 & 0.11 & 0.04 & 0.10 & 0.05 & 0.08 & \underline{0.19} & \underline{0.23} & 0.03 & 0.05 & 0.04 & 0.03 & 0.06 & 22.27 \\
deepIQA\_NR & \textbf{0.14} & 0.18 & 0.11 & 0.11 & 0.04 & 0.07 & 0.02 & 0.06 & 0.11 & 0.12 & 0.06 & 0.04 & 0.08 & 22.24 \\
NIQE        & 0.05 & \underline{0.19} & \underline{0.12} & 0.23 & 0.03 & 0.18 & \underline{0.19} & 0.22 & 0.03 & 0.15 & 0.08 & 0.05 & 0.15 & 22.06 \\
VIIDEO      & 0.12 & 0.12 & \underline{0.12} & \textbf{0.28} & 0.22 & \underline{0.30} & 0.04 & 0.15 & \textbf{0.29} & \textbf{0.31} & \underline{0.19} & \underline{0.13} & \textbf{0.23} & \textbf{21.72} \\
FastVQA     & \underline{0.13} & \textbf{0.33} & \textbf{0.29} & 0.21 & \textbf{0.28} & \textbf{0.31} & \textbf{0.38} & \textbf{0.38} & \underline{0.19} & \underline{0.17} & \textbf{0.25} & \textbf{0.16} & \underline{0.22} & \underline{21.78} \\ 
\bottomrule	
\end{tabular}
\end{table*}

The full-reference quality metrics evaluated in this experiment include the most commonly used metrics in the VFI literature, namely PSNR, SSIM~\cite{wang2004image} and LPIPS~\cite{zhang2018unreasonable}. While PSNR and SSIM are conventional methods, LPIPS is a deep learning-based method that measures distortion in deep feature space. Additionally, we evaluated other popular conventional image quality metrics including MS-SSIM~\cite{wang2003multiscale}, VIF~\cite{sheikh2005information}, VSI~\cite{zhang2014vsi}, FSIM~\cite{zhang2011fsim}, GMSD~\cite{xue2013gradient}, and deep learning-based ones: DISTS~\cite{ding2020image}, PieAPP~\cite{prashnani2018pieapp}, DeepIQA~\cite{bosse2017deep} and CONTRIQUE~\cite{madhusudana2022image}. Since these image metrics cannot capture distortions in the temporal domain, we also tested several video quality metrics, including FAST~\cite{wu2019quality}, SpEED~\cite{bampis2017speed}, ST-RRED~\cite{soundararajan2012video}, VMAF~\cite{li2016toward}, C3DVQA~\cite{xu2020c3dvqa}, {FovVideoVDP~\cite{mantiuk2021fovvideovdp},} FRQM~\cite{zhang2017frame}, ST-GREED~\cite{madhusudana2021st}, VSTR~\cite{lee2021space} and FloLPIPS~\cite{danier2022flolpips}. Among these, FAST, SpEED and ST-RRED are conventional metrics that are based on either spatio-temporal gradient information or natural scene statistics (NSS). VMAF is a popular learning-based metric that uses a support vector regressor (SVR) to combine hand-crafted features. C3DVQA is a deep learning-based metric that processes the reference and distorted videos with a 3D convolutional neural network. {FovVideoVDP considers the characteristics of the human vision system and accounts for both spatial and temporal artefacts.} FRQM, ST-GREED, and VSTR are specifically designed to capture temporal/spatial resolution-related artefacts. Lastly, FloLPIPS is a recently proposed bespoke metric for VFI that combines distortions in optical flow with LPIPS.

For completeness, we have also considered no-reference (NR) quality metrics. These include image models, NIQE~\cite{mittal2012making}, {BRISQUE~\cite{mittal2012no}} and deepIQA-NR~\cite{bosse2017deep}, and video models, VIIDEO~\cite{mittal2015completely}, VBLIINDS~\cite{saad2014blind}, VIDEVAL~\cite{tu2021ugc}, RAPIQUE~\cite{tu2021rapique}, {ChipQA~\cite{ebenezer2021chipqa}, TLVQM~\cite{korhonen2019two}, FastVQA~\cite{wu2022fast}} and FAVER~\cite{zheng2022faver}. {These NR metrics have been evaluated on both MOS (mean opinion score) and DMOS scores. Specifically, for each distorted video, the MOS was calculated by averaging all the opinion scores assigned to it. Similarly to our approach with DMOS, the MOS values were also processed by the P.910 subject screening procedure~\cite{p910} to obtain more reliable MOS data.}

In order to measure the performance of these quality metrics on predicting subjective quality, we employed four statistical measures: Spearman's Rank-order Correlation Coefficient (SRCC), the Kendall Rank-order Correlation Coefficient (KRCC), Pearson's Linear Correlation Coefficient (PLCC) and Root Mean Squared Error (RMSE). Better prediction performance is indicated by higher values of the three correlation coefficients and lower RMSE values. For the calculation of PLCC and RMSE, a logistic curve was fit between the ground-truth DMOS values and the metric scores following the procedure in \cite{zhang2015perception}.

\begin{table*}[t]
\scriptsize
\centering
\caption{F-test results between DMOS prediction residuals of selected quality metrics at a 95\% confidence interval. The value ``1'' indicates the metric in the row is superior to the metric in the column and ``0'' means the opposite, while a ``-'' denotes statistical equivalence. In each cell, there are 6 entries corresponding to 30fps, 60fps, 120fps, non-DL, DL and all videos respectively.}
\label{tab:ftest}
\begin{tabular}{c|c|c|c|c|c|c|c|c|c|c|c|c}
\toprule
Model       & VMAF   & SSIM   & STRRED & FloLPIPS & FSIM   & MSSSIM & GMSD   & VSI    & SpEED  & FovVideoVDP & PSNR   & FAST   \\ \midrule
VMAF        & \texttt{-}\texttt{-}\texttt{-}\texttt{-}\texttt{-}\texttt{-} & \texttt{-}\texttt{-}\texttt{-}\texttt{-}\texttt{-}\texttt{-} & \texttt{-}\texttt{-}\texttt{-}\texttt{-}1\texttt{-} & \texttt{-}\texttt{-}\texttt{-}\texttt{-}\texttt{-}\texttt{-}   & \texttt{-}\texttt{-}\texttt{-}\texttt{-}\texttt{-}\texttt{-} & \texttt{-}\texttt{-}\texttt{-}\texttt{-}\texttt{-}\texttt{-} & \texttt{-}\texttt{-}\texttt{-}\texttt{-}\texttt{-}\texttt{-} & \texttt{-}\texttt{-}\texttt{-}\texttt{-}\texttt{-}\texttt{-} & \texttt{-}\texttt{-}1\texttt{-}\texttt{-}\texttt{-} & \texttt{-}\texttt{-}\texttt{-}\texttt{-}\texttt{-}\texttt{-}      & \texttt{-}\texttt{-}\texttt{-}\texttt{-}\texttt{-}\texttt{-} & \texttt{-}\texttt{-}\texttt{-}\texttt{-}00 \\
SSIM        & \texttt{-}\texttt{-}\texttt{-}\texttt{-}\texttt{-}\texttt{-} & \texttt{-}\texttt{-}\texttt{-}\texttt{-}\texttt{-}\texttt{-} & 1\texttt{-}\texttt{-}\texttt{-}1\texttt{-} & \texttt{-}\texttt{-}\texttt{-}\texttt{-}\texttt{-}\texttt{-}   & \texttt{-}\texttt{-}\texttt{-}\texttt{-}\texttt{-}\texttt{-} & \texttt{-}\texttt{-}\texttt{-}\texttt{-}\texttt{-}\texttt{-} & \texttt{-}\texttt{-}\texttt{-}\texttt{-}\texttt{-}\texttt{-} & \texttt{-}\texttt{-}\texttt{-}\texttt{-}\texttt{-}\texttt{-} & 1\texttt{-}1\texttt{-}\texttt{-}\texttt{-} & \texttt{-}\texttt{-}\texttt{-}\texttt{-}\texttt{-}\texttt{-}      & \texttt{-}\texttt{-}\texttt{-}\texttt{-}\texttt{-}\texttt{-} & \texttt{-}\texttt{-}\texttt{-}\texttt{-}00 \\
STRRED      & \texttt{-}\texttt{-}\texttt{-}\texttt{-}0\texttt{-} & 0\texttt{-}\texttt{-}\texttt{-}0\texttt{-} & \texttt{-}\texttt{-}\texttt{-}\texttt{-}\texttt{-}\texttt{-} & 0\texttt{-}\texttt{-}\texttt{-}0\texttt{-}   & 0\texttt{-}\texttt{-}\texttt{-}0\texttt{-} & \texttt{-}\texttt{-}\texttt{-}\texttt{-}0\texttt{-} & 0\texttt{-}\texttt{-}\texttt{-}0\texttt{-} & 0\texttt{-}\texttt{-}\texttt{-}0\texttt{-} & \texttt{-}\texttt{-}\texttt{-}\texttt{-}0\texttt{-} & \texttt{-}\texttt{-}\texttt{-}00\texttt{-}      & 0\texttt{-}\texttt{-}\texttt{-}0\texttt{-} & 00\texttt{-}\texttt{-}00 \\
FloLPIPS    & \texttt{-}\texttt{-}\texttt{-}\texttt{-}\texttt{-}\texttt{-} & \texttt{-}\texttt{-}\texttt{-}\texttt{-}\texttt{-}\texttt{-} & 1\texttt{-}\texttt{-}\texttt{-}1\texttt{-} & \texttt{-}\texttt{-}\texttt{-}\texttt{-}\texttt{-}\texttt{-}   & \texttt{-}\texttt{-}\texttt{-}\texttt{-}\texttt{-}\texttt{-} & \texttt{-}\texttt{-}\texttt{-}\texttt{-}\texttt{-}\texttt{-} & \texttt{-}\texttt{-}\texttt{-}\texttt{-}\texttt{-}\texttt{-} & \texttt{-}\texttt{-}\texttt{-}\texttt{-}\texttt{-}\texttt{-} & 1\texttt{-}1\texttt{-}\texttt{-}\texttt{-} & \texttt{-}\texttt{-}\texttt{-}\texttt{-}\texttt{-}\texttt{-}      & \texttt{-}\texttt{-}\texttt{-}\texttt{-}\texttt{-}\texttt{-} & \texttt{-}\texttt{-}\texttt{-}\texttt{-}0\texttt{-} \\
FSIM        & \texttt{-}\texttt{-}\texttt{-}\texttt{-}\texttt{-}\texttt{-} & \texttt{-}\texttt{-}\texttt{-}\texttt{-}\texttt{-}\texttt{-} & 1\texttt{-}\texttt{-}\texttt{-}1\texttt{-} & \texttt{-}\texttt{-}\texttt{-}\texttt{-}\texttt{-}\texttt{-}   & \texttt{-}\texttt{-}\texttt{-}\texttt{-}\texttt{-}\texttt{-} & \texttt{-}\texttt{-}\texttt{-}\texttt{-}\texttt{-}\texttt{-} & \texttt{-}\texttt{-}\texttt{-}\texttt{-}\texttt{-}\texttt{-} & \texttt{-}\texttt{-}\texttt{-}\texttt{-}\texttt{-}\texttt{-} & 1\texttt{-}1\texttt{-}\texttt{-}\texttt{-} & \texttt{-}\texttt{-}\texttt{-}\texttt{-}\texttt{-}\texttt{-}      & \texttt{-}\texttt{-}\texttt{-}\texttt{-}\texttt{-}\texttt{-} & \texttt{-}\texttt{-}\texttt{-}\texttt{-}00 \\
MSSSIM      & \texttt{-}\texttt{-}\texttt{-}\texttt{-}\texttt{-}\texttt{-} & \texttt{-}\texttt{-}\texttt{-}\texttt{-}\texttt{-}\texttt{-} & \texttt{-}\texttt{-}\texttt{-}\texttt{-}1\texttt{-} & \texttt{-}\texttt{-}\texttt{-}\texttt{-}\texttt{-}\texttt{-}   & \texttt{-}\texttt{-}\texttt{-}\texttt{-}\texttt{-}\texttt{-} & \texttt{-}\texttt{-}\texttt{-}\texttt{-}\texttt{-}\texttt{-} & \texttt{-}\texttt{-}\texttt{-}\texttt{-}\texttt{-}\texttt{-} & \texttt{-}\texttt{-}\texttt{-}\texttt{-}\texttt{-}\texttt{-} & \texttt{-}\texttt{-}1\texttt{-}\texttt{-}\texttt{-} & \texttt{-}\texttt{-}\texttt{-}\texttt{-}\texttt{-}\texttt{-}      & \texttt{-}\texttt{-}\texttt{-}\texttt{-}\texttt{-}\texttt{-} & \texttt{-}\texttt{-}\texttt{-}\texttt{-}00 \\
GMSD        & \texttt{-}\texttt{-}\texttt{-}\texttt{-}\texttt{-}\texttt{-} & \texttt{-}\texttt{-}\texttt{-}\texttt{-}\texttt{-}\texttt{-} & 1\texttt{-}\texttt{-}\texttt{-}1\texttt{-} & \texttt{-}\texttt{-}\texttt{-}\texttt{-}\texttt{-}\texttt{-}   & \texttt{-}\texttt{-}\texttt{-}\texttt{-}\texttt{-}\texttt{-} & \texttt{-}\texttt{-}\texttt{-}\texttt{-}\texttt{-}\texttt{-} & \texttt{-}\texttt{-}\texttt{-}\texttt{-}\texttt{-}\texttt{-} & \texttt{-}\texttt{-}\texttt{-}\texttt{-}\texttt{-}\texttt{-} & 1\texttt{-}1\texttt{-}\texttt{-}\texttt{-} & \texttt{-}\texttt{-}\texttt{-}\texttt{-}\texttt{-}\texttt{-}      & \texttt{-}\texttt{-}\texttt{-}\texttt{-}\texttt{-}\texttt{-} & \texttt{-}\texttt{-}\texttt{-}\texttt{-}0\texttt{-} \\
VSI         & \texttt{-}\texttt{-}\texttt{-}\texttt{-}\texttt{-}\texttt{-} & \texttt{-}\texttt{-}\texttt{-}\texttt{-}\texttt{-}\texttt{-} & 1\texttt{-}\texttt{-}\texttt{-}1\texttt{-} & \texttt{-}\texttt{-}\texttt{-}\texttt{-}\texttt{-}\texttt{-}   & \texttt{-}\texttt{-}\texttt{-}\texttt{-}\texttt{-}\texttt{-} & \texttt{-}\texttt{-}\texttt{-}\texttt{-}\texttt{-}\texttt{-} & \texttt{-}\texttt{-}\texttt{-}\texttt{-}\texttt{-}\texttt{-} & \texttt{-}\texttt{-}\texttt{-}\texttt{-}\texttt{-}\texttt{-} & 1\texttt{-}1\texttt{-}\texttt{-}\texttt{-} & \texttt{-}\texttt{-}\texttt{-}\texttt{-}\texttt{-}\texttt{-}      & \texttt{-}\texttt{-}\texttt{-}\texttt{-}\texttt{-}\texttt{-} & \texttt{-}\texttt{-}\texttt{-}\texttt{-}0\texttt{-} \\
SpEED       & \texttt{-}\texttt{-}0\texttt{-}\texttt{-}\texttt{-} & 0\texttt{-}0\texttt{-}\texttt{-}\texttt{-} & \texttt{-}\texttt{-}\texttt{-}\texttt{-}1\texttt{-} & 0\texttt{-}0\texttt{-}\texttt{-}\texttt{-}   & 0\texttt{-}0\texttt{-}\texttt{-}\texttt{-} & \texttt{-}\texttt{-}0\texttt{-}\texttt{-}\texttt{-} & 0\texttt{-}0\texttt{-}\texttt{-}\texttt{-} & 0\texttt{-}0\texttt{-}\texttt{-}\texttt{-} & \texttt{-}\texttt{-}\texttt{-}\texttt{-}\texttt{-}\texttt{-} & \texttt{-}\texttt{-}00\texttt{-}\texttt{-}      & 0\texttt{-}0\texttt{-}\texttt{-}\texttt{-} & 0\texttt{-}0\texttt{-}0\texttt{-} \\
FovVideoVDP & \texttt{-}\texttt{-}\texttt{-}\texttt{-}\texttt{-}\texttt{-} & \texttt{-}\texttt{-}\texttt{-}\texttt{-}\texttt{-}\texttt{-} & \texttt{-}\texttt{-}\texttt{-}11\texttt{-} & \texttt{-}\texttt{-}\texttt{-}\texttt{-}\texttt{-}\texttt{-}   & \texttt{-}\texttt{-}\texttt{-}\texttt{-}\texttt{-}\texttt{-} & \texttt{-}\texttt{-}\texttt{-}\texttt{-}\texttt{-}\texttt{-} & \texttt{-}\texttt{-}\texttt{-}\texttt{-}\texttt{-}\texttt{-} & \texttt{-}\texttt{-}\texttt{-}\texttt{-}\texttt{-}\texttt{-} & \texttt{-}\texttt{-}11\texttt{-}\texttt{-} & \texttt{-}\texttt{-}\texttt{-}\texttt{-}\texttt{-}\texttt{-}      & \texttt{-}\texttt{-}\texttt{-}\texttt{-}\texttt{-}\texttt{-} & \texttt{-}\texttt{-}\texttt{-}\texttt{-}0\texttt{-} \\
PSNR        & \texttt{-}\texttt{-}\texttt{-}\texttt{-}\texttt{-}\texttt{-} & \texttt{-}\texttt{-}\texttt{-}\texttt{-}\texttt{-}\texttt{-} & 1\texttt{-}\texttt{-}\texttt{-}1\texttt{-} & \texttt{-}\texttt{-}\texttt{-}\texttt{-}\texttt{-}\texttt{-}   & \texttt{-}\texttt{-}\texttt{-}\texttt{-}\texttt{-}\texttt{-} & \texttt{-}\texttt{-}\texttt{-}\texttt{-}\texttt{-}\texttt{-} & \texttt{-}\texttt{-}\texttt{-}\texttt{-}\texttt{-}\texttt{-} & \texttt{-}\texttt{-}\texttt{-}\texttt{-}\texttt{-}\texttt{-} & 1\texttt{-}1\texttt{-}\texttt{-}\texttt{-} & \texttt{-}\texttt{-}\texttt{-}\texttt{-}\texttt{-}\texttt{-}      & \texttt{-}\texttt{-}\texttt{-}\texttt{-}\texttt{-}\texttt{-} & \texttt{-}\texttt{-}\texttt{-}\texttt{-}0\texttt{-} \\
FAST        & \texttt{-}\texttt{-}\texttt{-}\texttt{-}11 & \texttt{-}\texttt{-}\texttt{-}\texttt{-}11 & 11\texttt{-}\texttt{-}11 & \texttt{-}\texttt{-}\texttt{-}\texttt{-}1\texttt{-}   & \texttt{-}\texttt{-}\texttt{-}\texttt{-}11 & \texttt{-}\texttt{-}\texttt{-}\texttt{-}11 & \texttt{-}\texttt{-}\texttt{-}\texttt{-}1\texttt{-} & \texttt{-}\texttt{-}\texttt{-}\texttt{-}1\texttt{-} & 1\texttt{-}1\texttt{-}1\texttt{-} & \texttt{-}\texttt{-}\texttt{-}\texttt{-}1\texttt{-}      & \texttt{-}\texttt{-}\texttt{-}\texttt{-}1\texttt{-} & \texttt{-}\texttt{-}\texttt{-}\texttt{-}\texttt{-}\texttt{-} \\
\bottomrule	
\end{tabular}
\end{table*}

\begin{figure*}[!ht]
    \centering
    \subfloat[PSNR]{\includegraphics[width=0.240\linewidth]{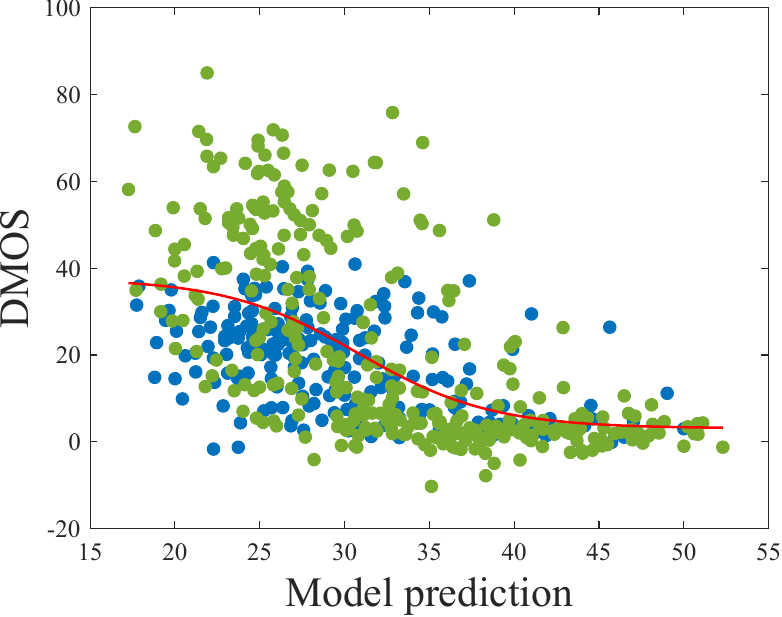}}\;\!
    \subfloat[SSIM]{\includegraphics[width=0.240\linewidth]{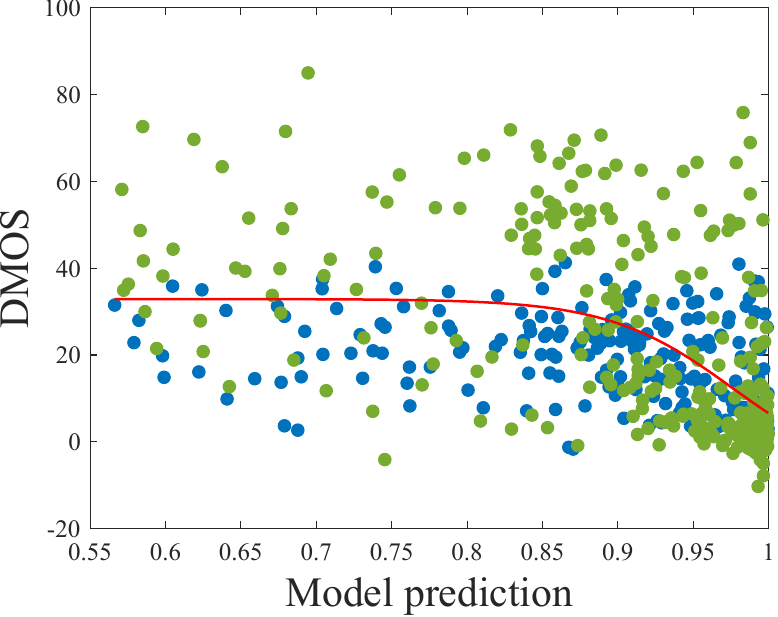}}\;\!
    \subfloat[LPIPS]{\includegraphics[width=0.240\linewidth]{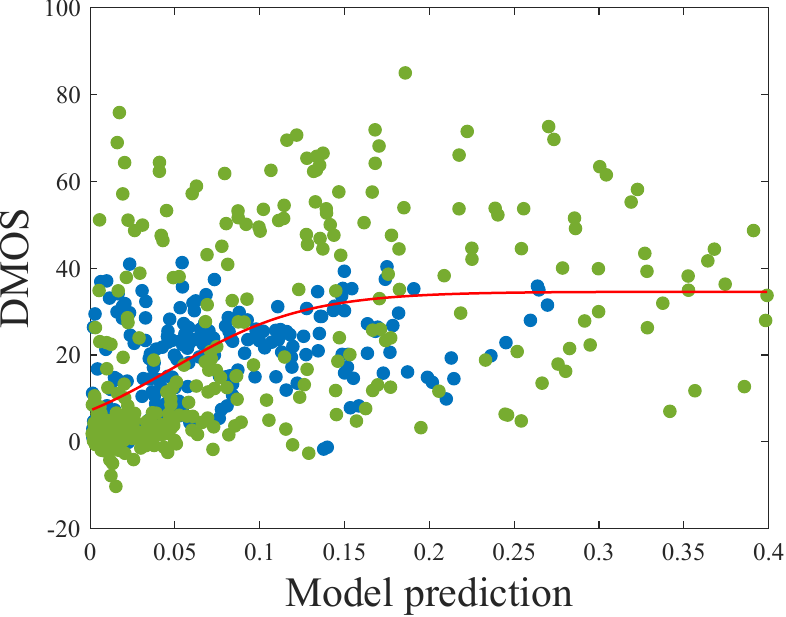}}\;\!
    \subfloat[FAST]{\includegraphics[width=0.240\linewidth]{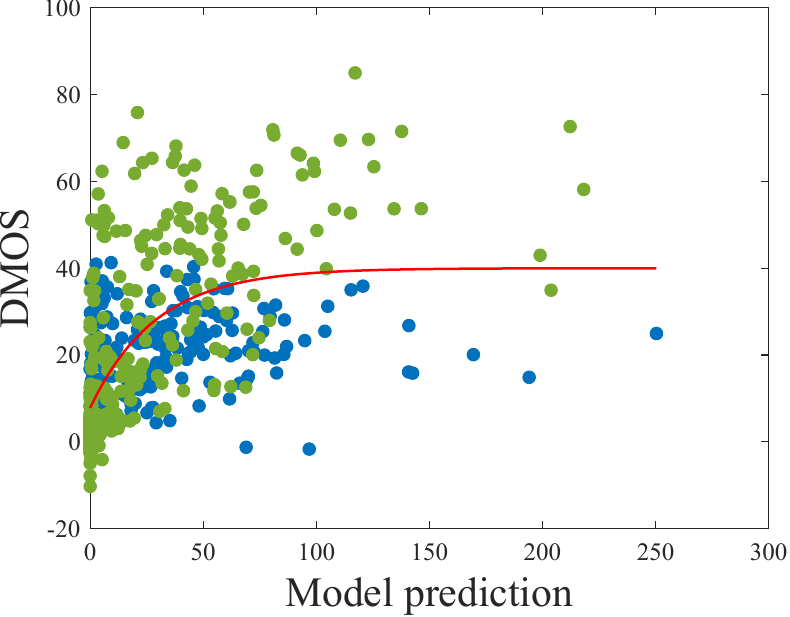}}\\
    \caption{Scatter plots of DMOS values against the predictions of selected quality models. The blue scatter points correspond to distorted videos generated by non-DL methods, while the green points denote DL-based methods. The red lines are the logistic functions fitted between the model predictions and DMOS values on the entire BVI-VFI database.}
	\label{fig:scatters}
\end{figure*}

For all the learning-based metrics (VMAF, ST-GREED, VIDEVAL, C3DVQA{, FastVQA, ChipQA, BRISQUE}), we first generated results based on their publicly available pre-trained models (as shown in Section~\ref{sec:overall}). However, such pre-trained versions were absent for some of the metrics, including VSTR, FAVER, RAPIQUE, {TLVQM} and VBLIINDS. To account for this, and also to allow the models to benefit from their learning-based nature, we also conducted cross-validation experiments (see Section~\ref{sec:cv}) on the BVI-VFI database. It is noted that the deep learning-based image quality metrics (i.e. LPIPS, DISTS, PieAPP, DeepIQA, CONTRIQUE{, BRISQUE, FastVQA}) were pre-trained on large-scale image{/video} quality databases. Since our database only contains subjective scores at the video level {and has a limited size (540 distorted videos)}, we did not re-train these models during cross validation.

\subsection{Overall Performance on BVI-VFI}\label{sec:overall}

We first evaluate the overall performance of all the tested quality assessment methods (except for VSTR, FAVER, RAPIQUE and VBLIINDS, as explained above) on the entire BVI-VFI database. Table~\ref{tab:overallfpsmethod} and \ref{tab:overallfpsmethod_nr} summarise these results as well as the performance at the three specific frame rates. It can be observed that none of the tested metrics exhibit satisfactory overall correlation with the subjective quality scores, with the best-performing metric, FAST, achieving a SRCC value of only  0.70. This may be due to the approach employed in FAST, which captures spatio-temporal quality degradation by comparing the moving parts of the reference and distorted frames along motion trajectories. It can also be seen that the metrics most commonly used in the VFI literature (PSNR, SSIM and LPIPS) all achieved SRCC values below 0.65. This indicates that there is still significant scope for improving the accuracy of quality metrics in the context of VFI. The last four rows of Table~\ref{tab:overallfpsmethod} report the performance of four no-reference quality assessment models. These methods significantly under-performed most of the full-/reduced-reference models. 

{The best performing metric, FAST, is a FR-VQA model that assumes the high relevance of moving visual content to the perceived video quality. It captures spatio-temporal quality degradation by comparing the moving parts of the reference and distorted frames along motion trajectories. Additionally, FAST considers discrepancy in the motion information by comparing optical flows, as well as spatial degradation estimated through GMSD, a gradient-based image quality model. The superior performance (although not optimal) of FAST implies that placing extra focus on moving regions of a video as well as global motion information can be beneficial for predicting VFI quality. This can especially be the case for assessing videos interpolated by DL-based VFI algorithms, which tend to encounter challenges and introduce salient artefacts in such moving parts of videos. It is also observed from Table II that although PSNR showed the second best overall SRCC performance, it is not significantly better compared to some other metrics, e.g., SSIM, SpEED and FovVideoVDP (see Table~\ref{tab:ftest}). All of these have SRCC values below 0.65.}

When comparing the variation of model performance across different frame rates (Table~\ref{tab:overallfpsmethod}), it can be seen that most models achieved highest correlation with opinion scores on videos at 60fps and the lowest correlation on 30fps sequences. This implies that these models could not properly account for the effect of frame rate on the human perception of interpolation artefacts, and that frame rate can be a useful factor to consider when designing a better quality assessment model for frame interpolated content.

\begin{table*}[t]
\centering
\scriptsize
\caption{{Cross validation results of the tested quality assessment methods on the DMOS values in BVI-VFI dataset across different frame rates and interpolation methods. Each cell presents the median and standard deviation of the statistical measures obtained over 1000 random train-test splits. In each column, the best and the second best models are \textbf{boldfaced} and \underline{underlined}.}}
\label{tab:cvfpsmethod}
\resizebox{\linewidth}{!}{\begin{tabular}{c||c|c||c|c||c|c||c|c||c|c||c|c}
\toprule
  & \multicolumn{2}{c||}{30fps} & \multicolumn{2}{c||}{60fps} & \multicolumn{2}{c||}{120fps} & \multicolumn{2}{c||}{non-DL} & \multicolumn{2}{c||}{DL} & \multicolumn{2}{c}{Overall}\\
\midrule
Model & SRCC & PLCC & SRCC & PLCC & SRCC & PLCC & SRCC & PLCC & SRCC & PLCC & SRCC & PLCC  \\
\midrule 
CONTRIQUE   & 0.16 (0.11)          & 0.38 (0.11)          & 0.19 (0.12)          & 0.46 (0.12)          & 0.26 (0.13)          & 0.52 (0.11)          & 0.13 (0.09)          & 0.25 (0.10)          & 0.40 (0.11)          & 0.48 (0.10)          & 0.24 (0.08)          & 0.41 (0.09)          \\
deepIQA\_FR & 0.26 (0.14)          & 0.43 (0.13)          & 0.43 (0.15)          & 0.49 (0.12)          & 0.42 (0.18)          & 0.47 (0.15)          & 0.38 (0.13)          & 0.48 (0.10)          & 0.48 (0.12)          & 0.52 (0.12)          & 0.46 (0.11)          & 0.47 (0.10)          \\
PieAPP      & 0.40 (0.15)          & 0.52 (0.12)          & 0.48 (0.16)          & 0.54 (0.14)          & 0.46 (0.17)          & 0.52 (0.15)          & 0.32 (0.15)          & 0.42 (0.12)          & 0.54 (0.12)          & 0.57 (0.13)          & 0.49 (0.12)          & 0.50 (0.11)          \\
VMAF        & 0.39 (0.16)          & 0.51 (0.13)          & 0.57 (0.14)          & 0.55 (0.12)          & 0.52 (0.17)          & 0.58 (0.17)          & 0.37 (0.15)          & 0.46 (0.13)          & 0.57 (0.12)          & 0.61 (0.11)          & 0.52 (0.10)          & 0.50 (0.10)          \\
VIF         & 0.37 (0.13)          & 0.47 (0.14)          & 0.55 (0.12)          & 0.54 (0.12)          & 0.53 (0.15)          & 0.54 (0.14)          & 0.38 (0.12)          & 0.47 (0.11)          & 0.56 (0.10)          & 0.58 (0.11)          & 0.53 (0.09)          & 0.50 (0.09)          \\
C3DVQA      & 0.28 (0.14)          & 0.43 (0.17)          & 0.57 (0.11)          & 0.51 (0.21)          & {\ul 0.64 (0.13)}    & 0.50 (0.25)          & 0.40 (0.12)          & 0.46 (0.18)          & 0.59 (0.09)          & 0.56 (0.14)          & 0.55 (0.08)          & 0.48 (0.10)          \\
DISTS       & 0.42 (0.11)          & 0.52 (0.10)          & 0.54 (0.10)          & 0.60 (0.10)          & 0.58 (0.12)          & 0.68 (0.11)          & 0.39 (0.09)          & 0.46 (0.09)          & 0.64 (0.07)          & 0.65 (0.08)          & 0.57 (0.07)          & 0.59 (0.07)          \\
FRQM        & 0.45 (0.13)          & 0.57 (0.12)          & 0.66 (0.11)          & 0.64 (0.10)          & 0.60 (0.15)          & 0.63 (0.14)          & \textbf{0.81 (0.07)} & \textbf{0.84 (0.06)} & 0.49 (0.06)          & 0.47 (0.07)          & 0.58 (0.06)          & 0.50 (0.06)          \\
LPIPS       & 0.41 (0.12)          & 0.52 (0.12)          & 0.57 (0.11)          & 0.59 (0.12)          & 0.58 (0.13)          & 0.63 (0.14)          & 0.43 (0.11)          & 0.49 (0.12)          & 0.61 (0.09)          & 0.63 (0.10)          & 0.58 (0.08)          & 0.57 (0.09)          \\
FloLPIPS    & 0.47 (0.12)          & 0.57 (0.11)          & 0.59 (0.11)          & 0.62 (0.11)          & 0.60 (0.13)          & 0.68 (0.13)          & 0.47 (0.11)          & 0.53 (0.12)          & 0.66 (0.07)          & 0.67 (0.08)          & 0.61 (0.07)          & 0.61 (0.08)          \\
STRRED      & 0.48 (0.12)          & 0.38 (0.20)          & 0.64 (0.09)          & 0.48 (0.22)          & 0.58 (0.15)          & 0.45 (0.22)          & 0.41 (0.11)          & 0.29 (0.18)          & 0.66 (0.07)          & 0.58 (0.20)          & 0.61 (0.07)          & 0.47 (0.24)          \\
SSIM        & 0.50 (0.12)          & 0.60 (0.12)          & 0.66 (0.11)          & 0.63 (0.10)          & 0.59 (0.17)          & 0.60 (0.14)          & 0.43 (0.12)          & 0.50 (0.12)          & 0.67 (0.09)          & 0.70 (0.10)          & 0.62 (0.08)          & 0.59 (0.09)          \\
FSIM        & 0.47 (0.12)          & 0.58 (0.11)          & 0.65 (0.11)          & 0.62 (0.10)          & 0.62 (0.14)          & 0.65 (0.13)          & 0.46 (0.10)          & 0.52 (0.10)          & 0.67 (0.08)          & 0.69 (0.09)          & 0.63 (0.07)          & 0.59 (0.08)          \\
MSSSIM      & 0.47 (0.12)          & 0.58 (0.11)          & 0.66 (0.10)          & 0.63 (0.09)          & 0.61 (0.16)          & 0.62 (0.14)          & 0.46 (0.12)          & 0.51 (0.11)          & 0.68 (0.08)          & 0.69 (0.08)          & 0.63 (0.07)          & 0.59 (0.07)          \\
FovVideoVDP & 0.46 (0.11)          & 0.55 (0.12)          & 0.66 (0.09)          & 0.62 (0.10)          & 0.60 (0.16)          & 0.59 (0.14)          & {\ul 0.57 (0.12)}    & {\ul 0.63 (0.10)}    & 0.66 (0.06)          & 0.64 (0.07)          & 0.64 (0.06)          & 0.59 (0.06)          \\
GMSD        & 0.51 (0.12)          & 0.61 (0.11)          & 0.67 (0.11)          & 0.64 (0.10)          & 0.63 (0.14)          & 0.63 (0.12)          & 0.42 (0.14)          & 0.52 (0.11)          & 0.69 (0.07)          & 0.72 (0.09)          & 0.64 (0.08)          & 0.61 (0.08)          \\
VSI         & 0.48 (0.12)          & 0.59 (0.11)          & 0.67 (0.10)          & 0.65 (0.09)          & 0.63 (0.14)          & 0.64 (0.12)          & 0.46 (0.12)          & 0.52 (0.10)          & 0.68 (0.07)          & 0.70 (0.09)          & 0.64 (0.08)          & 0.61 (0.07)          \\
ST-GREED       & \textbf{0.53 (0.14)} & {\ul 0.60 (0.12)}    & {\ul 0.71 (0.12)}    & \textbf{0.72 (0.12)} & 0.61 (0.17)          & {\ul 0.71 (0.15)}    & 0.41 (0.12)          & 0.46 (0.11)          & 0.70 (0.10)          & {\ul 0.77 (0.09)}    & 0.64 (0.10)          & {\ul 0.66 (0.11)}    \\
SpEED       & 0.50 (0.12)          & 0.39 (0.19)          & 0.68 (0.09)          & 0.51 (0.21)          & 0.64 (0.14)          & 0.49 (0.19)          & 0.45 (0.11)          & 0.31 (0.15)          & 0.69 (0.06)          & 0.61 (0.17)          & {\ul 0.65 (0.07)}    & 0.49 (0.25)          \\
VSTR        & 0.51 (0.12)          & 0.57 (0.10)          & 0.68 (0.09)          & 0.68 (0.13)          & 0.59 (0.14)          & 0.69 (0.14)          & 0.44 (0.12)          & 0.48 (0.10)          & 0.70 (0.08)          & 0.73 (0.09)          & 0.65 (0.08)          & 0.64 (0.09)          \\
PSNR        & {\ul 0.51 (0.11)}    & \textbf{0.60 (0.11)} & 0.69 (0.10)          & 0.66 (0.10)          & 0.63 (0.14)          & 0.63 (0.13)          & 0.46 (0.13)          & 0.54 (0.12)          & {\ul 0.71 (0.07)}    & 0.72 (0.09)          & 0.65 (0.08)          & 0.62 (0.07)          \\
FAST        & 0.50 (0.12)          & 0.60 (0.13)          & \textbf{0.74 (0.08)} & {\ul 0.69 (0.17)}    & \textbf{0.72 (0.14)} & \textbf{0.77 (0.19)} & 0.47 (0.12)          & 0.50 (0.12)          & \textbf{0.77 (0.06)} & \textbf{0.79 (0.07)} & \textbf{0.71 (0.07)} & \textbf{0.66 (0.08)} \\
\midrule
NIQE        & 0.20 (0.13)          & 0.32 (0.12)          & 0.00 (0.15)          & 0.28 (0.10)          & 0.10 (0.18)          & 0.26 (0.09)          & 0.20 (0.13)          & 0.30 (0.10)          & 0.06 (0.14)          & 0.30 (0.11)          & 0.01 (0.12)          & 0.22 (0.09)          \\
BRISQUE     & 0.18 (0.15)          & 0.31 (0.12)          & 0.08 (0.17)          & 0.28 (0.11)          & 0.03 (0.20)          & 0.25 (0.10)          & 0.18 (0.15)          & 0.30 (0.13)          & 0.10 (0.16)          & 0.29 (0.12)          & 0.04 (0.14)          & 0.22 (0.09)          \\
VIIDEO      & 0.02 (0.13)          & 0.31 (0.14)          & 0.03 (0.11)          & 0.34 (0.15)          & 0.16 (0.14)          & 0.37 (0.13)          & 0.01 (0.07)          & 0.17 (0.09)          & 0.24 (0.14)          & 0.33 (0.12)          & 0.10 (0.09)          & 0.28 (0.11)          \\
deepIQA\_NR & 0.14 (0.12)          & 0.25 (0.11)          & 0.14 (0.17)          & 0.27 (0.10)          & 0.06 (0.20)          & 0.23 (0.11)          & 0.05 (0.15)          & 0.23 (0.11)          & 0.12 (0.14)          & 0.25 (0.11)          & 0.10 (0.12)          & 0.20 (0.08)          \\
FastVQA     & 0.11 (0.15)          & 0.29 (0.13)          & 0.18 (0.13)          & 0.25 (0.11)          & 0.21 (0.17)          & 0.36 (0.16)          & 0.29 (0.11)          & 0.38 (0.10)          & 0.10 (0.12)          & 0.19 (0.10)          & 0.16 (0.10)          & 0.19 (0.09)          \\
RAPIQUE     & 0.11 (0.15)          & 0.12 (0.17)          & 0.24 (0.17)          & 0.41 (0.16)          & 0.22 (0.19)          & 0.44 (0.21)          & 0.20 (0.17)          & 0.22 (0.18)          & 0.28 (0.14)          & 0.41 (0.14)          & 0.23 (0.12)          & 0.35 (0.13)          \\
VIDEVAL     & 0.46 (0.12)          & 0.51 (0.11)          & 0.42 (0.13)          & 0.48 (0.13)          & 0.40 (0.19)          & 0.60 (0.17)          & 0.28 (0.15)          & 0.38 (0.13)          & 0.43 (0.15)          & 0.54 (0.14)          & 0.34 (0.13)          & 0.44 (0.13)          \\
ChipQA      & 0.40 (0.14)          & 0.46 (0.12)          & 0.46 (0.15)          & 0.51 (0.14)          & 0.38 (0.16)          & 0.54 (0.16)          & 0.26 (0.15)          & 0.36 (0.12)          & 0.52 (0.17)          & 0.58 (0.14)          & 0.41 (0.14)          & 0.46 (0.13)          \\
FAVER       & 0.36 (0.16)          & 0.44 (0.12)          & 0.33 (0.16)          & 0.45 (0.15)          & 0.32 (0.17)          & 0.50 (0.19)          & 0.56 (0.13)          & 0.61 (0.12)          & 0.44 (0.14)          & 0.53 (0.15)          & 0.43 (0.13)          & 0.48 (0.13)          \\
VBLIINDS    & 0.40 (0.13)          & 0.50 (0.12)          & 0.54 (0.13)          & 0.62 (0.13)          & 0.39 (0.19)          & 0.63 (0.17)          & 0.34 (0.14)          & 0.40 (0.12)          & 0.54 (0.14)          & 0.60 (0.13)          & 0.44 (0.13)          & 0.49 (0.13)          \\
TLVQM       & 0.27 (0.15)          & 0.41 (0.13)          & 0.44 (0.15)          & 0.52 (0.14)          & 0.44 (0.17)          & 0.65 (0.15)          & 0.30 (0.15)          & 0.38 (0.13)          & 0.58 (0.12)          & 0.64 (0.12)          & 0.45 (0.11)          & 0.50 (0.11)          \\
\bottomrule	
\end{tabular}
}
\end{table*}

Table~\ref{tab:overallfpsmethod} also presents the evaluation results for different VFI algorithms: non-deep learning-based (non-DL) or deep learning-based (DL). It can be observed that the wavelet-based VQA model FRQM achieved the highest SRCC value of 0.80 on sequences interpolated by non-DL methods, but its performance on DL-based group dropped significantly. Similarly, FAST, which achieved the highest SRCC value (of 0.77) on the DL group, saw a large performance degradation on the sequences in the non-DL group. This indicates that if only DL-based VFI methods are concerned, FAST is a more reliable quality metric for evaluating the perceptual quality of VFI content. The scatter plots of the model predictions and DMOS values, as well as the fitted logistic curves, are shown in Fig.~\ref{fig:scatters} for FAST and the three most popular metrics in VFI: PSNR, SSIM and LPIPS.

To confirm the statistical significance of difference between model performances, we also conducted F-tests among twelve metrics with the best overall SRCC values. For every pair of quality metrics, the F-test was performed on the residuals between the ground-truth DMOS values and the DMOS predicted by those metrics through the fitted logistic curves~\cite{lee2021subjective, madhusudana2021subjective}. Assuming that the residuals follow a zero-mean Gaussian distribution, the F-test is under the null hypothesis that two groups of residuals have equal variance at 95\% confidence level. Table~\ref{tab:ftest} summarises the F-test results, where it can be seen that FAST outperformed all the other eleven models significantly on videos in the DL group.

\subsection{Cross-Validation Results for Learning-based Models}\label{sec:cv}

As mentioned above, for some learning-based metrics, we also conducted cross-validation experiments where the models were trained and evaluated on randomly split non-overlapping data sets. Specifically, we performed five-fold cross validation as in~\cite{lee2021subjective}, where 80\% of the 540 distorted videos are used as the training set and the other 20\% as the test set. To better evaluate the generalisation ability of the learning-based metrics, we ensured the content-wise separation between the training and test sets, i.e. the source contents of all the training videos are distinct from those of the test videos. As a result, 435 distorted videos from 29 sources were in the training set, and the other 105 distorted videos from 7 sources fell in the test set. In each of such splits, we trained each of the learning-based models following their original training method and evaluated them on the test set. For those metrics that do not need re-training, we simply tested them on the test set. Such random training-test data splitting was repeated 1000 times, and the medians and standard deviations of the SRCC and PLCC values were calculated and reported in Table~\ref{tab:cvfpsmethod}.

\begin{table*}[t]
\centering
\scriptsize
\caption{{Cross validation results of the tested no-reference quality assessment methods on the MOS values in the BVI-VFI dataset across different frame rates and interpolation methods. Each cell presents the median and standard deviation of the statistical measures obtained over 1000 random train-test splits. In each column, the best and the second best models are \textbf{boldfaced} and \underline{underlined}.}}
\label{tab:cvfpsmethod_nr}
\resizebox{\linewidth}{!}{\begin{tabular}{c||c|c||c|c||c|c||c|c||c|c||c|c}
\toprule
  & \multicolumn{2}{c||}{30fps} & \multicolumn{2}{c||}{60fps} & \multicolumn{2}{c||}{120fps} & \multicolumn{2}{c||}{non-DL} & \multicolumn{2}{c||}{DL} & \multicolumn{2}{c}{Overall}\\
\midrule
Model & SRCC & PLCC & SRCC & PLCC & SRCC & PLCC & SRCC & PLCC & SRCC & PLCC & SRCC & PLCC  \\
\midrule 
BRISQUE     & 0.05 (0.19) & 0.29 (0.12) & 0.05 (0.19) & 0.30 (0.11) & 0.11 (0.20) & 0.25 (0.11) & 0.16 (0.13) & 0.28 (0.12) & 0.08 (0.15) & 0.26 (0.11) & 0.02 (0.12) & 0.21 (0.08) \\
NIQE        & 0.02 (0.19) & 0.30 (0.12) & 0.05 (0.18) & 0.30 (0.11) & 0.03 (0.22) & 0.28 (0.09) & 0.19 (0.13) & 0.32 (0.11) & 0.02 (0.13) & 0.27 (0.09) & 0.04 (0.12) & 0.22 (0.08) \\
deepIQA\_NR & 0.19 (0.17) & 0.32 (0.12) & 0.16 (0.18) & 0.28 (0.10) & 0.07 (0.19) & 0.22 (0.10) & 0.03 (0.13) & 0.22 (0.10) & 0.15 (0.12) & 0.26 (0.10) & 0.09 (0.11) & 0.20 (0.08) \\
VIIDEO      & 0.10 (0.13) & 0.27 (0.14) & 0.11 (0.12) & 0.32 (0.15) & 0.20 (0.14) & 0.37 (0.14) & 0.05 (0.05) & 0.22 (0.08) & 0.29 (0.14) & 0.36 (0.11) & 0.20 (0.08) & 0.28 (0.10) \\
FastVQA     & 0.11 (0.17) & 0.36 (0.16) & 0.28 (0.15) & 0.33 (0.12) & 0.26 (0.18) & 0.38 (0.14) & 0.38 (0.10) & 0.43 (0.10) & 0.17 (0.12) & 0.27 (0.10) & 0.24 (0.11) & 0.28 (0.09) \\
RAPIQUE     & 0.23 (0.18) & 0.34 (0.19) & 0.32 (0.18) & 0.45 (0.14) & 0.31 (0.18) & 0.47 (0.19) & 0.29 (0.18) & 0.29 (0.21) & 0.30 (0.15) & 0.39 (0.13) & 0.25 (0.14) & 0.33 (0.12) \\
VIDEVAL     & \textbf{0.55 (0.15)} & \textbf{0.58 (0.11)} & 0.51 (0.13) & 0.55 (0.12) & \underline{0.53 (0.20)} & 0.61 (0.13) & \underline{0.47 (0.13)} & \underline{0.51 (0.12)} & 0.51 (0.14) & 0.56 (0.13) & 0.45 (0.13) & 0.49 (0.12) \\
ChipQA      & 0.49 (0.16) & \underline{0.54 (0.13)} & \underline{0.54 (0.17)} & \underline{0.57 (0.14)} & 0.43 (0.18) & 0.56 (0.15) & 0.27 (0.15) & 0.35 (0.12) & 0.56 (0.15) & \underline{0.60 (0.14)} & 0.47 (0.13) & 0.50 (0.12) \\
VBLIINDS    & \underline{0.52 (0.13)} & \textbf{0.58 (0.11)} & \textbf{0.60 (0.13)} & \textbf{0.67 (0.12)} & 0.49 (0.17) & \textbf{0.67 (0.14)} & 0.40 (0.12) & 0.44 (0.11) & \textbf{0.59 (0.11)} & \textbf{0.63 (0.11)} & \underline{0.48 (0.11)} & \underline{0.51 (0.11)} \\
TLVQM       & 0.40 (0.18) & 0.49 (0.14) & 0.49 (0.15) & 0.54 (0.13) & \textbf{0.53 (0.19)} & \underline{0.67 (0.15)} & 0.42 (0.14) & 0.47 (0.13) & \underline{0.59 (0.12)} & \textbf{0.63 (0.11)} & 0.48 (0.12) & \textbf{0.52 (0.12)} \\
FAVER       & 0.44 (0.15) & 0.50 (0.12) & 0.41 (0.16) & 0.50 (0.14) & 0.39 (0.17) & 0.54 (0.17) & \textbf{0.62 (0.13)} & \textbf{0.66 (0.11)} & 0.52 (0.14) & 0.58 (0.13) & \textbf{0.49 (0.13)} & \textbf{0.52 (0.12)} \\
\bottomrule	
\end{tabular}
}
\end{table*}

We can observe from Table~\ref{tab:cvfpsmethod} and \ref{tab:cvfpsmethod_nr} that FAST remained the best-performing metric in terms of SRCC, whereas the learning-based method ST-GREED now ranked second in terms of PLCC. This implies the effectiveness of the entropic difference features extracted in ST-GREED in capturing temporal artefacts. It is also noticeable that the learning-based no-reference model TLVQM achieved competitive performance compared to some of the full-reference models. The cross-validation results for different types of VFI algorithms are also presented in Table~\ref{tab:cvfpsmethod}, which shows that on DL-interpolated videos, FAST and ST-GREED achieved higher correlation compared to all the other models, while FRQM and FovVideoVDP outperforms others for the non-DL group.

\subsection{Summary}

From the evaluation results presented above, we observed that firstly, none of the 33 tested quality models showed consistent performance across different frame rates, {while an ideal perceptual VFI quality metric should be robust at different frame rates.} Secondly, the evaluated metrics were also not robust against various types of interpolation algorithms, with most models except FRQM performing better on sequences interpolated by DL-based VFI methods. Finally, our results showed that the motion saliency-based full reference VQA model, FAST, achieved the best overall performance on predicting the perceptual quality of interpolated videos, in particular on those interpolated by DL-based VFI methods. However, the overall SRCC value achieved by FAST was 0.70, which is far from satisfactory. Therefore, a much more accurate quality metric is urgently needed which can better predict the perceptual quality of interpolated videos.

\section{Conclusion and Future Work}\label{sec:conclusion}

In this paper, we presented a video quality database, BVI-VFI, which was employed to perform a subjective study on the perceptual quality of video frame interpolated content. The database comprises 108 reference and 540 distorted sequences with frame rates of 30, 60 and 120 fps. The distorted sequences were generated from 36 diverse source sequences at three spatial resolutions using five representative video frame interpolation methods. Based upon this database, we conducted a large scale psychophysical experiment to collect more than 10800 subjective opinion scores from 189 participants. The collected subjective data shows that the state-of-the-art deep learning (DL)-based VFI algorithms achieve better overall interpolation performance over simple frame averaging and repeating, in particular when the frame rate is low. As the frame rate increases, this improvement becomes less evident. 

Using this database, we have also benchmarked the performance of 33 image/video quality assessment metrics. The results indicate that all the three commonly used quality metrics for VFI, PSNR, SSIM and LPIPS, achieve poor correlation with subjective opinions scores on this database. Even the best-performing metric, FAST, was unsatisfactory with an overall SRCC value of only 0.70. This clearly highlights the urgent need for a new quality metric capable of accurately predicting the perceptual quality of VFI content. 

To the best of our knowledge, BVI-VFI is the first video quality database that contains purely VFI-induced artefacts, which allows us to thoroughly study the subjective quality of frame interpolated content and measure the performance of existing video quality metrics on assessing VFI quality. It provides a valuable resource for developing and validating new perceptual quality metrics for VFI, and it can be also employed to evaluate the performance of VFI algorithms. 
{Considering the unsatisfactory performance of existing image/video quality metrics on frame interpolated videos, future work should focus on developing a perceptual quality metric for VFI, which can accurately reflect the perceived quality of interpolated videos, and ideally be used as a loss function for training deep learning-based VFI methods.}

\footnotesize
\bibliographystyle{IEEEtran}
\bibliography{IEEEabrv,ref}

\begin{IEEEbiographynophoto}{Duolikun Danier} (Student Member, IEEE) received the B.Eng (First Class Honours) in Electronic and Electrical Engineering and M.Sc (Distinction) in Machine Learning from University College London in 2019 and 2020 respectively. He is currently pursuing a Ph.D degree with the Visual Information Laboratory (VILab), University of Bristol. His research interests include video frame interpolation, video compression and video quality assessment.
\end{IEEEbiographynophoto}

\begin{IEEEbiographynophoto}{Fan Zhang} (Member, IEEE) received the B.Sc. and M.Sc. degrees from Shanghai Jiao Tong University, Shanghai, China, in 2005 and 2008, respectively, and
the Ph.D. degree from the University of Bristol, Bristol, U.K., in 2012. He is currently a Lecturer of visual communications with the Department of Electrical and Electronic Engineering, University of Bristol. His research interests include deep video coding, video quality assessment, perceptual video compression, and creative technology.
\end{IEEEbiographynophoto}

\begin{IEEEbiographynophoto}{David R. Bull} (Fellow, IEEE) received the B.Sc. degree from the University of Exeter, Exeter, U.K., in 1980, the M.Sc. degree from the University of Manchester, Manchester, U.K., in 1983, and the Ph.D. degree
from the University of Cardiff, Cardiff, U.K., in 1988. He was previously a Systems Engineer with Rolls Royce, Bristol, U.K., and a Lecturer with the University of Wales, Cardiff, U.K. In 1993, he joined the University of Bristol, Bristol, U.K., and is currently its Chair of Signal Processing and the Director of Bristol Vision Institute. He is also the Director of the recently announced £46 m UKRI ‘MyWorld’ Strength in Places Programme. In 2001, he Co-Founded a university spin-off company, ProVision Communication Technologies Ltd., specialising in wireless video technology. He has authored more than 450 papers on the topics of image and video communications and analysis for wireless, Internet and broadcast applications, together with numerous patents, several of which have been exploited commercially. He is the author of three books, and has delivered numerous invited/keynote lectures and tutorials. He was the recipient of the two IET Premium Awards for his work. Dr. Bull is a Fellow of the Institution of Engineering and Technology.
\end{IEEEbiographynophoto}

\vfill

\end{document}